# In the realm of the hybrid brain: Human Brain and AI

Hoda Fares[1], Margherita Ronchini[1], Milad Zamani[1], Hooman Farkhani[1], Michela Chiappalone[2], Emre Neftci[3], and Farshad Moradi[1]

[1]ICELab, IbrAIn center, Department of electrical and Computer Engineering, Aarhus University, Aarhus, Denmark
[2]Department of Informatics, Bioengineering, Robotics and Systems Engineering (DIBRIS), University of Genova, Genova, Italy
[3] Peter Grünberg Institute, Forschungszentrum Jülich, Germany
**\*Correspondence:**
Corresponding Authors: Farshad Moradi and Hoda Fares
Email: moradi@ece.au.dk, hfares@ece.au.dk

Abstract

With the recent developments in neuroscience and engineering technology, it is now possible to record brain signals and decode them. In parallel, a growing number of stimulation methods are being utilized to modulate and influence brain activity. These advancements opened the door for innovative neurotechnologies that directly interface with the human brain. Although current brain-computer interface (BCI) technology is mainly focused on therapeutic outcomes, it already demonstrated its efficiency as an assistive and rehabilitative technology for patients with severe motor impairments. At the same time, artificial intelligence (AI) and machine learning (ML) have been recently used to understand the enormous multimodal neural data and to decode brain signals. Beyond this progress, interconnecting AI with advanced brain-computer interfaces in the form of implantable neurotechnologies grant unique possibilities for the diagnosis, prediction, and treatment of neurological and psychiatric disorders. In this context, we envision the development of a closed-loop intelligent, low-power, and miniaturized neural interface that uses brain-inspired techniques to process data from the brain; referred as Brain Inspired- Brain Computer Interface/Implant (BI-BCI). Such a neural interface would offer access to deeper brain regions for a better understanding of brain's functions, thus improving BCIs operative stability and system's efficiency. On one hand, brain inspired-AI algorithms represented by spiking neural networks (SNNs) would be used to interpret the multimodal neural signals in the BCI system. On the other hand, due to the ability of SNNs to capture the rich dynamics of biological neurons and to represent and integrate different information dimensions such as time, frequency, and phase, they would be used to model and encode complex information processing in the brain and to provide feedback to the users. In this paper, we provide an overview of different methods to interface with the brain and discuss the merger of AI with BCI, as BI-BCI systems for present and future applications.

## 1    Neurotechnology: The future game changer

Neurodegenerative disorders such as Parkinson's Disease (PD), Epilepsy, Multiple Sclerosis (MS), Alzheimer, and Dementia, are incurable and debilitating conditions caused by gradual damage or loss of the nervous system structure and function. They lead to cognitive, sensory, and motor dysfunction. As the world's population ages and life expectancy increases, age-related neurodegenerative diseases are becoming more prevalent and the risk of being affected by them is increasing dramatically [2].

Such diseases are responsible for the greatest economic burden and influence the lives of millions of people worldwide, for instance, in 2010, more than 179 million people in Europe were affected by brain disorders with an associated bill of around 800 billion euros [3]. According to the Global Burden of Disease Injuries, and Risk Factors Study (GBD) in 2016, neurological disorders were reported as the top leading causes of disability in the globe with 11.6% Disability-Adjusted Life Years (DALYs) (~276 million per year), and second leading cause of deaths after cardiovascular diseases with 16.5% of all deaths (~9 million) [3]. A general summary of the most common neurological disorders, their effects and economic burden is listed in Table 1. Currently, there is no effective therapeutics to cure such disorders, except for some traditional pharmaceutical drugs that could reduce the symptoms severity such as dopaminergic treatment for PD and movement disorders, cholinesterase for cognitive disorders, anti-inflammatory and analgesic for neuronal infections and pain, antipsychotic for dementia, etc.[4], [5]. To this end, a large body of research is focusing on establishing novel therapeutic tools and strategies by targeting the nervous system, as in the case of Deep Brain Stimulation (DBS) [6]–[8] , as alternative treatment to the traditional pharmaceutical approaches.

In the last 20 years, neurotechnologies aimed at interfacing the brain with machines and computers (i.e., BMI/BCI Brain Machine Interface/ Brain- Computer Interface) emerged as interesting tools to allow paralyzed people to communicate and interact with the external world. At the same time, they started to be used to investigate brain functions in different experimental conditions. Neurotechnologies or specifically neural interfaces cover any method or electronic device (e.g., electrodes, computers, robotic arm, etc.) that interface with the nervous system to monitor or alter neural activity. They can either record and decode the brain signals into control commands or electrically stimulate the brain to modulate its activity. Several neurotechnologies have been developed in the past few decades which proved to be useful for both assistive and rehabilitative applications, for example in cochlear implants for restoring hearing [9], retinal implants for restoring vision [10], [11] , and brain-computer interfaces (BCIs) for brain-controlled applications [12] . More recently, the advances in neuroscience and engineering technologies, along with the development of Artificial intelligence (AI) and machine learning-related techniques have allowed neurotechnologies to become intelligent for achieving a better performance [13]–[16].

Nowadays, researchers consider neurotechnologies to be the next game-changer for diagnosis, treatment and even prediction of neurological and psychiatric disorders [13], [17], [18]. However, most of the current ones are still limited to laboratories and their performance needs to be improved so that they can be used in real life scenarios [18].

**Table 1: Top leading neurodegenerative diseases based on world health organization (WHO) reports** [2], [3], [19]

| Neurodegenerative Diseases | Facts and Symptoms | Percentages and economic Burden |
|---|---|---|
| **Dementia and Alzheimer's disease** | - Dementia causes symptoms that affect memory, thinking, and social abilities severely enough to interfere with a patient's daily life.<br>- Memory Loss, planning difficulties, mood changes, personality changes, Confusion about time and place. | - Over 50 million people worldwide were living with dementia in 2020 (will double every 20 years).<br>- ~10 million new cases every year (one every 3 seconds).<br>- 7[th] leading cause of death.<br>- In 2018, it costed one trillion USD (it will be around two trillion by 2030). |
| **Parkinson's disease (PD)** | - Rigidity, postural disturbance, rest tremor, slow movement, anosmia in early stages. | - 10 million patients affected globally (1.5x more likely men than women)<br>- The prevalence ranges from 41 per 100,000 among people in their thirties to more than 1,900 per 100,000 among those who are over 80.<br>- In 2016, it caused 3.2 million DALYs and 211.96 deaths.<br>- In 2021, in the USA it costed 51.9 billion USD (double previous estimates). |
| **Multiple sclerosis (MS)** | - Multiple sclerosis is a disease with unpredictable symptoms that can also vary | - ~around 2.8 million people worldwide registered.<br>- Women four times more likely to have MS than men. |

| | | in intensity. Different symptoms can manifest during relapses or attacks.<br>- Pain from spasticity, impaired ambulation, depression, cognitive impairment, ataxia, and tremor. | - Mean costs of MS ~37100 USD annually per patient with moderate disease in EU. |
|---|---|---|---|
| Epilepsy | | - Recurrent seizures, which are brief episodes of involuntary movement that may involve a part of the body or the entire body and are sometimes accompanied by loss of consciousness and control of bowel or bladder function. | - ~ 50 million people worldwide have epilepsy (most common neurological disease globally).<br>- Up to 70% of people living with epilepsy could be seizure-free if properly diagnosed and treated.<br>- Premature death in people with epilepsy is up to three times higher than for the general population.<br>- The estimated proportion of the general population with active epilepsy (i.e., continuing seizures or with the need for treatment) at a given time is between 4 and 10 per 1000 people. |

The human brain is an extremely complex system and thus an active area of research for neuroscientists and clinicians in designing treatments of (non)-neurological disorders, and for engineers for its capability to perform complex tasks by means of ultra-energy-efficient computing. Therefore, knowing how the brain works can be beneficial for both communities. In support of this, vast resources have been assigned to study, model and map the brain and its fundamental mechanisms along with neurotechnology development. The BRAIN initiative in 2013 supported by US government, Brain/MINDS (Brain Mapping by Integrated Neurotechnologies for Disease Studies) project launched in 2014 by Japan, and the Human Brain Project (HBP) funded by the European commission (($703 million) are a few examples. In Dec 2020, the HBP launched its EBRAINS platform, which grants access to datasets and digital tools for analysis and experiment conduction [20]. Due to its high potential for treating neurological disorders, neurotechnology research has significantly became an interesting attraction for industry in the past decade (e.g., © Neuralink [21], ©Paradromics [22], ©Synchron [23], ©Blackrock Neurotech [24], ©Neurable [25], ©Thync [26], ©Medtronic [27], ©kernel [28], etc.). For instance, ©NeuroPace developed a brain-responsive neurostimulator called RNS System for treating adults with drug resistant focal epilepsy, using feature thresholding over 4 channels to detect seizures [29], [30]. Also, ©Medtronic developed Percept PC DBS system by that implements 4 Channels [31] and ©Neuralink developed a 1024 channel closed-loop Brain Machine Interface (BMI) implantable chip integrating neural recoding, spike detection circuitry while using external devices for motor intention decoding [32].

In this perspective paper, we aim to provide an overview of the current state of applied research in neurotechnology including neural interfaces, neuroprostheses, BMIs/BCIs and surmise about future developments and clinical application that may arise from it. Furthermore, we will delve into the co-integration of AI-based processing and neural interfaces. The paper is structured as follows: section 2 presents a comprehensive synopsis about methodologies used to extract and transmit information from and to the brain. Section 3 reviews the new generation of AI systems called spiking neural networks or spiking neuromorphic architectures. Section 4 discusses the use of SNNs in neurotechnology. Finally, the last section offers our closing remarks and our vision about merging brain-inspired computing with neural interfaces to achieve Brain Inspired- Brain Computer Interfaces/Implants (BI-BCIs) that would be the new generation of low-power, smart, and miniaturized therapeutic devices for a wide range of neurological and psychiatric disorders.

In this context of this opinion, we use the "BCI" term for any technology that communicates directly with the brain, either to extract information from it, or to feed information into it by means of brain stimulation. Also, BCIs, or Brain-Computer Interfaces, are also commonly referred to as Brain-Machine Interfaces and Neural Interfaces.

## 2 Neural Interfaces: The Brain Editors

### 2.1 Connecting Brain to Computers: Neural interfaces History and outlook

BCI is a general term for any technology that directly communicates with the brain and extracts information from it either by observing its unperturbed electrical signal or by eliciting a measurable neural response (evoked potential) through sensory stimulation [8][9] (Fig. 1). This terminology was introduced by Vidal in 1970s, who first attempted to create a system capable of translating EEG signals (i.e., Electroencephalography non-invasive method that records neural activity from the scalp) into computer control signals [33]. Research applications of BCI technology have evolved substantially over the past two decades [34]. Research and development of BCI technologies was boosted by the technological advancement of microelectrode and single neuron recordings technologies, both in rodents [35] and non-human primates [36], [37]. Researchers used electrode arrays and implanted them in the parietal or motor cortex of patients with severe paralysis and tetraplegia to perform skilled motor movements with a robotic arm [38]–[41]. BCIs could be used to restore (e.g., unlock patients with locked-in syndrome), replace (e.g., BCI-controlled neuroprosthesis), enhance (e.g., user experience enhancement through computer games), supplement (e.g., VR, virtual reality, and AR, augmented reality, glasses), improve (e.g., lower limb rehabilitation after stroke), and as a research tool (e.g., coding, and decoding brain activity with real-time feedback) [34], [42], [43]. BCIs can record and decode cortical activity while performing or imagining performing a task. The neural signals related to the intended movement can be transformed into visual [44], auditory [45]–[47], or haptic feedback of the movement [34], [48]. Figure. 1 illustrates the generalized schematic for BCIs and common state-of-art (SoA) applications.

BCIs can be classified based on the way they interact with the brain. First category is the **active BCIs** that either use the users consciously induced brain activity such as Motor Imagery (MI) [46], [49], [50] or induced brain activity by external stimuli (e.g., visual, auditory, or somatosensory stimuli) [34], [49], [51], [52] . While the second category called **passive BCIs** decode brain's unconscious psychological states and do not require an active participation from the user [53], [54]. They have been used to monitor users' cognitive states such as intentions, emotional states, situational interpretations [45], [54], and drowsiness [55], [56].

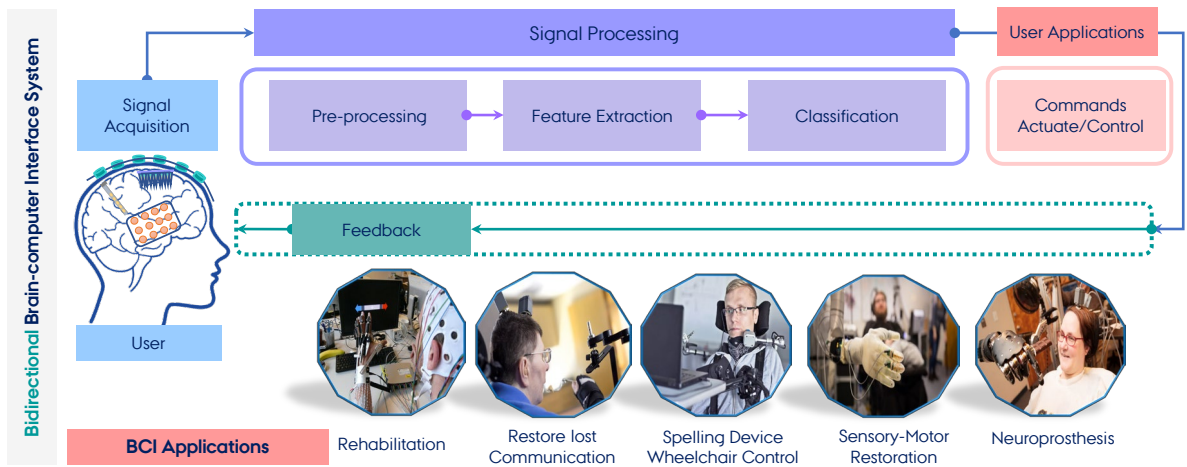

**Figure 1 Generalized schematic for Bidirectional brain computer Interface (BBCI)**. The produced Brain signals are recorded from the scalp, cortical surface or from within the brain by electrodes. These signals are processed to extract the correlated features with user's intentions. The extracted features are translated into commands to control / actuate a wide range of applications. (Could be used to control devices, artificial limbs, or obtain knowledge of user's intentions). Then, sensory information is fed back to the user either invasively or non-invasively.

## 2.2 From the brain to external devices: Recording and Decoding

### 2.2.1 *Recording techniques*

Several techniques are employed to gather metabolic and electrophysiological signals from the brain, each offering distinct temporal and spatial resolution. Intracellular recordings measure the voltage across the cell membrane of a single neuron by placing electrodes inside and outside the membrane, also they capture the sub-threshold variations from resting potential. Extracellular recordings capture the summation of signals by nearby neurons, and they provide lower neural signal amplitudes in comparison to intracellular recordings, but they cover larger neural areas [57]. Based on electrode location, extracellular recordings techniques can be categorized into either **invasive** or **non-invasive** methods [12], [34]. The most relevant non-invasive and invasive recording techniques respectively and their BCI applicability are summarized in Table 2 and 3.

*Non-invasive* methods used for neural signals recording comprise **electroencephalography** (EEG) [31], **magnetoencephalography** (MEG) [58], [59], and metabolic signals recorded either by **functional near-infrared spectroscopy** (fNIRS) [52], [60], or **functional magnetic resonance imaging** (fMRI) [61]. EEG is the most employed technique in clinical setups for diagnosis purposes due to its non-invasive nature and ease of use. However, this technique captures only collective information from the top cortical layers of the brain, and it suffers from low spatial resolution, poor contact between the electrode and the scalp and low signal quality [62]. Wet EEG electrodes are typically made of metals and gels and mounted in elastic caps to enhance the signal quality. Dry electrodes (i.e., without gels) are more favorable and have a comparable performance with wet electrodes, yet they are less robust to moving artifacts and show higher electrode-tissue impedance [63], [64]. To address these challenges, active electrodes with integrated preamplifiers have been developed, also new materials has been used to design EEG electrodes such as polymer foam electrodes, soft conductive textiles electrodes, etc. fMRI, as another non-invasive the uses blood-oxygen-level-dependent (BOLD) signals that reflect changes in deoxyhemoglobin driven by localized changes in brain blood flow and blood oxygenation, which are coupled to underlying neuronal activity by a process termed neurovascular coupling. But it is more expensive method in clinical setups, offers a much higher spatial resolution metabolic signals (~1mm) and is more sensitive to subcortical regions than electrophysiological signals. fMRI is heavily used in cognitive research [65]. Researchers were able to reconstruct perceived visual images, just by analyzing fMRI signals collected from visual cortex [66]. With similar approaches, it has been demonstrated that patients in a vegetative or minimally conscious state understand and respond to instructions [67], [68] (Table 2).

Alternatively, *invasive* methods such as **electrocorticography** (ECoG or µEoG) [57], [69], [70] and **intracortical recordings** (IR) [71]–[73] provide higher signal-noise ratio and higher-frequency signal bands, as well as better localization of brain activity as they enable more direct interaction with the brain. For instance, flexible µECoG electrodes have pushed spatial resolution down to 1mm even sub-mm range [74] unlike conventional ECoG electrodes which have a pitch of around 1 cm [75]. Transistor multiplexed ECoG arrays managed to increase the electrode density and channel count and reduce the area for routing wires [76], [77]. Both ECoG and µECoG are used in preclinical and clinical research settings [78], [79]. Lately, bundled arrays of microwires were used to interface with up to 1 million neurons through a neural input-output bus (NIOB) funded by DARPA [80]. Despite the improved performance in spike sorting and mechanical stability offered by microwires, this method still faces challenges in signal attenuation and cross talk. To overcome these limitations, silicon-based needle shaped microelectrodes enabling multisite recording were proposed [81]–[83].

Yet, rigid probes may lead to tissue damage and inflammation, which may degrade the recorded neural signal [84] Polymer-based flexible electrodes ensure tight and conformable geometries which make them more suitable for chronic long-term implants [85] Despite their flexibility, their insertion can be challenging as they are prone to bending and deflecting. To address this issue, a robotic insertion method for polymer electrodes has been introduced [86], also carbon fiber high density arrays have been developed [87]. Flexible intra-fascicular electrodes such as LIFE (longitudinally intra-fascicular electrodes) [88] and TIME (Transverse intra-fascicular multi-channel electrodes) [89] have been used on peripheral nervous system and they are showing a promising future for neuroprosthetic applications [90]. The latest advances in CMOS enabled the fabrication of high-density micro-electrode arrays/probes, allowing the simultaneous recording from hundreds of neurons in humans and monkeys [78]. Unlike conventional electrodes, the novel electrode technologies based on organic material, multifunctional flexible polymer fibers and meshes, offer increased spatial integration (e.g., Neuropixles [91] and NeuroGrid [92]), long-term temporal stability (e.g., mesh electronics), and improved biocompatibility [93]. Alternatively, the advances in optical cellular imaging have shifted the neural recording numbers by several orders of magnitude. For example, Kim et al. used such technique to optically record the activity of up to one million neurons in a single rodent [94]. While numerous studies utilize transgenic mice with constitutive GECI (genetically encoded calcium indicators) expression, modified viruses have been employed to introduce a genetic construct inducing expression in non-transgenic animals. Likewise, viral delivery methods have been used in humans for gene therapy. Despite the challenges posed by such technologies (e.g., optical access to neural tissues, need of high precision microscopes), the combination of these and other methods holds the potential to facilitate the recording of millions of neurons in humans in the future [95] (Table 3). Figure 2 presents the commonly used neural recording electrodes to extract neural signals.

A major application of invasive BCIs is assisting paralyzed people, for example by implanting subdural electrodes over the cortex [79]. Higher bandwidths could be achieved by implanting electrode arrays deeper into different areas of the cortex to record neuronal spiking activity [38], [96]. The recorded spiking activity can be used to decode user's intention to move hand in certain direction or control a robotic arm for skilled movements [38], [96]. Bouton et al. demonstrated that a quadriplegic patient with a 96-electrode array implanted in the motor cortex's hand area managed to utilize cortical signals to electrically activate/stimulate muscles in his paralyzed forearm. This process enabled him to execute six distinct movements involving the wrist and hand.[97]. Also, Lajoie, G., et al. used bidirectional BCI to artificially induce task-related neuroplasticity [98]. In summary, such BCI methods mainly target brain regions that represent low-level and high-level motor commands to restore motor control in patients. In addition, these methods could be used to read the neural codes associated with perceptions [99], attention [100], and decisions [101]. Recently, invasive BCIs enabled researchers to decode the thoughts of a person from the activity of their "concept cells", which were discovered for the first time in the temporal lobe of patients implanted with electrodes to identify brain regions responsible for intractable epilepsy [102]. They also found out that these cells represent abstract concepts about specific places and people, and they get activated when the person thinks, sees, or retrieves memories about such concepts [103], [104]. Researchers would likely be able to monitor thoughts with greater clarity if large numbers of concept neurons were recorded at one time, given that this study recorded only a single cell or a few neurons at one time. In fact, extracting complex intentions could revolutionize communication BCIs used to decode the intentions of locked-in syndrome patients (ALS) [12], [105] (Table 3).

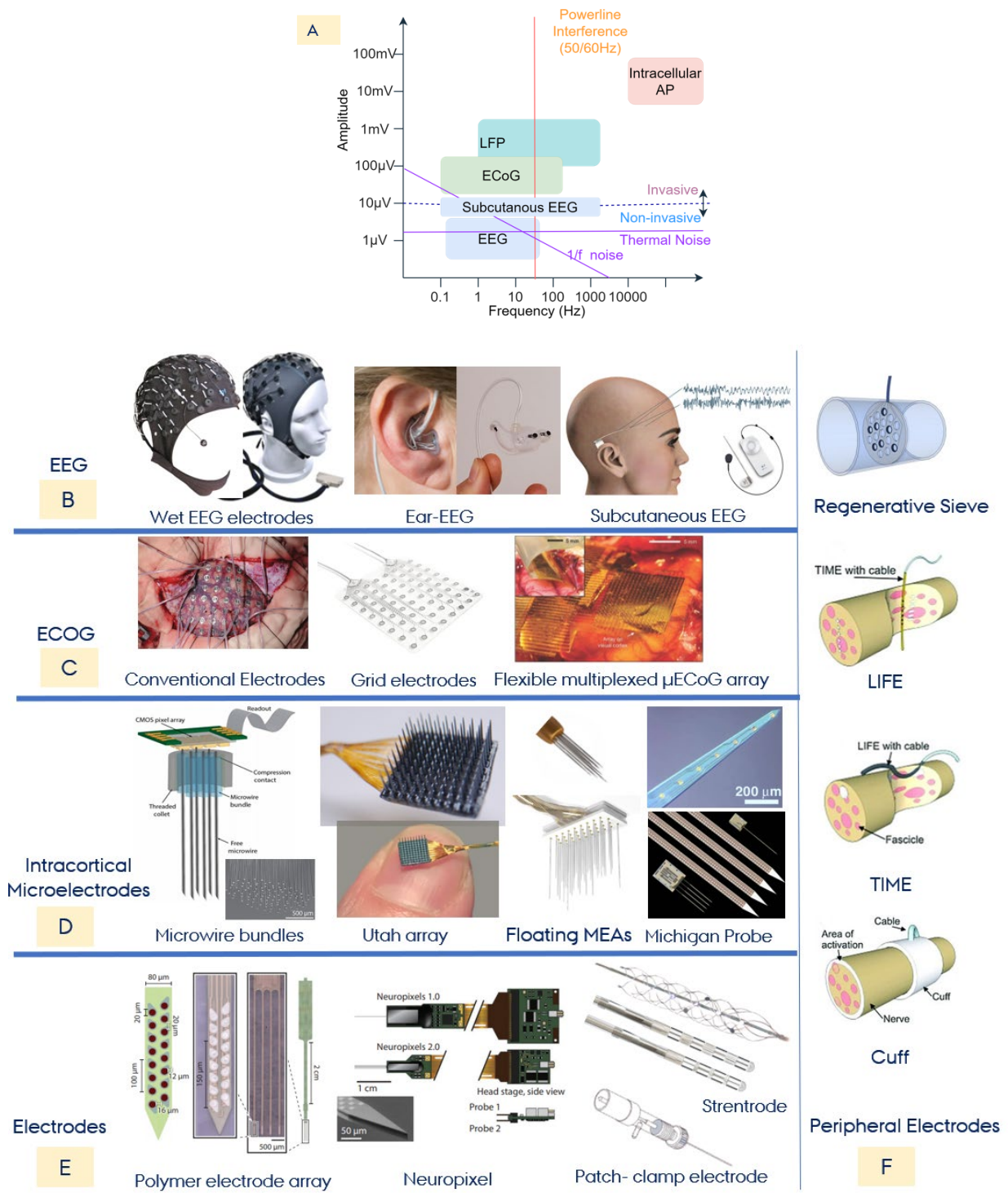

**Figure 2: Neural signals and Neural electrodes. A: Common extracted neural signals**, the vertical and horizontal solid lines represent the amplitude (V) and frequency (Hz) range of neural extracted neural signals. The purple lines represent the thermal and 1/f noise from the circuit itself, while the red line represents the external powerline interference which may degrade then neural signal. The shaded areas represent the span of each recording technique. AP, action potential; LFP, local field potential; ECoG, electrocorticography; EEG, electroencephalography [351] . **B-F: Commonly used recording neural electrodes** to interface with the brain and the peripheral nervous system. **B- Typical non-invasive electrodes** wet EEG electrodes (128- Channels Quick cap by Neuvo), Ear EEG [352], subcutaneous EEG [353]. **C- Conventional invasive non-penetrating electrodes for ECoG recording, e.g.,** Flexible grid ECoG electrode array [354], [355], transistor- multiplexed μECoG array [77][77], [356]. **D- Intracortical penetrating needle shaped microelectrodes for recording** e.g., microwire bundles integrated with CMOS chips for bundles [80], Utah array by Blackrock neurotech [82], floating microelectrodes array (MEA), planar Michigan probe [83]. **E- Electrodes:** High density polymer-based electrode array for electrophysiological recordings [85], Neuropixel for long term Brian recoding [91], [138] , Stentrode - minimally invasive endovascular arrays for chronic recordings [130], patch-clamp electrode. **F- Peripheral nerve electrodes** [357]**,** e.g., TIME [89] , LIFE [88]

**Table 2: Non-Invasive Recording Methodologies and neurotechnological applications**

| Techniques Definition | Type | Resolution Temporal | Resolution Spatial | Portable | Application | Publication |
|---|---|---|---|---|---|---|
| **EEG** A method to record the electrical activity of the brain with electrodes placed on the scalp. It represents the macroscopic activity of the surface layer of the brain underneath. | Electrical | ~ 0.05 s | ~10mm | Yes | Epilepsy | [106] |
| | | | | | Motor learning & plasticity induction | [107] |
| | | | | | Multimodal BCI for psychological prediction (attention, motivation, memory load, fatigue) | [50], [108], [109] |
| | | | | | Motor recovery after stroke | [110] |
| | | | | | Transfer learning (Inter-subject BCI) | [111] |
| | | | | | Motor rehabilitation (Parkinson's disease) | [112] |
| **MEG** A neuroimaging method that uses SQUID to measure weak magnetic fields outside the head. It reflects the magnetic changes arising from cortical neural activity. | Magnetic | ~ 0.05 s | ~ 5 mm | No | Multimodal BCI for motor rehabilitation | [113] |
| | | | | | Assistive technology (BMI for paralysis) | [113] |
| | | | | | MEG-based brain–computer interface (BCI) | [58] |
| | | | | | Real time control of neuroprosthetic hand | [114] |
| **fMRI** Uses magnetic resonance imaging to detect local brain activity by measuring the changes in the BOLD signal. | Metabolic | ~ 1 s | ~ 1 mm | No | Multimodal BCI for Psychological prediction (attention, motivation, memory load, fatigue) | [108], [115] |
| | | | | | Brain activity Regulation | [44] |
| | | | | | Neuroplasticity Rehabilitation of attention deficit | [116] |
| **fNIRS** Measures the concentration variation of oxygenated and deoxygenated hemoglobin respectively HbO and HbR in brain tissue depending on changes of the exiting photon intensity and incident -photon intensity, then characterizes the local neural activity | Metabolic | ~ 1 s | ~ 5 mm | No | Multimodal BCI for Gait rehabilitation | [60] |
| | | | | | Robotic control | [117] |
| | | | | | Motor Rehabilitation | [118] |
| | | | | | Intention detection | [119] |
| **PET** A imaging procedure. It is a combination of nuclear medicine and biochemical analysis. PET studies evaluate the metabolism of a particular organ or tissue, so that information about the physiology (functionality) and anatomy (structure) of the organ or tissue is evaluated, as well as its biochemical properties. | Metabolic | ~ 1-2 min | ~ 4 mm | No | Limit potential Diagnostic tools of neurological disorders such as dementia, epilepsy | [120] |
| **MRI** An imaging technique that combines strong magnetic fields, electrical gradients, and radio waves to measure the biological tissue composition and derive its structure. | Hybrid (magnetic + electrical+ radio waves) | ~ 1 s | ~ 1 mm | No | Locked In syndrome. (Somatosensory Rehabilitation in stroke and phantom limb pain) | [121] |
| | | | | | Neurofeedback | [65] |

*Abbreviations: EEG (Electroencephalography), MEG (Magnetoencephalography), fMRI (Functional Magnetic Resonance Imaging), fNIRS (Functional near-infrared spectroscopy), MRI (Magnetic Resonance Imagining), BMI (Brain machine interface), BOLD (blood oxygenation level dependent, SQUID (superconducting quantum interference device)*

**Table 3: Invasive Recording Methodologies and neurotechnological applications**

| Techniques Definition | Type | Resolution | | Portable | Application | Publication |
|---|---|---|---|---|---|---|
| | | Temporal | Spatial | | | |
| **ECoG** Uses flexible subdural grid or strip electrodes that directly interface with the brain surface to measure cortical activity. It exhibits higher spatio-temporal resolution than EEG, larger bandwidth, and excellent signal-to-noise ratio (SNRs) | Electrical | ~ 0.003 s | ~ 1mm (subdural ECoG ~1.25mm) (epidural ECoG ~1.4 mm) | Yes | Assistive technology neuroprosthetic control | [69], [122] |
| | | | | | Motor learning and rehabilitation | [70] |
| | | | | | Intracranial BCI for severely motor-impaired patients | [79], [123] |
| | | | | | Neural decoding and encoding (speech synthesis, text translation) | [124], [125] |
| **IRI** (Intracortical recording interfaces) are critical components of BCIs and consist of arrays of penetrating electrodes that are implanted into the motor cortex of the brain. | Electrical | ~ 0.003 s | LFP ~ 0.5mm MUA ~ 0.1 mm SUA ~ 0.05 mm) | Yes | BCI for ALS, Diagnostics, Therapeutic Treatments | [79], [126] |
| | | | | | Neural decoding and stimulation, robotic prosthetics control | [127], [128] |
| | | | | | Assistive technologies and clinical BCI | [40], [72] |
| | | | | | Restoration of mobility and communication, cursor control, epilepsy monitoring | [40], [72] |
| **Stentrodes** | Catheter angiography guided implantation | - | - | Yes | Minimally invasive BCIs (ongoing Human trails) Motor neuroprosthesis | [129], [130] |
| **Neural Dusts** | - | - | - | | Optogenetics (biomedical/ Brain implants) Parkinson (STARDUST) | [131]–[137] |
| **Neuropixel** | High density probe for stable long-term brain recording | - | - | - | Recording thousands of individual neurons in living brain (Freely moving animals and recently tested in humans) | [91], [138]–[140] |
| **Neural lace** | - | - | - | - | | |

*Abbreviations: ECoG (electrocorticography), PET (Positron emission tomography), IRI (Intracortical recording interfaces), iMEA, (intracortical microelectrode array), ALS (Amyotrophic lateral sclerosis)*

## 2.2.2 Decoding neural signals for BCI applications

Neural decoding methods focus on extracting information from neural activity, either to reconstruct the event or stimulus that generated it, or to predict the actions that it might elicit. For classical BCI systems, a decoder typically includes three main procedures: *signal preprocessing, feature extraction,* and *pattern classification* as shown in Figure 1. Signal preprocessing such as artifact reduction methods aim to eliminate the noise from the recorded neural signals and extract useful /relevant components. Feature extraction entails identifying the most relevant features linked to the subject's intention from the neural activity. These features could include spectral power, event-related potentials (ERPs), firing rates of individual neurons etc. Pattern classification, on the other hand, differentiates the diverse classes of user's intentions based on the extracted features. Among these processes, pattern classification stands as the pivotal algorithm in decoding brain signals [79], [96], [141] .Various decoding approaches have been explored for neural decoding, including linear models, Kalman filters [142], state-space models , Bayesian decoding , information theory [143] and more advanced techniques based on artificial neural networks like deep learning models such as recurrent neural networks (RNNs) or convolutional neural networks (CNNs) [144] to produce accurate predictions using spikes [145] or intracortical LFPs [145], [146]. Several attempts have been made to use non-linear methods for neural decoding like particle filter and unscented Kalman filter. Most of the non-invasive BCIs are based on classification of mental states rather than on decoding Kinematic parameters as is it the case for invasive BCIs. F Lotte et al. present an updated review of EEG-based BCIs classification algorithms and distribute them over four main classes: adaptive classifiers, matrix and tensor classifiers, transfer learning and deep learning, plus a few other classifiers [1].

AI and ML methods have been heavily used in brain signal decoding and classification [100], [101]. Despite the success of traditional linear decoding approaches, recent advances in machine learning have led to even more effective neural network decoding approaches [144]. For example, Deep Neural Networks (DNNs) have been used to interpret activity patterns elicited by visual stimuli and predict it with a remarkable precision around 90%. Mostly, lower layers in DNNs represent simple features while higher layers represent conceptual information which in turn will be mapped to different brain regions [146]. Therefore, the development of novel technologies for recording and decoding neural activity could enable researchers to read the human brain and expose subject's intentions, preferences, thoughts, and emotions. Nevertheless, this raises issues of privacy, security, and mental ownership as well.

## 2.3 From external device to the Brain: Encoding and Stimulating

### 2.3.1 Stimulation techniques

Several technologies have been developed to influence neural activity Modulating neural activity could be done either by brain stimulation (i.e., by applying physical energy e.g., electrical, magnetic, acoustic directly to brain regions) or by neurofeedback (i.e., without direct brain stimulation). Several non-invasive methods has been used to such as **transcranial magnetic stimulation** (TMS) [147], **transcranial direct current stimulation** (tDCS) [148], **transcranial electrical stimulus** (TES) [148], **focused ultrasound stimulation** (FUS), and transcranial focused ultrasound (tFUS) [149], as well, invasive methods such as DBS [6], surface cortical and intracortical stimulation [71], all belong to direct brain stimulation technologies. Non-invasive methods have limited spatial resolution compared to the invasive methods, and their effects on the brain activity are still not clarified.

Conventional invasive stimulation technologies are classified into two broad classes according to the location of implantable electrodes 1) cortical stimulation or intracortical micro-stimulation (ICMS) which focus on the cortex and 2) deep brain stimulation (DBS) that focus on deep brain tissue. ICMS has been applied for elicit sensory feedback with high spatiotemporal accuracy [150]–[152], while DBS has been used to treat neurological and neuropsychiatric disorders through regulating or controlling an internal brain state instead of controlling the movement of external actuator [153], [154].Table 4 depicts the developed brain stimulation technologies and their BCI applicability.

### 2.3.2 Encoding neural/ Brain signals for BCI applications

Researchers have been investigating methods for encoding and delivering information in a biomimetic or artificial manner through stimulation to neuronal networks within the brain and other parts of the nervous system. These efforts encompass various sensory modalities, including auditory, visual, proprioceptive, and tactile perception.

As regards the restoration of sensory modalities, the first attempts were done with cochlear implants, that electrically stimulate the inner ear and meanwhile became the common stream treatment for deafness [9]. Also, restoring vision in patients with damaged retina has been realized, either by electrical stimulation of living cells through retinal chips [155] or by chemical or optogenetic stimulation that re-establish retina's light sensitivity [156], [157]. Scientists implanted stimulation electrodes in the visual cortex to activate the neurons nearby (tens of micrometers around the electrode) then evoke visual percepts in the brain. Such a visual cortical prosthesis uses a camera mounted on a pair of glasses to capture visual information. The information is processed by a BCI system and translated into stimulation patterns that are delivered to the cortex through microelectrodes. Each stimulation electrode evokes a phosphene (i.e., the percept of light) to later build a visual image (pixel by pixel). Stimulating higher visual cortical areas elicits more detailed percepts like motion or depth, as well as shapes of faces [158] and spatial layout of scenes [159] by stimulating the temporal cortex. Further progress has been done on restoring sense of touch to upper limb prostheses [160], [161] and discriminable tactile information has been elicited by stimulating the somatosensory cortex of non-human primates [162].Yet, the main challenge still is the need of detailed knowledge of how complex thoughts are encoded in brain activity as well as the technical capability to evoke activity patterns required to directly communicate information to higher order cortices, such as the parietal and temporal cortex.

As regards the restoration of motor capabilities, many advances have been made as well. For instance, spinal cord implants allow patients to relieve chronic pain by sending low level electrical impulses directly into the spinal cord [163], [164]. Scholars have demonstrated the therapeutic potential of stimulation techniques in treating neurological and neuropsychiatric disorders. For instance, in Parkinson, DBS of the subthalamic nucleus eliminates essential tremors. Also, it has been clinically tested for treating psychiatric disorders including depression and obsessive-compulsive disorder (OCD) [165], [166]. However, the currently used stimulation patterns are still considered rudimentary due to electrodes' large contact point surface areas. Future technological advancement is needed to augment the electrodes' precision and maximize the therapeutic output while minimizing risks [167].

Optogenetics, a newly developed invasive method for modulating neuron's activity through light and genetic modification. This technique involves the implantation of an optical fiber to deliver light to specific areas within the brain, which is known for its highly scattering tissue. It has enabled neuroscientists to explore neural microcircuits in mice that were previously uncharted. It mainly depends on bulky photo microscopes, which make it impractical for clinical. Till now, it is only utilized in animal studies [135], [168], [169]. Researchers believe that increasing optical methods precision, quality, and bandwidth could shift the brain stimulation techniques therapeutic potential significantly by utilizing its ability in activating specific subsets of neurons [170], [171]. Researchers demonstrated that the activation of higher brain regions by stimulation may employ a change the subject's behavior and emotional state [172]. For instance, A recent study shows that optogenetic stimulation could steer complex behaviors such as attacking a prey [173], eating, drinking and sexual behavior in rodents and other animals [174]. Also, functions like memory and attention can be influenced [175] . For instance, the stimulation of parietal cortex or frontal eye field modulates visuo-spatial attention [176], [177], likewise electrical stimulation of the temporal lobes induces vivid recollection of memories of a patient's past [178] . Several BCI applications instead target behavior change to avoid certain diseases, for example Lipsman et al. used for the treatment of refractory anorexia nervous [179]. Researchers could also suppress or reinforce certain behavior by inhibiting or activating set of neurons (e.g., activation of dopamine neurons in the ventral tegmental, activation of circuits that mediate aversion in the lateral habenula) [180], [181].

Despite its high potential, optogenetic stimulation is still not ideal for neuromodulation in humans. Several efforts have been made to develop a non-invasive cell type selective stimulation method. Researchers investigated alternative stimulation methods that are more sensitive than optogenetic mediators and less penetrating than light. Sonogenetics - a noninvasive stimulation method- employs ultrasound waves to activate exogenous ultrasound-sensitive mediators or actuators within specific neurons or cell types, allowing for the manipulation and control of their activity. Its therapeutic ability has been tested in animals. The results show that sonogenetic stimulation of the visual cortex of the rodents induces a behavioral response associated with light perception [182], [183]. This discovery could be used to restore vision in blind people. Also, it has been clinical tested for treatment of neurodegenerative diseases such as Parkinson [182], [183].

Figure 3 presents the SoA of existing invasive and non-invasive neural interfaces for stimulation and recording.

## 2.4 From Open-loop to Closed-loop intelligent

Classical BCIs are one-way feedforward systems that mostly used in control and communication applications [12]. Early BCIs generally rely on different types of electroencephalograms signals such as, **slow cortical potentials** (SCP) [184], **sensorimotor rhythms** (SMRs) [185], **P300 event-related potentials** (ERPs), and **steady-state visual evoked potentials** (SSVEPs) [186], to realize brain–computer communication and they feedback the results on a computer screen (as presented in Table 5 which summarizes the different EEG-based open-loop BCIs and their applications).To improve the performance of such systems, new BCIs paradigms have emerged and introduced such as hybrid neural interfaces/BCIs that either rely on combining other physiological signals with classical EEG-based BCIs or integration of. In parallel, various open-loop invasive BCIs have been developed based on spikes and intracortical LFP signals have been developed with the same goal of controlling external devices/actuators and modulating neural activities based on external signals. Such classical systems enable generation, acquisition and decoding of brain signals. However, when the feedback is used to alter the neural or behavioral activities as well as the external devices, they become closed-loop brain computer interaction systems. These latter systems are based on a brain-in-the-loop control paradigm and combine both decoding and encoding pathways to form a *bidirectional brain computer interface (BBCI)* [187]. Hence, as illustrated in Figure 4-A, in BBCI system (i.e., closed-loop there are two complementary routes: from the brain to the device (control route in purple which is alike classical BCI system) and from the device to the brain (feedback route in green). BBCIs could be extremely influential, as they would enable *real-time* and *co-adaptive feedback* between the brain and the external devices (i.e., the encoding stimulation could be conditioned on the current state of the brain). For instance, neuroprosthesis is an illustration of BBCI system that integrates both motor outputs and sensory input. It not only reads neural activity from the primary motor cortex and translates it to movement commands but also provides somatosensory feedback, through sending external sensory information from the neuroprosthesis back to the somatosensory cortex through by electrical micro-stimulation [71], [152]. This stimulation improved control's precision and provided feedback about touched objects. Also, BBCIs aim to change the brain's state to augment human performance (e.g., modulation of brain activity for treating neurological disorders or boost mental capacity of healthy subjects) [187].

Closed- loop DBS is another example of bidirectional neurotech system, where stimulation relies on simultaneous monitoring of brain activity. In Parkinsonian patients, the recorded LFP oscillations of the subthalamic nucleus provide real-time information about their clinical state, this in turn can be utilized to control the electrical stimulation settings and limit its application to specific time slots when it is needed. Limiting the application time expands the battery life and decreases the occurrence of potential adverse effects. Similarly, closed-loop brain stimulation has been used to detect early epileptic seizures then interrupt them by electrically stimulating the anterior thalamus or deep cerebellum nuclei before the seizures progress [188]. Lately, electrical stimulation was applied to the temporal cortex to improve memory encoding in users exhibiting weak memory [175]. Based on what was priorly discussed, to realize robust and effective closed-loop neurotechnologies, it is necessary to have three key elements: *neuromodulation, real-time co-adaptive interaction,* and *close-loop construction,* so the brain adapts to external stimulus and continuously optimizes the task execution, and the decoders/actuators learn to tailor their responses according to the changes in neural activity and user's intentions, as illustrated in Figure 4-B. However, the major challenge is how to send the feedback directly to the brain. Also, there are still several problems that need to be resolved, such as implanted electrodes longevity, electrical stimulation artifacts, electrochemical safety of electrode tissue interface, etc. [189].

Concurrently, as BCI technology advances swiftly alongside the progress in AI and ML fields, scientists have initiated efforts to foster collaborative synergies between these domains, giving rise to what is commonly referred to as "intelligent BCIs". On the one hand, AI can be used for interpreting the wide variety of the recorded multimodal neural signals in BCIs, on the other hand, AI-based intelligent devices can encode and feed information back to the users. This synergy could improve the performance and expand the applicability of BCI systems. Moreover, they postulate that AI can complement human cognitive abilities, which in turn will enable the development of *hybrid intelligence* driven by the direct interface with the brain. To realize close-loop intelligent BCIs, it is necessary to couple human cognitive ability to AI computing systems and exploit its fast operations and storage capabilities. Also, it should establish human-AI co-adaptive learning to ensure more adaptive dynamic and personalized interaction between the brain and the BCI system [190] as illustrated in Figure 4-B. In a similar vein, we envision the development of a closed-loop intelligent BCI system called BIBCI (Brain Inspired- Brain Computer Interface) that merges the latest iterations of AI with advanced BCI technologies (as detailed in section 4).

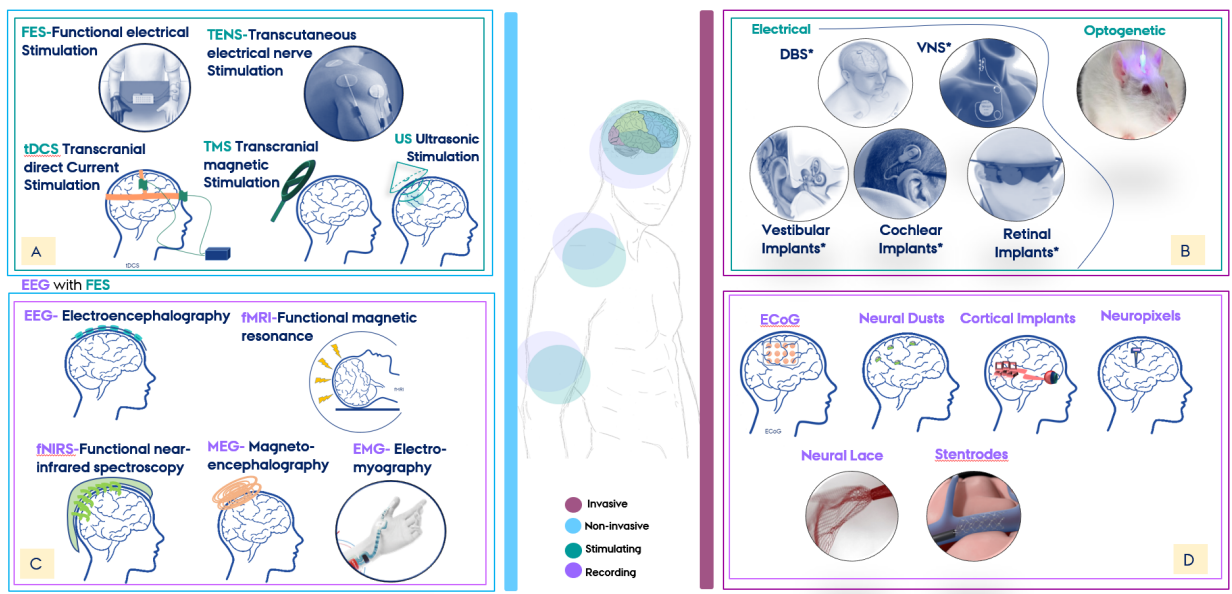

**Figure 3: Neural stimulation and recording methodologies: Invasive vs. Non-invasive.** On the left in blue represents the non-invasive interfaces, while the right in red represents the invasive interfaces. **Stimulation vs. Recording interfaces**: A, B: represent some of the stimulation interfaces. C, D: represent some of the recording neural interfaces.

Ultimately, regardless of whether BCI systems are used for recording brain activity, stimulating /modulating it in an open-loop fashion, or combining both in a closed-loop system, their fundamental objective remains the assistance of patients. Nevertheless, the notion of enhancing human cognition for healthy subjects through directly interfacing with the brain is still merely an abstract and fictional idea. We will delve into the potential advantages and drawbacks of employing brain-computer interfaces for augmenting human cognition in later discussions.

**Table 4: Stimulation Techniques and BCIs**

| Techniques Definition | Invasiveness | Context of Application | Publication |
|---|---|---|---|
| **DBS** Surgical procedures chronically implant electrodes into the brain to allow stimulation of deep structure. | Invasive | Chronic pain treatments | [191] |
| | | Treatment of resistant movement and neuropsychiatric disorders | [166] |
| | | Parkinson disease | [7] |
| | | Obsessive-compulsive disorder (OCD) | [165] |
| | | Treatment of psychiatric disorders (depression) | [8], [166] |
| **ICMS** Intracortical Microsimulation | Invasive | Modulation of cortical activity | [192] |
| | | Restoration of tactile feedback | [71], [151] |
| | | Restoration of vision | [193], [194] |
| **FUS** Non-Invasive neuromodulation method that focuses on a beam of high-frequency soundwaves to synchronize at specific location of the brain to influence the neuronal activity. | non-Invasive | Modulation of brain and behavior | [149] |
| | | Brain modulation (Human somatosensory cortex activity modulation) | [149], [195] |
| **Sonogenetics** Method used to selectively control neural activity through genetically encoded ultrasound -responsive mediators or actuators. It is feasible for both invitro and invivo and have sub-mm spatial resolution and sub-sec temporal resolution | Non-Invasive | Neuromodulation | [182], [183], [196]–[198] |
| | | Drug delivery | |
| **OS** Method used to control cellular activity through light. It involves genetically modified neurons to express light-sensitive ion channels or pumps which can be opened or closed with light of specific wavelengths. | Invasive (minimally invasive) | Choice biasing in primates | [176] |
| | | Restoration of light sensitivity of the retina | [11], [156] |
| | | Behavioral control (e.g., pursuit of prey) | [173] |
| | | Optogenetics, electrophysiology and pharmacology with an ultrasonically powered DUST for Parkinson's disease | [132], [133], [135] |
| **tDCS** Non-invasive brain stimulation method in which a low constant direct current is applied to electrodes on the skull to elicit current flow in the underlying brain tissue. | Non-invasive | Behavior modulation, neuroplasticity | [148], [199], [200] |
| | | Hyper interaction (Brain to brain interface) | [201], [202] |
| **TMS** Non-invasive brain stimulation technique that uses a changing magnetic field outside the skull to generate a localized electric current in the brain via electromagnetic induction. | Non-invasive | Cognitive and clinical neuroscience | [147], [203], [204] |

*Abbreviations: FES (functional electrical stimulation); TMS (transcranial magnetic stimulation), DBS (deep brain stimulation), ICMS Intracortical microstimulation, FUS (focused ultrasound stimulation), OS (optogenetic stimulation), tDCS (transcranial direct current stimulation)*

Table 5: EEG- based Recording Technologies and BCIs applications

| EEG- based BCIs Definition | Modulation | Application | | Publication |
|---|---|---|---|---|
| **P300 ERP-based BCI** P300- event related potential based BCI: Based on P300 event-related potential which is a positive deviation occurs at approximately 300 ms after a rare and relevant stimulus happens. P300 signals could have higher amplitudes when a specific stimulus acquires higher attention. | P300 signals could have higher amplitudes when a specific stimulus acquires higher attention | vibrotactile stimulation | Spinal cord injury rehabilitation | [48], [205] |
| | | Driving scenario in virtual reality | Drowsiness detection | [55], [56] |
| | | Amyotrophic lateral sclerosis (ALS) | Visual P300 speller, Brain painting | [206], [207] |
| | | Cerebral palsy | Cognitive assessment | [208] |
| | | Brain fingerprinting | Lie Detection | [209] |
| | | Assistive technology | Wheelchair control | [210] |
| **SMR-based BCI** Sensorimotor rhythms Based BCI: Based on mu (8-12Hz) and Beta (18-26Hz) oscillations in EEG signals recorded over the sensorimotor cortex. | The amplitude of SMRs could be modulated using mental strategy of motor imagery | Motor learning Brain plasticity | Motor Rehabilitation Lower limb rehabilitation | [46], [211] |
| | | Robotics and Assistive technology (ALS, Stroke) | Hand prosthesis control | [51], [52], [113] |
| | | Assistive technology | Wheelchair navigation | [185] |
| | | Motor training with proprioceptive feedback | Upper limb rehabilitation | [212] |
| | | Plasticity Induction | Stroke rehabilitation | [107] |
| | | Neuroanatomical predictor | Motor rehabilitation | [211] |
| **SCP- based BCI** Slow-cortical potential-based BCI: Based on very slow variation of the cortical activity. | Positive SCPs correlate with mental inhibition and relaxation, while negative SCPs correlate with mental preparation. | | Biofeedback in epilepsy | [184], [213] |
| **SSVEPs-Based BCI** Steady-state visually evoked Potential-based BCI: Based on periodic brain responses induced by repeated visual stimulation. | SSVEPs appear as an increase in brain activity at the stimulation frequency and its harmonics | Dry and non-contact sensors | Typical BCI applications | [186], [214] |
| | | Motor plasticity | Upper extremity Rehabilitation combined with FES | [215] |
| | | Assistive technology | Video Games, Text speller | [216], [217] |
| **VEP** Visual evoked potential | | | Epilepsy, Hybrid and multimodal BCI applications | [45], [218] |

*Abbreviations: P300 (an event-related potential), SSVEP (steady-state visual evoked potential), VEP (visual evoked potential), SMR (sensorimotor rhythms), SCP (slow-cortical potential),*

## 3 Brain-inspired Intelligence: Merging Neuroscience with AI

### 3.1 Neuromorphic computing

With the rapid growth of AI and development of neural networks, AI technologies nowadays display outstanding abilities in multiple cognitive tasks such as large language models (LLMs) like OpenAI's GPT series that generate text in human like fashion and Alphazero that overcome human players at several strategic games like boardgame Go, chess, etc.[219] . Though such performance is outstanding, the key question is still how to reduce the computational cost of these algorithms and how to get brain-like efficiency? The human brain is one the most fascinating organs. It is a remarkable information storage and processing system with impressive computation-per-volume efficiency. The raw computational power of the human brain ranges between $10^{13}$ to $10^{16}$ operations/sec [220], [221]. It performs diverse operations such as recognition, reasoning, control movement with a power budget of about 20 W (like a lightbulb) [222] and power density of 1.1-1.8 $\times 10^4$ W/m$^3$ at an operating temperature of 37 °C [221]. In contrast, running AlphaGo requires power of approximately 170 KW (it used at around 1202 CPUs (central processing units) and 176 GPUs (Graphic processing unit)). Neurons and synapses constitute the fundamental computational units and storage of the human brain [223]. Brain neural networks are formed by billions of neurons (~$9\times10^9$ neurons) interconnected with trillions of synapses (~$3\times10^{14}$ synapses). Neurons are responsible for transferring information through discrete action potentials or 'spikes, while synapses are the intrinsic elements for temporal information processing, long-term and short-term memory storage, and deletion. Also, synapses act as signal transducers and plasticity mediators. Apart from neurons and synapses, studies have also revealed that several elements such as dendritic trees, axons, proteins, and neural microtubules contribute to the brain storage and computation capabilities [224]. Based on the discovery that dendrites generate 10 times more spikes than neurons and that they are hybrids that could process both analog and digital signals, the estimated human brain computational capacity rises 10 times higher than previously thought (i.e., from $1.48 \times 10^{11}$ bits/secs to $3.2\times10^{29}$ bits/sec) [220], [221], [224].

These intertwined networks of neurons and synapses along with the temporal spiking processing enable the fast and efficient transfer of information between the brain's various areas. State-of-the-art artificial intelligence is intrinsically based on neural networks, which are inspired by the brain's hierarchical structure and neuro-synaptic architecture. Deep Neural Networks (DNN) are hierarchical structures composed of multiple layers or transformations that represent variable features within input data. For example, deep convolutional neural networks are multilayered models inspired by the primate visual cortex [225], [226] and utilize synaptic storage and neuronal nonlinearity to learn representative features. These neural networks are powered by conventional hardware computing systems based on CMOS transistors. Billions of transistors can be integrated on a single silicon chip for enormous computing platforms. Such platforms have been a key enabler in the current machine learning revolution. Today's DNNs are trained on powerful cloud servers, yielding incredible accuracy although incurring huge energy consumption. For instance, the deployment of deep network on an embedded smart-glass processor would drain its battery (2.1 Wh) within 25 minutes [227]. It may be possible to enable energy-efficient machine intelligence by designing full-custom hardware that mimics the dynamics and architecture of biological brains, achieving greater power efficiency. Nevertheless, obtaining a high degree of connectivity in neuronal networks and replicating their time-dependent plasticity of synapses are the biggest challenges in this endeavor.

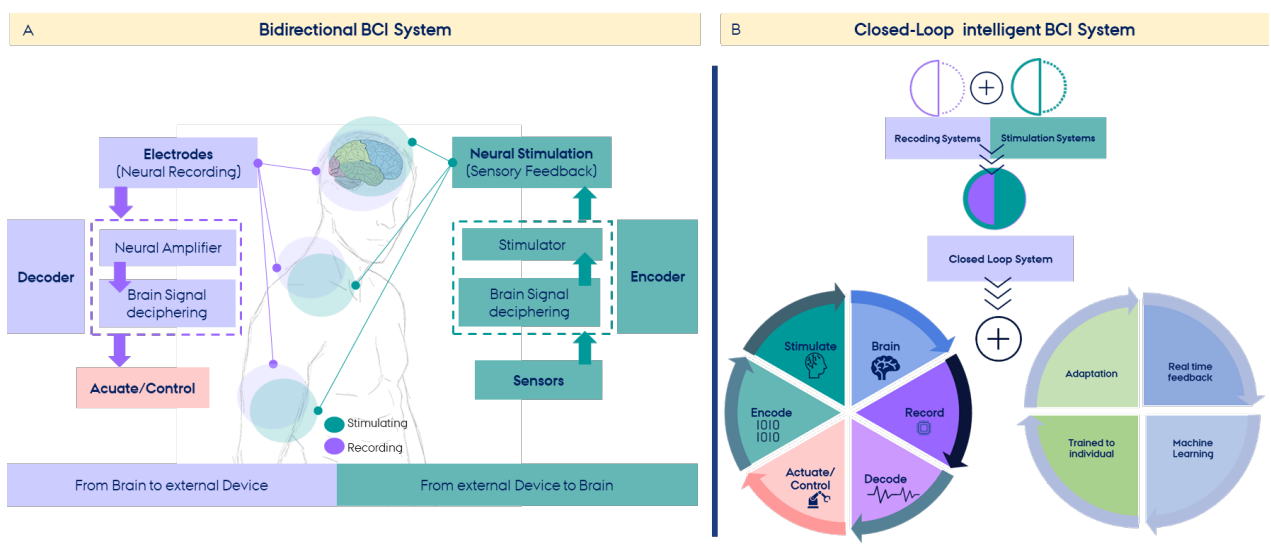

**Figure 4: Bidirectional closed loop brain interface: A: Bidirectional BCI systems**: Two complementary routes: in purple the controlled actuator (from brain to device) and in green the modulated brain state (from device to brain), **B: Closed-loop intelligent BCI system**: the recording and the stimulation interfaces will be combined to formulate closed-loop system. To implement closed-loop intelligent interactive BCI system several features need be added such as: adaption, real-time feedback, AI and to be trained for individual use.

In 1980s, neuromorphic computing was invented to mimic the human brain functionality, by exploiting the similarity between ionic transport across the neuron membrane and carrier diffusion in the transistor channel [228]. A key characteristic of neuromorphic computing that distinguishes it from conventional computing is its spike/event-driven nature of information representation and communication [228]. Currently, researchers are exploring the advantages of using spike-driven computations to promote scalable, energy efficient spiking neural networks (SNNs). In this regard, the neuromorphic computing field could be described as a synergetic domain that focuses on both hardware and software tools to enable spike-based artificial intelligence [229]. It has become an appealing paradigm to overcome the von-Neumann bottleneck and accelerate computing efficiency[229] . Brain-inspired computing systems and conventional digital (von-Neumann) computing architectures present the following key contrasting characteristics: (1) Neuromorphic computing systems (NCSs) exhibit highly parallel operation; thus, neurons and synapses operate in parallel, even though the performed computations by neurons and synapses are still simple in comparison to von Neumann systems [230]–[232]. (2) Both the processing unit (i.e., neurons) and memory (i.e., synapses) are co-located on the same hardware unit which reduces the data communication and improves throughput. (3) NCSs are inherently scalable. In other words, adding more neuromorphic chips could increase the number of neurons and synapses implemented without incurring significant technical obstacles. Several neuromorphic chips can also be combined and treated as a single neuromorphic implementation, such as in SpiNNaker [233], [234], Loihi [235], and mixed signal chips [236], [237] . (4) Some NCSs display stochasticity which may be exploited for computation. (5) Moreover, NCSs use asynchronous, event-driven computations (i.e., it computes only when the data are available), which allows them to be more energy efficient and lower latency than conventional systems [238]. Table 6: presents a short comparison between Von-Neuman architecture and neuromorphic ones.

**Table 6: Fundamental differences between conventional architecture and neuromorphic architecture**

|  | Conventional architecture | Neuromorphic architecture |
| --- | --- | --- |
| **Organization** | Separated Computation and Memory units | Collocated processing and memory |
| **Operation** | Sequential processing | Parallel processing |
| **Timing** | Synchronous (clock-driven) | Asynchronous (event- driven) |
| **Communication** | Binary | Spikes |
| **Programming** | Digital (code with binary instructions) | Spiking neural network (SNN) |

## 3.2 Hardware Implementation: CMOS Neuromorphic chips

Neuromorphic computing systems could become natural platforms for future AI and machine learning applications, as they inherently operate at extremely low-power and implement neural network computation style. Both industry and academic researchers have been keenly interested in developing and implementing neuromorphic systems. Some industrial neuromorphic chips including IBM's TrueNorth [239] and Intel's Loihi [235]. In academia, there are many works aiming to build large-scale neuromorphic chips, for instance, BrainScales [240], [241] , SpiNNaker [233], [234] were realized as a part of the European Union Human Brain project for neuroscience simulations, NeuroGrid [242], IFAT [243] and DYNAPs [236]. These chips have their own specific end- to end software toolchains and applications. Several tasks for instance keyword spotting, medical image analysis and object detection have been successfully applied on existing platform like Intel's Loihi and IBM's TrueNorth [244], [245]. Alternatively, there is research emerging to create general-purpose neuromorphic platforms that connect hardware and software frameworks for wider classes applications [229], [230]. Tianjic chip is a hybrid neuromorphic chip that was developed to support both neuromorphic SNNs and traditional ANNs [246]. The previously mentioned large-scale neuromorphic chips are silicon-based and implemented with conventional Complementary metal oxide semiconductor (CMOS) technology either in digital (synchronous or asynchronous), analog (subthreshold or super-threshold), or mixed signal (where in general neurons are implemented in analog and synapses and learning are implemented in digital domain) [229], [230].

Despite remarkable progress in CMOS based-neuromorphic computing systems, they are far from the brain's energy- and area efficiency which limits the scalability of such networks [247]. This has driven a significant effort to investigate non-CMOS implementations of ANNs using emerging technologies such as memristors [248], [249] , magnetic tunnel junctions (MTJs) and spin Hall nano-oscillators (SHNOs) [250], etc.,  as synapses, and MTJs [251], [252], spin-torque nano-oscillators (STNOs)  [253]–[255] , SHNOs [256], phase change materials (PCM), ferroelectric, topological insulators, biomolecular memristor, no-filamentary, etc. [257], [258]  as artificial neurons. Meanwhile, memristive technologies are used to build resistive memories to collocate both processing and memory units (in-memory computing) [230], [259]. On the other hand, spintronics is a strong contender as it is CMOS compatible, multifunctional, and extremely versatile with features like non-volatility, plasticity and oscillatory behavior which can be exploited to implement both artificial neural components (i.e., neuron and synapse) to develop energy efficient ANNs. Each of these technologies have shown unique features that significantly improved energy efficiency with comparable footprints as biological counterparts. Figure 5 (adapted from [231]) represents the structural organization of the central nervous system from carriers (1A°) until the brain and most common methods used to study it (top-down analysis), alternatively it shows the equivalent neuromorphic systems used to mimic the brain (bottom-up analogy). Also, Table 7 presents a summary for designing neuromorphic systems from materials to applications [229], [230], [258]  .

## 3.3 About Spiking Neural Networks SNNs: analogy, overview, and perspectives

SNNs are considered as the third generation of ANNs [260]. Bearing in mind that neurons communicate via electric pulses or action potentials (AP) called spikes. It was only in the early nineties that neuroscientists discovered that biological brains use the exact timing of spikes to encode information [261]. This in turn boosted the development of spike-based neural networks to further understand the information processing skills of the brain. SNNs provide a more biologically realistic, brain-like approach compared to ANNs by incorporating spatial and temporal considerations through neural connectivity and plasticity. Their ability to deal with precisely timed spikes makes them competitive with traditional ANNs in terms of accuracy and computational power, and in some cases, better suited for hardware implementation [262], [263].

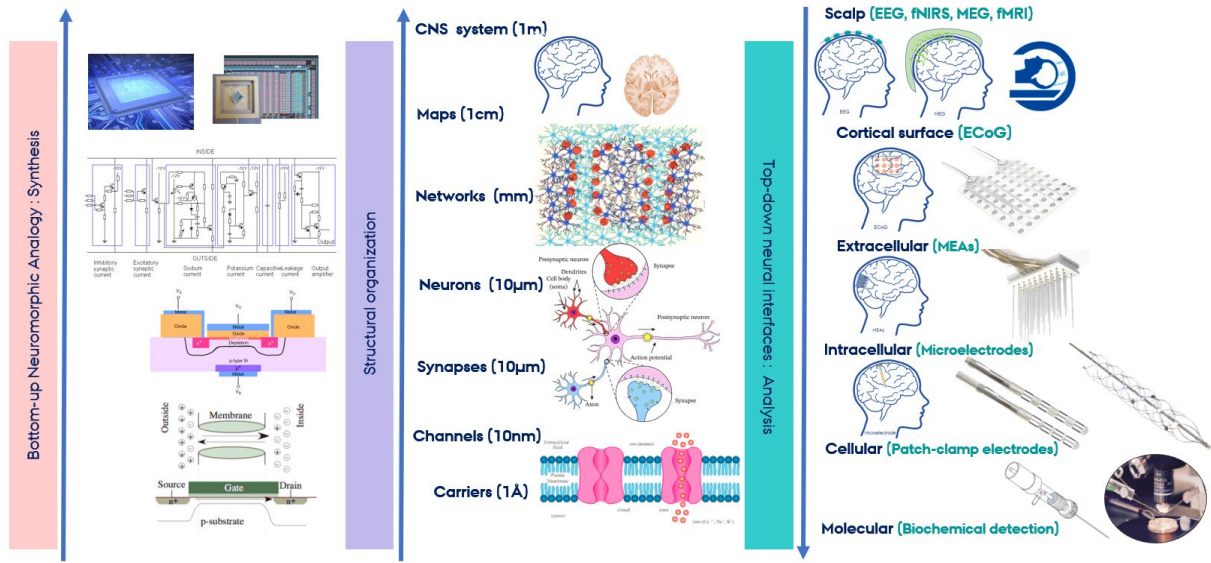

**Figure 5** Structural organization of the central nervous system, the corresponding neuromorphic equivalent systems (synthesis) and the neural interfaces for investigating it (analysis) adapted from [297]

**Table 7: Summary of state-of-art for designing neuromorphic systems from materials to applications.**

| Materials | Technology | Circuits Micro-architecture System architecture | Algorithms | Applications |
|---|---|---|---|---|
| 1. **Phase-change materials** 2. **Ferromagnetic materials** 3. **Ferroelectric materials** 4. **Non-filamentary RRAM materials** 5. **Topological insulator materials** 6. **Channel-doped bio-membrane** | 1. Electrochemical transistors 2. Spintronic 3. Memristors 4. Optical devices 5. Charge-trapping transistors 6. Phase-change memories 7. Ferroelectric transistors 8. Threshold switching devices | 1. Digital 2. Analogue 3. Mixed signal | 1. Feed-forward neural network 2. Recursive neural network 3. Reservoir computing 4. Spike-based backpropagation 5. Mapping (conversion-based) 6. STDP 7. Graph-based 8. Evolutionary based | 1. Control 2. Classification 3. Security 4. Benchmarks 5. Neural signal processing 6. Forecasting 7. Edge-computing |

*Neuron*--The basic computational unit for ANNs is the artificial neuron and its input is processed by an activation function *f(-)*, while the SNN unit is the spiking neuron which is expressed by a set of differential or difference equations ($x_t=f(x_{t-1})$). SNNs employ several spiking neuron models to simulate the nervous system and the properties of the neurons that generate electrical potentials across their cell membrane. Neuron models implemented in SNNs in the literature range from simple linear models with a fixed threshold and non-linear models with a spiking mechanism, to more complex and biologically plausible models [264] (Figure. 6). The most representative neuron models are: i) Leaky Integrate-and-Fire (*LIF*) [264], ii) Hodgkin–Huxley [265], iii) Izhikevich [266], iv) FitzHugh Nagumo [267]. In the LIF model, a charge is integrated over time until a threshold value is reached. In brief, a neuron emits a spike each time its membrane potential gets to a specific threshold, after which it enters a hyperpolarization state during which it is impossible to emit another spike for a certain time (i.e., absolute refractory period). It can be modeled by an electrical circuit consisting of a capacitor in parallel with resistor and driven by current I(t). The Hodgkin–Huxley model also approximates the neuron action potential generation mechanism by a simulation of ion channel dynamics. On the other hand, the Izhikevich neuron claimed to be as biologically plausible as Hodgkin–Huxley model and computationally effective as LIF model [229], [264]. SNNs use mainly event-driven and clock-driven dynamics, in contrast with ANNs which use step-by-step stimulation process [263]. Figure. 7 presents the most representative neurons models (adapted from [264]).

*Spike encoding*-- To apply the neuron model, the input data must be encoded into spike trains before presenting it to SNNs. Neuroscience is still grappling with several significant questions about the encoding part including: what is the information included in these spatio-temporal spike patterns? What is the used code to transmit information by neurons? Also, how do other neurons receive this information and communicate? It is necessary to create spike patterns that preserve most of the task-related information in the input stimuli. Early studies have found that such information is merely embedded in the mean firing rate of neurons [269].

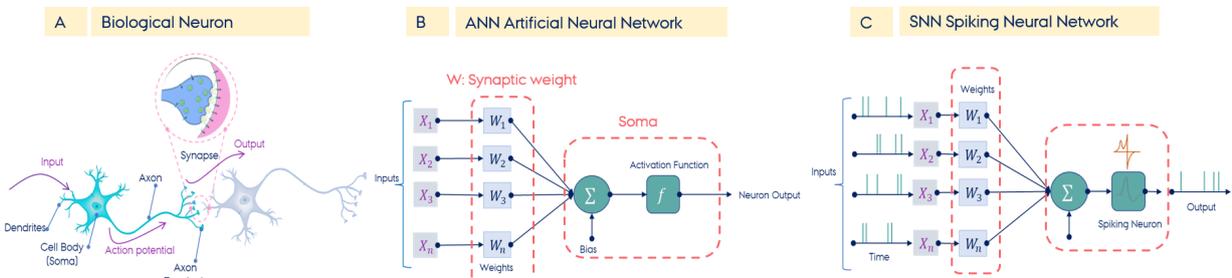

**Figure 6: A. Simplified diagram of a typical biological neural cell.** The soma receives synaptic signals from other neurons through its dendrites, and axon propagates signals to other neurons. A synapse is a contact between the axon of one neuron and the dendrite of another. The soma maintains a voltage gradient across neuron membrane. If the voltage changes by a large enough amount, an action potential (spike) may be elicited. The action potential then travels along the axon, and eventually activates synaptic connections with other cells. **B. Artificial neural network (ANNs),** and **C. Artificial spiking neural network (SNNs),** are only active when it receives or emits spikes which make it energy efficient over a given period.

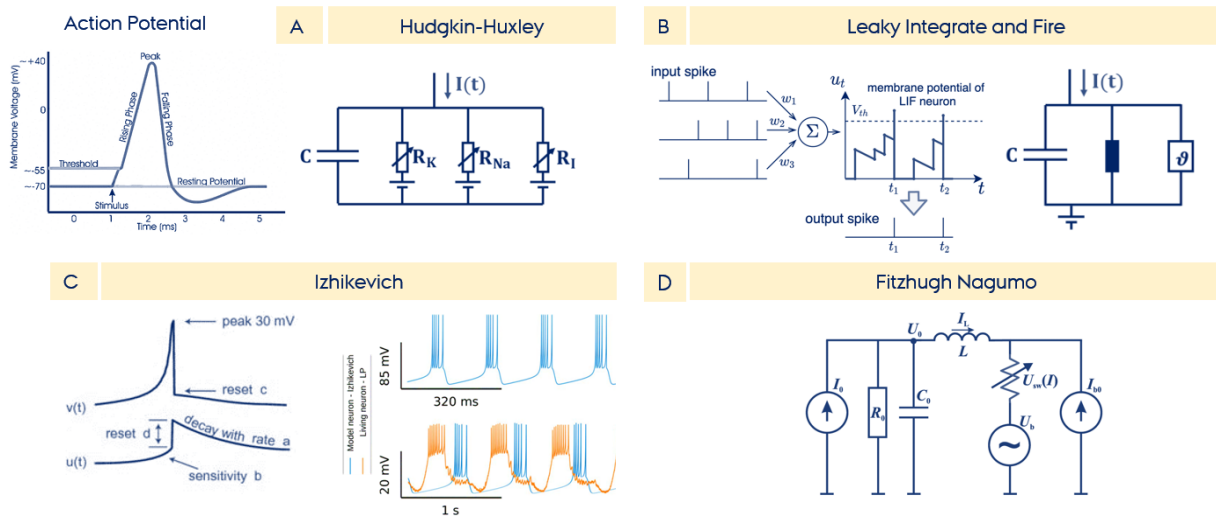

**Figure 7: Representative neurons models A. Hudgkin-Huxely, B. Leaky Integrate and Fire, C. Izhikevich, and D. Fitzhugh Nagumo**

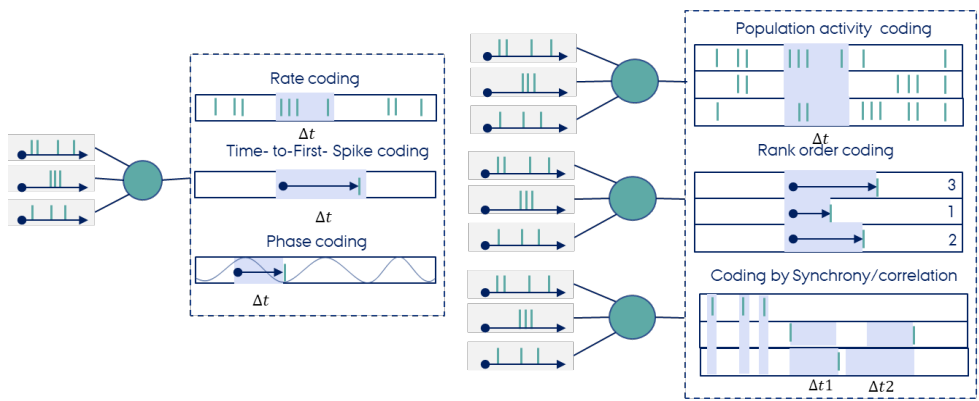

**Figure 8: Rate-based encoding versus temporal encoding:** Schematic representation of some neural codes, neurons are the green circles the simplified spike coding types and the actual coding schemes are shown respectively on the left and the right © computation by Time by Walter et al.[268]

In the literature, there are two main encoding schemes a) *rate-based encoding* and b) *temporal encoding* (see Figure 6) [268], [270]. The first scheme is based on a spiking characteristic within an interval of time (such as frequency), while the latter is based on spike timing. Several neuron models use rate codes to explain computational processes in the brain. Rate-based schemes include three different notions of mean firing rate "rate as a spike count", "rate as a spike density" and "rate as a population activity". However, spiking neuron models can model more complex processes that depend on the relative timing between spikes or timing relative to a reference signal such network oscillation. Taking as a reference the encoding mechanism of biological neurons to specific stimulus signals, researchers have come up with many temporal encoding strategies, such as "time-to-first-spike", "latency phase," population encoding", "correlations and synchrony", and "Ben's spiker algorithm" [268], [269]. Often decisions must be made before a reliable estimate of a spike rate can be computed, therefore temporal codes are highly interesting where even a single spike or small-scale temporal variation in the firing time of a neuron may trigger a different reaction. Figure. 8 shows the difference between the two encoding schemes.

Table 8: Comparison between ANNs and SNNs [260], [271]

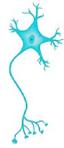

|  | ANNs 1st gen. | ANNs 2nd gen. | SNNs 3rd gen. | Biological systems |
|---|---|---|---|---|
| **Computational units** | Perceptron neuron | Artificial Neuron | Spiking Neuron |  |
| **Function** | Performs thresholding in a digital (1,0) output | Sigmoid unit or a rectified linear unit (ReLU), adds continuous nonlinearity to the neural unit, which enables it to evaluate continuous set of output values | Mainly based on integrate and fire type that exchange information via spikes |  |
| **Neuron Model** | $y = f\left(\sum_{i=1}^{n} w_i x_i - \theta\right)$ $= \begin{cases} 0, \wedge \sum_{i=1}^{n} w_i x_i - \theta < 0 \\ 1, \wedge \sum_{i=1}^{n} w_i x_i - \theta \geq 0 \end{cases}$ | $y = f\left(\sum_{i=1}^{n} w_i x_i\right) + b$  $f$ $y$: neuron output $x_i$: neuron input $n$: number of neurons $w_i$: synaptic weight $\theta$: Activation threshold $b$: Bias | $\begin{cases} \frac{dX}{dt} = f(X) \\ X \leftarrow g_i(X) \end{cases}$  $X$: vector of state variables of neurons $f$ for state of variable evolution $g_i$ variables due spikes a neuron |  |
| **Encoding Scheme** | Rate encoding | Rate encoding | Temporal encoding | |
| **Information representation** | Scalar values | Scalars values | Spike trains | |
| **Computation mode** | Activation function | Activation function | Differential equations | |
| **Network simulation** | Step-by-step | Step-by-step | Clock driven and event driven | |
| **Input vs Output** | Binary, Binary | Real, Real | Real, Real | |
| **Neural Networks architecture** | 1. Perceptron<br>2. Multilayer Perceptron (MLP) | 1. Conventional Neural networks (CNNs)<br>2. Recurrent neural network (RNNs) | Spiking Neural Network (SNNs) | |
| **Pros** | 1. Intuitive interpretation as spikes<br>2. Few parameters to be optimized | 1. Higher biological plausibility than 1st gen.<br>2. Large scale gradient base optimization techniques<br>3. Higher computational power than 1st gen.<br>4. Implicit time notion, codes fire rates | 1. Higher biological plausibility than 2nd gen.<br>2. Massive parallelization and powerful computation<br>3. Successful hardware implementation (Neuromorphic VLSI)<br>4. Excels with Spatio and Spectro-temporal data treatment (SSDT)<br>5. Explicit time notion, codes fire rates | |
| **Cons** | 1. Only binary outputs<br>2. Limited analog implementation<br>3. No time notion (coding spikes) | 1. No direct Hebbian learning, It only compute fire rates instead of fires.<br>2. No intuitive interpretation for spikes<br>3. Moderate parallelization and spatio and spectro-temporal data treatment (SSDT)<br>4. VLSI Hardware implementation<br>5. More parameters to be optimized and high sensitivity to its values | 1. Biological insights still not sufficient to develop rigid theory.<br>2. Unknown behavior with spatio-temporal data<br>3. More parameters to be optimized and high sensitivity to its values.<br>4. Lack of common framework for the different models and coding schemes | |

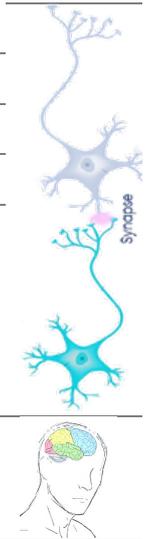

## 3.4 Learning methods in SNNs

Synaptic plasticity refers to the ability of synaptic connections to change their strength (i.e., modulation of synaptic weights) over time based on network activity. Synaptic plasticity is believed to be a key mechanism of learning and memory processes in the brain. There are various schemes of synaptic plasticity such as Hebbian [270], non-Hebbian [272] and neo-Hebbian[273], [274] schemes, differing primarily in their time scales and induction conditions. The timescales of synaptic plasticity scale from dozens of msec to hours (e.g., Long-Term Potentiation (LTP) or Long- Term-Depression (LTD)). Depending on induction conditions, some synaptic plasticity rules depend on the relative timing of the pre- neuron and post-neuron synaptic spikes, the temporal order spikes, and on certain factors like specific chemical ion concentrations. While others rely only on the past pre-synaptic stimulation (disregarding the postsynaptic response). The neural circuits either get inhibited or excited according to the type of the synaptic input they receive [264], [275]. Choosing a proper spiking neuron model with an appropriate synaptic plasticity, along with exploiting event-based, data-driven updates (with event-based sensors) are essential factors to enable computationally efficient intelligence applications such as inference and recognition. Yi et al. presented a review on learning rules in SNNs [276]. SNNs have used two major learning schemes so far: *conversion-based* and *spike-based schemes* which are explained below:

*Conversion-based schemes* -- are based on converting a trained equivalent, DNN into SNNs using weight rescaling and normalization to adapt the attributes of nonlinear continuous output of the artificial neurons with that of the spiking neuron (e.g., leak time constant, refractory period, membrane threshold, firing rate, etc.) [278]–[280]. They tend to achieve better accuracies on large scale spiking networks in image classification like ImageNet dataset [279], [281], [282], but are less resource-efficient compared to spike-based training schemes. Commonly, DNNs are trained on frame-based data such as TensorFlow [283] which gives them a high training-associated flexibility. Conversion needs parsing of the trained DNNs on event-based data (generated by rate-coding of static image dataset) and then applying parameter conversions. Conversion-based scheme has the merit of removing training burden in the temporal domain. It achieves the same accuracy for image-recognition tasks as that obtained from image classification in traditional deep learning neural networks (DNNs) [280], [282], [284], however, it shows intrinsic limitations as follows. The non-linear neuron output value could take both positive and negative values, while the rate of a spiking neuron can only be positive. Therefore, all the negative values are excluded which lessens the accuracy of converted SNNs. Another issue with conversion-based is the difficulty of achieving an optimal firing rate at each layer without sacrificing performance. Researchers recently proposed new methodologies to optimize the firing rates, for example leaky ReLUs was introduced during DNNs training to better match the spiking neuron's firing rate [285]. Furthermore, the inference time for converted SNNs is generally large, resulting in higher latency and lower energy efficiency [280]. A new mapping strategy was proposed by Stockl et al., in which SNNs use Few Spikes neuron (FS-neuron) model to represent complex activation functions with two spikes at most. The proposed strategy exhibited similar deep learning accuracy with fewer time steps per inference in comparison to typical conversion- based techniques [286]. Applications such as keyword detection, medical image analysis, and object detection have used those mapping strategies and were implemented on existing neuromorphic hardware platforms (e.g., Intel's Loihi, IBM's True North) [245], [246], [287]. However, conversation- based methods- although they are used as approximators of ANNs- still missing the main goal of SNNs by achieving more computationally efficient.

*Spike-based* schemes-- fall into three main categories including supervised (training with labeled data), unsupervised (training without labeled data), and reinforcement learning [287]. Supervised Hebbian learning (SHL) rule is one of the most straightforward spike-based learning processes, it is usually supervised by an extra -teaching signal- that drives the post-synaptic neuron to fire at targeted time intervals and to remain silent at other times [276], [287], [288]. Remote Supervised Method (ReSuMe) [289], and tempotron [290] are two of the most early representative works in supervised learning for single layered SNN to perform classification. Researchers have been focusing on integrating spike-based quasi-backpropagation error gradient descent to deploy supervised learning in multi-layered SNNs [291], [292] . For instance, SpikeProp [293] -a learning rule based on gradient descent for training SNNs- and some of created a new backpropagation rule for SNNs by fixing a target spike train at the output layer [292], [293]. Recent works adapted deep-learning training style using surrogate gradient and smoothed activation function to compute the error gradients when adjusting weights in each of the successive layers [294], [295]. There have also been some approaches that perform stochastic gradient descent on real-valued membrane potentials to get more random spikes from the correct output neuron [296], [297]. These few demonstrations show close to state-of-art-classification performance on the Modified National Institute of Standards and Technology (MNIST) handwritten digits dataset [298]. Backpropagation through time and real time recurrent learning approaches have been applied in neuromorphic datasets such as the Spiking Heidelberg Digits (SHD) and the Spiking Speech Command (SSC) datasets[299], [300].

In unsupervised learning, the neural connections are reorganized depending on the modification of synaptic weights of the Hebbian processes [271], leading to new functions for example input clustering, pattern recognition, source separation, dimensionality reduction, associative memory formation, etc. Spike-timing-dependent plasticity (STDP) – a learning strategy based on varying the synaptic weights according to the relative spiking timing from pre- and postsynaptic neurons - is the most widely implemented synaptic plasticity in neuromorphic literature. It assimilates more brain-like architectures, due to the possibility of bringing both memory and computation units closer. This in turn induces more energy-efficient on-chip implementations. Diele et al. were from the first groups that demonstrated fully unsupervised learning on an SNN, leading to analogous accuracy to deep learning on the MNIST database [295], [301]. Deep SNNs - Spiking Convolutional Neural network (SCNN)- are one of the recent scenarios that have shown that adding random error signals through feedback connections enhance learning [302], [303]. It depends on training multilayer SNN network with local spike-based learning per layer then follow it up with global backpropagation for classification. An additional class of SNNs are the recurrent networks with delays and synaptic plasticity and used for modelling dynamical systems. Alemi and colleagues applied a local learning rule with recurrent SNNs with less spikes to demonstrate non-linear dynamical systems [304]. These recurrent SNNs display greater classification capacity with winner-take-all models [305], [306]. An alternative algorithm used in SNNs is reservoir computing or liquid state machines (LSM). It is considered as Spiking Recurrent Neural networks (SRNNs), as it uses sparse and recurrent connections with synaptic delays in spiking neural networks to shape the input into a higher dimensional space spatially and temporally [307]. In addition to liquid or reservoir which is the SNN component and untrained, the reservoir computing methods includes a readout mechanism which is trained to realize the output of reservoir. The main advantage of spike-based reservoir computing is the elimination of training in the SNN component, and it has shown its effectiveness at processing temporally varying signals in a wide range of application such as bio-signal processing and prosthetic applications [308], [309]. Table 9 and 10 summarize the learning methods in both ANNs and SNNs and give comments on their usage. In another flip, neuromorphic systems have recently been considering few non-machine learning based learning algorithms such as those rising from graph theory [310] and Markov chains [311]. For instance, neuromorphic computing together with graph theory was used as a tool for analyzing Covid-19 disease spread [312]. Neuromorphic deployment of discrete time Markov chains was used by Smith et al., to estimate particle transport problems and heat flow on complex geometries [313]. Figure. 9 depicted the different SNNs algorithms.

SNNs are promising candidates for processing data in a low energy mode although there are much more to be investigated to process the applicability of SNNs for general applications. Due to the feature of the SNNs as a brain-inspired computing or processing technique, the idea of bringing SNN to the BCIs, whether implantable or wearable, to communicate with the brain can be a potential game changer in this field with a high impact. In the next section, a discussion on the intelligent tools (ANN and SNN) for interfacing the brain signal as well as our perspective on brain inspired BCIs are included.

**Table 9: Learning Methods for both ANNs and SNNs based on** [277]

| Learning models | ANNs | SNNs | Comments |
| --- | --- | --- | --- |
| **Self-supervised supervised** | 1. | 1. | Learning without labels and works better than unsupervised |
| **Supervised Learning** | 1. Gradient descent | 2. **SPAN** Spike Pattern Association Neuron<br>3. Surrogate Gradient Descent<br>4. SRM Spike Response Model (Gradient descent supervised learning) | 2. Requires labeled data to fit a mapping of the input features to the output by minimizing a loss function (error function)<br>3. Two main categories: Classification and regression.<br>4. Labels in ANNs are represented as integers (classification) or real numbers (regression), while in SNNs it is represented as spike trains with spatio-temporal properties |
| **Unsupervised Learning** | 1. **Autoencoder**: Neural network model that encodes the features of the input data in a latent space (encoding) and uses latent vectors to reconstruct the original inputs (decoding).<br>2. **Generative adversarial network (GAN)** consists of two networks, a generative network and a discriminative network. The two networks are trained to "fool" the discriminative network.<br>3. **Self-organizing map (SOM)**: A method of dimension reduction using competitive learning in which output neurons compete for activation, with a subset of neurons being activated at any given time, e.g. through winner-takes-all neuron. | 1. **Hebbian Learning**:<br>2. **STDP Spike-Timing Dependent Plasticity**: A synaptic plasticity rule that captures the spike timing effects in synaptic plasticity. It may lead to either long-term-travel potentiation (LTP) or Long-term-travel depression (LTD) of the weights.<br>3. **Triplet STDP**: LTP is constructed as combination of one presynaptic and two postsynaptic spikes. While, LTD is based on two presynaptic and one postsynaptic spike. It takes in account the spiking timing interaction | 1. Unsupervised learning works with unlabeled data. |
| **Reinforcement learning** | 1. **Value-based**: Learns the state or state-action value. Q-learning is the most classic value-based algorithm. DeepMind proposed a combination between RL and deep neural networks (Deep Q-Network algorithm)<br>2. **Policy-based**: Maps the state space to the action space, then taking best action to maximize its return. | 1. **Three-factor learning rules**: Synaptic updates depend on three factors: a presynaptic trace, a postsynaptic scalar or trace and an extrinsic reward (third factor).<br>2. **ANNs to SNNs**: Conversion based learning approaches. It matches the firing frequency of the firing neurons and the successive analog neurons. The training phase is done on ANNs ad then converted to SNNs. | 1. RL was inspired by reward-based learning in animals.<br>2. In ANNs, RL based algos learns from feedback through iterative trails that are simultaneously sequential. It uses nonlinear function approximations.<br>3. RL is biologically interpretable.<br>RL in deep learning is time consuming |

**Table 10: Neural networks in SNNs**

| Neural Networks | Feedforward Neural network (FNN) | Convolutional Neural networks (CNN) | Recurrent Neural network (RNN) |
|---|---|---|---|
| **Definition** | Feedforward Neural networks or Multilayer perceptron (MLP) map input x to the output y(x) through a series of non-linear transformations. Elements of network: input layer (1st layer), output layer (last layer), hidden layers (in between layers), perceptron, | CNNs represent the spatial patterns in an image. A convolutional layer convolves the input with the cross-correlation operation followed with nonlinear activation function. The pooling layer down samples the spatial dimensions and so reduces the number of parameters. Generally, final fully connected layers transform the output of the feature extraction layer into class representations. The last layer classifies the output by a SoftMax function.<br><br>They are used to process visual information, especially images. Multiple layers are used to process the grided data such as convolutional, pooling and fully connected layers. | RNNs process the information in the most recent time step in the learning process based on an internal state. The internal state is updated to memorize task relevant information.<br><br>RNN-based Gated recurrent unit (GRU) and long short-term memory (LSTM) have been used in real-life applications.<br><br>They are used to process sequential data or time series data, and to solve ordinal or temporal problems, such as language translation, speech recognition, etc. |

| SNNs | Spiking feedforward Neural network (SFNN) | Spiking Convolutional Neural network (SCNN) | Spiking Recurrent Neural networks (SRNNs) |
|---|---|---|---|
| | SNNs based on STDP learning and Back propagation- based supervised learning used for pattern recognition. Using a two-layer SNN, based on the biological properties of excitatory type neurons and inhibitory neurons as the processing layer, using lateral inhibition as well as winner-take-all properties, enabling the neurons in the processing layer to extract features with significant characteristics from the input signal based on STDP learning rules or Surrogate gradients, with optimal performance of 95% on the MNIST dataset. | Converted SCNNs are close in performance to CNNs and could perform inference tasks on neuromorphic hardware and consume less time and energy. Difference-of-Gaussian kernel for the input image, followed by unsupervised STDP-based training of the convolutional layer as well as the pooling layer, and finally, the extracted features are passed into the classifier | SRNNs have complex nonlinear dynamics and are usually used to study biological neural networks in specific microcircuits of the brain. Excitatory and inhibitory neurons connect to form neural network that is chaotic yet in equilibrium state machine. LSM Liquid state machine is used for computational modeling. It is made of three layers: the input layer, the reservoir or the liquid layer and the memory less readout layer. It transforms the time varying input information into higher dimensional space to express temporal and spatial properties of neuronal dynamics, thus memorizes the input information. Eprop that perform RNNs with surrogate gradients |

## SNNs in BCIs: Spiking neural networks in brain-computer interfaces

ML algorithms commonly learn to identify categories or predict unknown future conditions starting from data. ML methods allow the prediction and progression of brain degenerative disorders as Alzheimer's disease, dementia, schizophrenia, multiple sclerosis, cancer, etc.[14]. As an example, an ML approach combining multiple biomarkers of tremor in LFP like multi-band spectral power, phase-amplitude coupling, and high-frequency oscillations ratio, with a smoothing Kalman filter achieved 89.2% sensitivity in detecting rest state tremor in Parkinson disease patients [14], [314], [315]. Further exemplars were reported in the detailed systematic review paper on AI for Brain diseases published by Segato et al. [316]. In addition, approaches for segmentation and detection of brain structures, as well as pathological tissues, are also widely studied. For instance, a subject-specific logistic regression model was applied to predict memory encoding state from brain-wide ECoG recordings and activate closed-loop stimulation to improve memories anchoring in humans [176]. Nevertheless, it is worth noting that, because of the complexity and the amount of brain data, ML methodologies usually comprise several steps to perform a task. For example, image pre-processing, feature selection and ranking, and dimensionality reduction are often required as initial stages to improve performance to adequate levels. Several low-power, area- efficient digital/mixed signal systems on-chips (SoC) with embedded ML have been reported in literature for neural signal acquisition and treatment. Zhang et al. developed the first- in- literature real-time SoC with both online tuning and one-shot learning for patient-specific closed-loop epilepsy tracking system [317]. The work [318] developed a neural interface processor for brain-state classification (NuriP), it implemented an exponentially decaying-memory support vector machine (EDM-SVM) classifier combined with a neural network autoencoder to lower the dimensionality of input data. Alternatively, Cheng et al. presented a low power closed-loop neuromodulation chipset for epilepsy with high common mode interference tolerance that integrates a two-level classifier [319]. Most of these systems were verified offline on human epilepsy data, and in closed-loop seizure control in animal models of epilepsy. ML SoCs also used DDNs (deep neural networks) for emotion detection of autistic children [320], and CNNs (Convolutional Neural Networks) with online training for emotion recognition from EEG- based data [321].

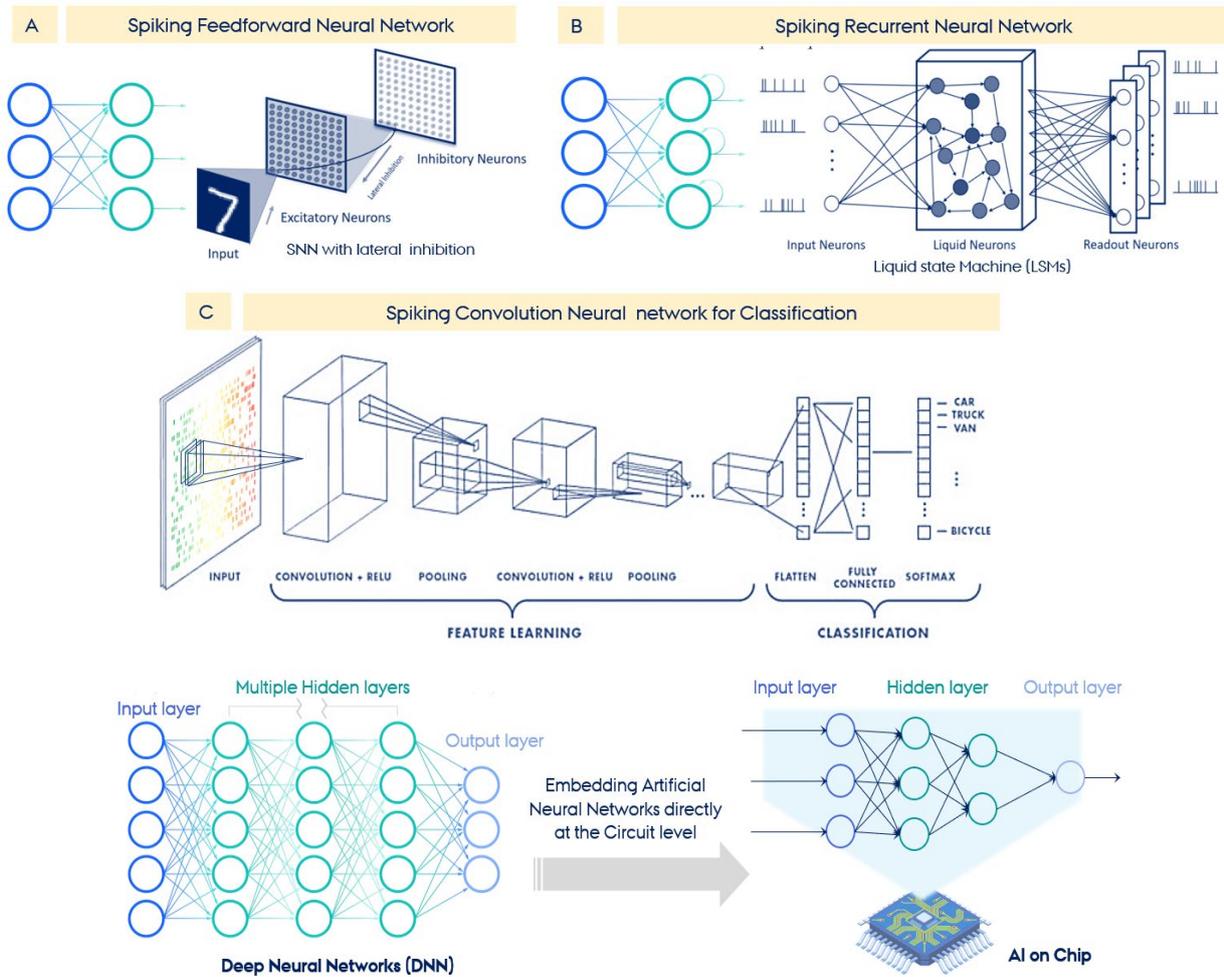

Figure 9: **Neural networks in SNNs**: A. **Spiking feedforward Neural network** (SFNN), B. **Spiking Recurrent Neural networks** (SRNNs), C. **Spiking Convolutional Neural network** (SCNN), on-chip Machine learning (ML) implementation.

Brain-inspired neuromorphic architectures have been used in several fields of applications such as image recognition, decision making and action selection, spatial navigation, and environment exploration, rehabilitation and motor control, robotic control based [322], etc. For example, the iCub Humanoid robot was able to develop brain-inspired cognitive abilities like memory and learned to interact and respond to a dynamic environment through SNNs-based controllers [323], [324]. Additionally, SNNs have also been used for brain disease diagnosis and prognosis, motor imagery signal classification and cognitive process measurement. For example, Capecci et al. have proposed a method based on NeuCube spiking neural network to classify brain EEG data from patients of Alzheimer's Disease and subjects diagnosed with mild cognitive impairment and analyze the functional changes in their brain activity [325]. Also, Ghosh-Dastidar et al. data investigated SpikeProp, QuikProp and RProp SNN's classification algorithms to detect epileptic seizures from EEG [291], [326]. Wang et al. suggested an alternative approach for multiple motor imagery decoding based on SNNs. They used a filter with one-vs-rest (OVR) strategy were employed to extract the spatio-temporal-frequency features of multiple imagery after preprocessing. Then, they applied F-score to optimize and select these features which in turn were fed for SNN for classification [327]. Some SNN accelerators take advantage of weight sparsity to efficiently reduce the model size, computation, and data transfer energy. They are commonly implemented in Application Specific Integrated Circuit (ASIC) with either offline [230], [231], [237] or on chip learning [328]–[330].

Thanks to their low power, high adaptability, and ability to emulate the nervous system functionality in analog, digital or mixed-signal CMOS hardware, neuromorphic designs are receiving more attention in BCIs and neural prostheses systems [232]. For instance, the neuromorphic system in ref [281] used an analog Spiking Neural Network (SNN) classifier and demonstrated an STDP rule for spike sorting in BCI applications [223]. Similarly, combinations of spiking reservoirs and STDP have been used in a SNN architecture called NeuCube [331], which was used to process electroencephalograms (EEG) signals and functional magnetic resonance imaging (fMRI) signals in applications such as sleep state detection [332] and prosthetic controllers [333]. Indiveri et al. developed an event-based neuromorphic system with on-line learning for classifying auditory stimuli [334] . Likewise, the neuromorphic processor was implemented in modular closed-loop BCI for decoding motor intentions and delivering sensory stimuli to the brains of anesthetized rats [238], [335]. Another method based on spiking activity used LFP features to evoke somatosensory feedback in closed loop BCI in rodents [73]. A related method used wireless battery-powered neural implant to stimulate the somatosensory cortex in response to spikes detected in the premotor cortex in a rat model with brain injury [336]. Neural chips can be applied too in a bidirectional closed loop prostheses for brain disorder treatment as epilepsy [336]. Recently, Moradi et al. built a CMOS-based neuromorphic device for the detection of epileptic seizures from local field potential (LFP) signals [337]. Also, a mixed-signal multi-core neuromorphic processor (DYNAPs) exploiting an event-based communication was used to detect High- Frequency Oscillations (HFO) as biomarkers of seizure events [105], [237]. A high-density retinal implant with in-pixel neuromorphic image processing and temperature-regulation circuits, mimics human retinal operation [338]. Neuromorphic platforms such as IBM's TrueNorth processor were used to implement CNNs that treats electrophysiological signals [339], [340]. (Figure. 10)

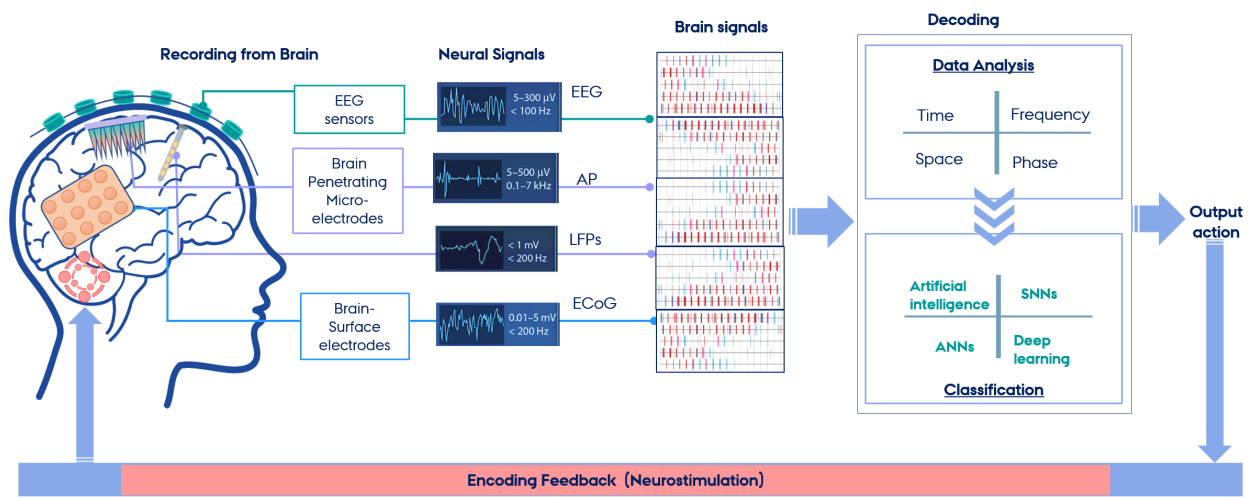

**Figure 10: SNNs In BCIs for communication.** The user's intention can be converted into commands to control the external devices after the process of decoding, transmitting, and encoding. Several extracted neural signals recorded by either invasive or non-invasive BCIs: Electroencephalography (EEG) signals recorded from the scalp, ECoG from brain surface, action potential (AP or spikes) and Local field potential (LEPs) from Brain penetrating microelectrodes. The extracted signals will be decoded and transformed into spikes and later translated into output action to control external devices such robotic arm, neuroprostheses, etc. The missing block is encoding multimodal information and feeding it back to brain through neurostimulation.

# 4 Toward Brain inspired- Brain computer interfaces (BI-BCIs)

Based on the work presented, there is evidence that neuromorphic computing embedded into neural interfaces can play an important role in the future landscape of technologies for treating neurological disorders. Although the feasibility of those systems has been demonstrated, there are still major questions that need to be resolved at the hardware and software levels.

Currently, the existing encoding systems [11], [33], [42], [341]–[343] suffer from high latency due to the large amount of data fed into these systems. To improve the latency of such systems is to reduce the volume of unnecessary information to be processed. Although, recently, the encoding speed as well as the accuracy of BCIs have been improved due to the advances in ML/AI algorithms and more optimized hardware [230]–[232], [344]–[349], we are still far from the realization of a real-time BCI interfacing the brain and communicating with it. SNNs provides a powerful tool for modeling complex information processing in the brain, due to their ability to simulate the rich dynamics of the biological neurons and to represent and integrate different information dimensions, as time, frequency, and phase. It leverages spike information representation (binary events) which is like the action potentials in the brain. Besides, SNNs use biologically plausible local learning rules such as STDP, Hebbian learning and three-factor rules, which allow for fast real-time learning and low computational complexity. SNNs as neuromorphic computing architecture offers several advantages such as: flexible structure, incremental life-long learning, temporal, or spatio-temporal associations between input variables are learned, event-based or asynchronous learning leading to less volume of data, facilitate interpretability of the model, low power and computational demand, more energy-efficient communication through spikes, and fault tolerance. By taking the advantages of the synergy and complementarity between SNNs and human intelligence, we postulate that brining SNNs into BCIs, in implantable or wearable form, to communicate with the brain would radically change the neuroscience research field and push it further to attain better results than any classical system using new BCIs [350].

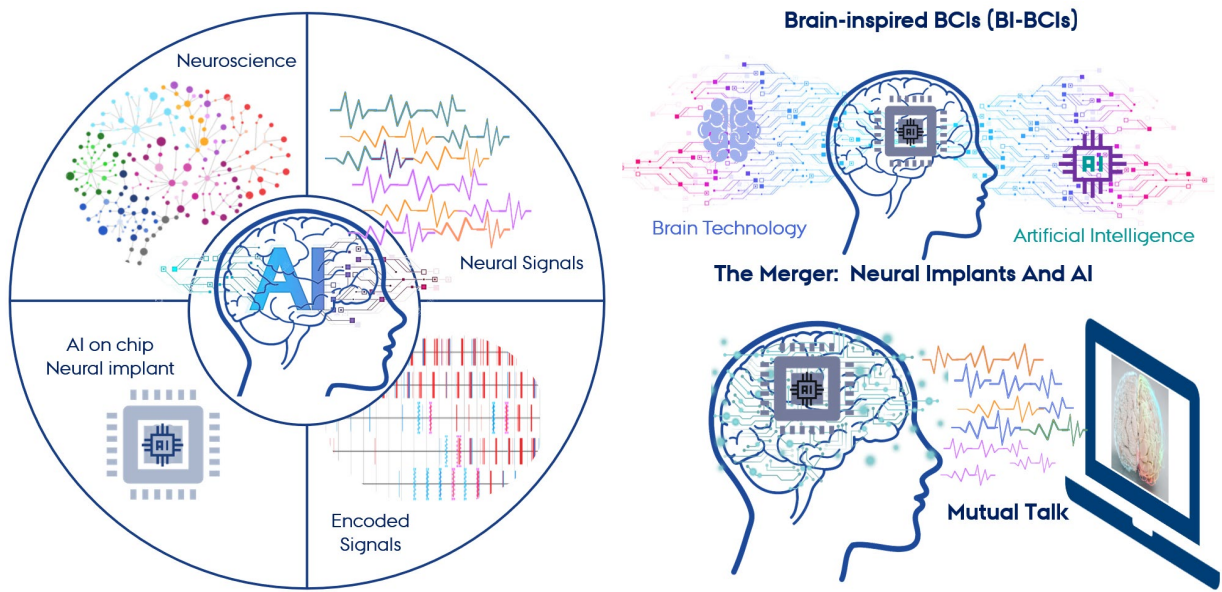

**Figure 11: Brain inspired-BCIs: Left:** BCI as an interdisciplinary field that combine neuroscience and engineering, **Right:** Merging neural implants interface with AI to obtain intelligent BCI that would interact with external devices.

Our future vision is to create an intelligent BCI system that would merge AI with neural interface technologies in what so-called Brain Inspired- BCI (BI-BCI) (Figure. 11) which interacts with the brain in the most natural way as it should. This paper reviewed the most recent developments in this domain, with a focus on brain-inspired computing techniques and their implementations. Merging brain-inspired neuromorphic computing with BCIs creates a human-in-loop system, in which both technologies interwind to alleviate disabilities and impairments and to restore human performance. The joint interaction between the human and the machine could lead us to realize augmented human intelligence, which in turn is one of the main endeavors for future BCI research. Moreover, we consider that these systems could lead to a whole new generation of intelligent brain interfaces with unprecedented therapeutic efficiency for a wide range of neurological and mental disorders as illustrated in Figure. 12.

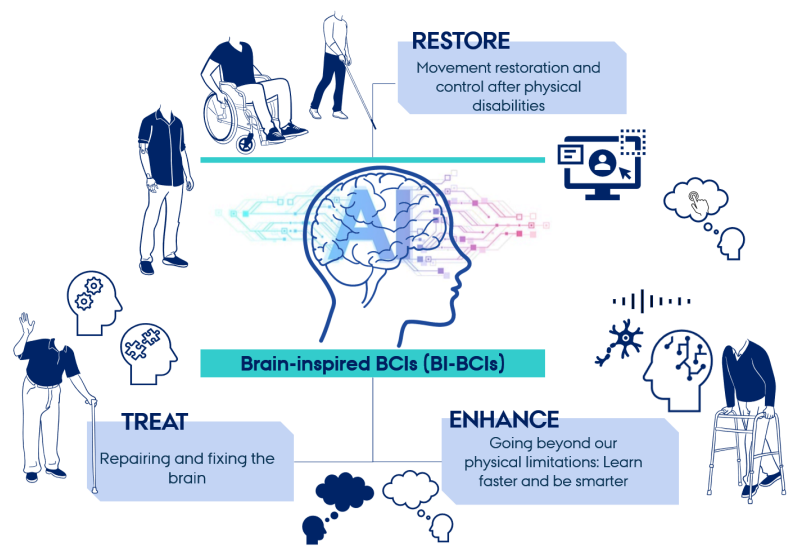

**Figure 12:** **Brain Inspired- Brain Computer Interfaces (BI-BCIs)** that links the human intelligence with AI-neural Implants.

Finally, BCI is an interdisciplinary field of research, and its advancement depends on the collaboration between neuroscience and engineering technologies. From the neuroscience standpoint, we need to understand better the function and the working mechanisms of the brain. While from the engineering standpoint, we need to create new develop intelligent miniaturized low-power neural implants that allows us to access deep brain structures, brain inspired neural algorithms to analyze neural activity and encode it efficiently to acquire real-time interaction, and spike-based energy efficient hardware.

**Acknowledgement:**
This work was supported by *Neuro-Sense* project, that received funding from the Lundbeck Foundation (LundbeckFonden) under the Reference number R402-2022-1413.

## 5 References

[1] S. S. Tarun Dua, Marco Garrido Cumbrera, Colin Mathers, "Global burden of neurological disorders: estimates and projections," *NEUROLOGICAL DISORDERS public health challenges*, pp. 27–39, 2007.

[2] G. Deuschl *et al.*, "The burden of neurological diseases in Europe: an analysis for the Global Burden of Disease Study 2017," *Lancet Public Health*, vol. 5, no. 10, pp. e551–e567, Oct. 2020, doi: 10.1016/S2468-2667(20)30190-0.

[3] X. Chen and W. Pan, "The Treatment Strategies for Neurodegenerative Diseases by Integrative Medicine," *Integr Med Int*, vol. 1, no. 4, pp. 223–225, Apr. 2014, doi: 10.1159/000381546.

[4] M. K. Poddar, A. Chakraborty, and S. Banerjee, "Neurodegeneration: Diagnosis, Prevention, and Therapy," *Oxidoreductase*, Jan. 2021, doi: 10.5772/INTECHOPEN.94950.
[5] J. K. Krauss et al., "Technology of deep brain stimulation: current status and future directions," *Nature Reviews Neurology 2020 17:2*, vol. 17, no. 2, pp. 75–87, Nov. 2020, doi: 10.1038/s41582-020-00426-z.
[6] G. Deuschl et al., "A Randomized Trial of Deep-Brain Stimulation for Parkinson's Disease," *New England Journal of Medicine*, vol. 355, no. 9, pp. 896–908, Aug. 2006, doi: 10.1056/NEJMOA060281.
[7] H. S. Mayberg et al., "Deep Brain Stimulation for Treatment-Resistant Depression," *Neuron*, vol. 45, no. 5, pp. 651–660, Mar. 2005, doi: 10.1016/J.NEURON.2005.02.014.
[8] A. I. Journal and D. L. Sorkin, "Cochlear implantation in the world's largest medical device market: Utilization and awareness of cochlear implants in the United States," *https://doi.org/10.1179/1467010013Z.00000000076*, vol. 14, no. SUPPL. 1, Mar. 2013, doi: 10.1179/1467010013Z.00000000076.
[9] A. K. Ahuja et al., "Blind subjects implanted with the Argus II retinal prosthesis are able to improve performance in a spatial-motor task," *British Journal of Ophthalmology*, vol. 95, no. 4, pp. 539–543, Apr. 2011, doi: 10.1136/BJO.2010.179622.
[10] I. Tochitsky et al., "Restoring Visual Function to Blind Mice with a Photoswitch that Exploits Electrophysiological Remodeling of Retinal Ganglion Cells," *Neuron*, vol. 81, no. 4, pp. 800–813, Feb. 2014, doi: 10.1016/J.NEURON.2014.01.003.
[11] M. J. Vansteensel and B. Jarosiewicz, "Brain-computer interfaces for communication," *Handb Clin Neurol*, vol. 168, pp. 67–85, Jan. 2020, doi: 10.1016/B978-0-444-63934-9.00007-X.
[12] C. Cinel, D. Valeriani, and R. Poli, "Neurotechnologies for human cognitive augmentation: Current state of the art and future prospects," *Front Hum Neurosci*, vol. 13, no. January, 2019, doi: 10.3389/fnhum.2019.00013.
[13] R. Bruffaerts, "Machine learning in neurology: what neurologists can learn from machines and vice versa," *J Neurol*, no. November 2018, 2018, doi: 10.1007/s00415-018-8990-9.
[14] J. Watts, A. Khojandi, O. Shylo, and R. A. Ramdhani, "Machine learning's application in deep brain stimulation for parkinson's disease: A review," *Brain Sci*, vol. 10, no. 11, pp. 1–16, Nov. 2020, doi: 10.3390/BRAINSCI10110809.
[15] M. A. Myszczynska et al., "Applications of machine learning to diagnosis and treatment of neurodegenerative diseases," *Nat Rev Neurol*, vol. 16, no. 8, pp. 440–456, 2020, doi: 10.1038/s41582-020-0377-8.
[16] M. C. Lo and A. S. Widge, "Closed-loop neuromodulation systems: next-generation treatments for psychiatric illness," *https://doi.org/10.1080/09540261.2017.1282438*, vol. 29, no. 2, pp. 191–204, Mar. 2017, doi: 10.1080/09540261.2017.1282438.
[17] T. Stieglitz, "Why Neurotechnologies? About the Purposes, Opportunities and Limitations of Neurotechnologies in Clinical Applications," *Neuroethics*, vol. 14, no. 1, pp. 5–16, 2021, doi: 10.1007/s12152-019-09406-7.
[18] V. L. Feigin et al., "Global, regional, and national burden of neurological disorders, 1990-2016: a systematic analysis for the Global Burden of Disease Study 2016," *Lancet Neurol*, vol. 18, no. 5, pp. 459–480, May 2019, doi: 10.1016/S1474-4422(18)30499-X.
[19] A. Abbott, "Bilion-Dollar Brain maps: What we've learnt," Oct. 07, 2021. Accessed: Jun. 20, 2022. [Online]. Available: https://media.nature.com/original/magazine-assets/d41586-021-02661-w/d41586-021-02661-w.pdf
[20] "Home - Neuralink." https://neuralink.com/ (accessed Sep. 21, 2022).
[21] "Paradromics." https://www.paradromics.com/ (accessed Sep. 21, 2022).
[22] "Synchron | Unlocking the natural highways of the brain." https://synchron.com/ (accessed Sep. 21, 2022).
[23] "Blackrock Neurotech - Blackrock Neurotech." https://blackrockneurotech.com/ (accessed Sep. 21, 2022).
[24] "Neurable." https://neurable.com/ (accessed Sep. 21, 2022).
[25] "Thync | Neurostimulation For Everyone – Thync Global, Inc." https://thync.com/ (accessed Sep. 21, 2022).
[26] "Spinal Cord Stimulation Systems | Medtronic." https://europe.medtronic.com/xd-en/healthcare-professionals/products/neurological/spinal-cord-stimulation-systems.html (accessed Sep. 21, 2022).
[27] "kernel | Home." https://www.kernel.com/ (accessed Sep. 21, 2022).
[28] "NeuroPace | A Smarter Way to Treat Epilepsy." https://neuropace.com/ (accessed Sep. 21, 2022).
[29] B. Jarosiewicz and M. Morrell, "The RNS System: brain-responsive neurostimulation for the treatment of epilepsy," *https://doi.org/10.1080/17434440.2019.1683445*, vol. 18, no. 2, pp. 129–138, 2020, doi: 10.1080/17434440.2019.1683445.
[30] S. Stanslaski et al., "A Chronically Implantable Neural Coprocessor for Investigating the Treatment of Neurological Disorders," *IEEE Trans Biomed Circuits Syst*, vol. 12, no. 6, pp. 1230–1245, Dec. 2018, doi: 10.1109/TBCAS.2018.2880148.
[31] E. Musk and Neuralink, "An Integrated Brain-Machine Interface Platform With Thousands of Channels," *J Med Internet Res 2019;21(10):e16194 https://www.jmir.org/2019/10/e16194*, vol. 21, no. 10, p. e16194, Oct. 2019, doi: 10.2196/16194.
[32] J. J. Vidal, "Toward direct brain-computer communication," *Annu Rev Biophys Bioeng*, vol. 2, pp. 157–180, 1973, doi: 10.1146/ANNUREV.BB.02.060173.001105.
[33] S. Saha et al., "Progress in Brain Computer Interface: Challenges and Opportunities," *Front Syst Neurosci*, vol. 15, no. February, pp. 1–20, 2021, doi: 10.3389/fnsys.2021.578875.
[34] J. K. Chapin, K. A. Moxon, R. S. Markowitz, and M. A. L. Nicolelis, "Real-time control of a robot arm using simultaneously recorded neurons in the motor cortex," *Nature Neuroscience 1999 2:7*, vol. 2, no. 7, pp. 664–670, Jul. 1999, doi: 10.1038/10223.

[35] D. M. Taylor, S. I. H. Tillery, and A. B. Schwartz, "Direct cortical control of 3D neuroprosthetic devices," *Science (1979)*, vol. 296, no. 5574, pp. 1829–1832, Jun. 2002, doi: 10.1126/SCIENCE.1070291/SUPPL_FILE/1070291S3.MOV.

[36] G. Santhanam, S. I. Ryu, B. M. Yu, A. Afshar, and K. V. Shenoy, "A high-performance brain–computer interface," *Nature 2006 442:7099*, vol. 442, no. 7099, pp. 195–198, Jul. 2006, doi: 10.1038/nature04968.

[37] M. Velliste, S. Perel, M. C. Spalding, A. S. Whitford, and A. B. Schwartz, "Cortical control of a prosthetic arm for self-feeding," *Nature 2008 453:7198*, vol. 453, no. 7198, pp. 1098–1101, May 2008, doi: 10.1038/nature06996.

[38] L. R. Hochberg *et al.*, "Neuronal ensemble control of prosthetic devices by a human with tetraplegia," *Nature 2006 442:7099*, vol. 442, no. 7099, pp. 164–171, Jul. 2006, doi: 10.1038/nature04970.

[39] J. P. Donoghue, A. Nurmikko, M. Black, and L. R. Hochberg, "Assistive technology and robotic control using motor cortex ensemble-based neural interface systems in humans with tetraplegia," *J Physiol*, vol. 579, no. 3, pp. 603–611, Mar. 2007, doi: 10.1113/JPHYSIOL.2006.127209.

[40] M. A. L. Nicolelis, "Actions from thoughts," *Nature 2001 409:6818*, vol. 409, no. 6818, pp. 403–407, Jan. 2001, doi: 10.1038/35053191.

[41] C. Brunner *et al.*, "BNCI Horizon 2020: towards a roadmap for the BCI community," *http://dx.doi.org/10.1080/2326263X.2015.1008956*, vol. 2, no. 1, pp. 1–10, Jan. 2015, doi: 10.1080/2326263X.2015.1008956.

[42] U. Chaudhary, "Brain-computer interfaces for communication and rehabilitation," *Nat Rev Neurol*, vol. 12, no. 9, pp. 513–525, 2016, doi: 10.1038/nrneurol.2016.113.

[43] A. Caria, R. Sitaram, and N. Birbaumer, "Real-time fMRI: A tool for local brain regulation," *Neuroscientist*, vol. 18, no. 5, pp. 487–501, Oct. 2012, doi: 10.1177/1073858411407205.

[44] Y. Wang, R. Wang, X. Gao, B. Hong, and S. Gao, "A practical VEP-based brain-computer interface," *IEEE Transactions on Neural Systems and Rehabilitation Engineering*, vol. 14, no. 2, pp. 234–239, Jun. 2006, doi: 10.1109/TNSRE.2006.875576.

[45] M. A. Romero-Laiseca *et al.*, "A Low-Cost Lower-Limb Brain-Machine Interface Triggered by Pedaling Motor Imagery for Post-Stroke Patients Rehabilitation," *IEEE Transactions on Neural Systems and Rehabilitation Engineering*, vol. 28, no. 4, pp. 988–996, Apr. 2020, doi: 10.1109/TNSRE.2020.2974056.

[46] F. Nijboer *et al.*, "An auditory brain–computer interface (BCI)," *J Neurosci Methods*, vol. 167, no. 1, pp. 43–50, Jan. 2008, doi: 10.1016/J.JNEUMETH.2007.02.009.

[47] Z. R. Lugo *et al.*, "A vibrotactile P300-based brain-computer interface for consciousness detection and communication," *Clin EEG Neurosci*, vol. 45, no. 1, pp. 14–21, Jan. 2014, doi: 10.1177/1550059413505533.

[48] S. Marchesotti, M. Bassolino, A. Serino, H. Bleuler, and O. Blanke, "Quantifying the role of motor imagery in brain-machine interfaces," *Scientific Reports 2016 6:1*, vol. 6, no. 1, pp. 1–12, Apr. 2016, doi: 10.1038/srep24076.

[49] L. Acqualagna, L. Botrel, C. Vidaurre, A. Kübler, and B. Blankertz, "Large-Scale Assessment of a Fully Automatic Co-Adaptive Motor Imagery-Based Brain Computer Interface," *PLoS One*, vol. 11, no. 2, p. e0148886, Feb. 2016, doi: 10.1371/JOURNAL.PONE.0148886.

[50] U. Chaudhary, "Brain–computer interfaces for communication and rehabilitation," *Nature Reviews Neurology 2016 12:9*, vol. 12, no. 9, pp. 513–525, Aug. 2016, doi: 10.1038/nrneurol.2016.113.

[51] U. Chaudhary, B. Xia, S. Silvoni, L. G. Cohen, and N. Birbaumer, "Brain–Computer Interface–Based Communication in the Completely Locked-In State," *PLoS Biol*, vol. 15, no. 1, p. e1002593, Jan. 2017, doi: 10.1371/JOURNAL.PBIO.1002593.

[52] T. O. Zander, C. Kothe, S. Welke, and M. Roetting, "Utilizing secondary input from passive brain-computer interfaces for enhancing human-machine interaction," *Lecture Notes in Computer Science (including subseries Lecture Notes in Artificial Intelligence and Lecture Notes in Bioinformatics)*, vol. 5638 LNAI, pp. 759–771, 2009, doi: 10.1007/978-3-642-02812-0_86/COVER/.

[53] P. Arico, G. Borghini, G. Di Flumeri, N. Sciaraffa, and F. Babiloni, "Passive BCI beyond the lab: current trends and future directions," *Physiol Meas*, vol. 39, no. 8, p. 08TR02, Aug. 2018, doi: 10.1088/1361-6579/AAD57E.

[54] C. S. Wei, Y. Te Wang, C. T. Lin, and T. P. Jung, "Toward Drowsiness Detection Using Non-hair-Bearing EEG-Based Brain-Computer Interfaces," *IEEE Transactions on Neural Systems and Rehabilitation Engineering*, vol. 26, no. 2, pp. 400–406, Feb. 2018, doi: 10.1109/TNSRE.2018.2790359.

[55] Z. Gao *et al.*, "EEG-Based Spatio-Temporal Convolutional Neural Network for Driver Fatigue Evaluation," *IEEE Trans Neural Netw Learn Syst*, vol. 30, no. 9, pp. 2755–2763, Sep. 2019, doi: 10.1109/TNNLS.2018.2886414.

[56] G. Buzsáki, C. A. Anastassiou, and C. Koch, "The origin of extracellular fields and currents — EEG, ECoG, LFP and spikes," *Nature Reviews Neuroscience 2012 13:6*, vol. 13, no. 6, pp. 407–420, May 2012, doi: 10.1038/nrn3241.

[57] J. Mellinger *et al.*, "An MEG-based brain–computer interface (BCI)," *Neuroimage*, vol. 36, no. 3, pp. 581–593, Jul. 2007, doi: 10.1016/J.NEUROIMAGE.2007.03.019.

[58] B. K. Min, M. J. Marzelli, and S. S. Yoo, "Neuroimaging-based approaches in the brain–computer interface," *Trends Biotechnol*, vol. 28, no. 11, pp. 552–560, Nov. 2010, doi: 10.1016/J.TIBTECH.2010.08.002.

[59] H. Khan, N. Naseer, A. Yazidi, P. K. Eide, H. W. Hassan, and P. Mirtaheri, "Analysis of Human Gait Using Hybrid EEG-fNIRS-Based BCI System: A Review," *Front Hum Neurosci*, vol. 14, p. 613254, Jan. 2020, doi: 10.3389/FNHUM.2020.613254.

[60] J. D. Haynes, "A Primer on Pattern-Based Approaches to fMRI: Principles, Pitfalls, and Perspectives," *Neuron*, vol. 87, no. 2, pp. 257–270, Jul. 2015, doi: 10.1016/J.NEURON.2015.05.025.

[61] R. Abiri, S. Borhani, E. W. Sellers, Y. Jiang, and X. Zhao, "A comprehensive review of EEG-based brain–computer interface paradigms," *J Neural Eng*, vol. 16, no. 1, p. 011001, Jan. 2019, doi: 10.1088/1741-2552/AAF12E.
[62] M. A. Lopez-Gordo, D. Sanchez Morillo, and F. Pelayo Valle, "Dry EEG Electrodes," *Sensors 2014, Vol. 14, Pages 12847-12870*, vol. 14, no. 7, pp. 12847–12870, Jul. 2014, doi: 10.3390/S140712847.
[63] J. Xu, S. Mitra, C. Van Hoof, R. F. Yazicioglu, and K. A. A. Makinwa, "Active Electrodes for Wearable EEG Acquisition: Review and Electronics Design Methodology," *IEEE Rev Biomed Eng*, vol. 10, pp. 187–198, 2017, doi: 10.1109/RBME.2017.2656388.
[64] J. H. Lee, J. Kim, and S. S. Yoo, "Real-time fMRI-based neurofeedback reinforces causality of attention networks," *Neurosci Res*, vol. 72, no. 4, pp. 347–354, Apr. 2012, doi: 10.1016/J.NEURES.2012.01.002.
[65] Y. Miyawaki et al., "Visual Image Reconstruction from Human Brain Activity using a Combination of Multiscale Local Image Decoders," *Neuron*, vol. 60, no. 5, pp. 915–929, Dec. 2008, doi: 10.1016/J.NEURON.2008.11.004.
[66] A. M. Owen and M. R. Coleman, "Functional neuroimaging of the vegetative state," *Nature Reviews Neuroscience 2008 9:3*, vol. 9, no. 3, pp. 235–243, Mar. 2008, doi: 10.1038/nrn2330.
[67] A. M. Owen, M. R. Coleman, M. Boly, M. H. Davis, S. Laureys, and J. D. Pickard, "Detecting awareness in the vegetative state," *Science (1979)*, vol. 313, no. 5792, p. 1402, Sep. 2006, doi: 10.1126/SCIENCE.1130197/SUPPL_FILE/OWEN.SOM.PDF.
[68] G. Schalk and E. C. Leuthardt, "Brain-computer interfaces using electrocorticographic signals," *IEEE Rev Biomed Eng*, vol. 4, pp. 140–154, 2011, doi: 10.1109/RBME.2011.2172408.
[69] T. Kaiju et al., "High spatiotemporal resolution ECoG recording of somatosensory evoked potentials with flexible micro-electrode arrays," *Front Neural Circuits*, vol. 11, p. 20, Apr. 2017, doi: 10.3389/FNCIR.2017.00020/BIBTEX.
[70] S. N. Flesher et al., "Intracortical microstimulation of human somatosensory cortex," *Sci Transl Med*, vol. 8, no. 361, Oct. 2016, doi: 10.1126/SCITRANSLMED.AAF8083/SUPPL_FILE/8-361RA141_SM.PDF.
[71] C. Pandarinath et al., "High performance communication by people with paralysis using an intracortical brain-computer interface," *Elife*, vol. 6, Feb. 2017, doi: 10.7554/ELIFE.18554.
[72] S. Venkatraman, K. Elkabany, J. D. Long, Y. Yao, and J. M. Carmena, "A system for neural recording and closed-loop intracortical microstimulation in awake rodents," *IEEE Trans Biomed Eng*, vol. 56, no. 1, pp. 15–22, Jan. 2009, doi: 10.1109/TBME.2008.2005944.
[73] E. Tolstosheeva et al., "A multi-channel, flex-rigid ECoG microelectrode array for visual cortical interfacing," *mdpi.com*, vol. 15, pp. 832–854, 2015, doi: 10.3390/s150100832.
[74] E. Leuthardt, G. Schalk, … J. W.-J. of neural, and undefined 2004, "A brain–computer interface using electrocorticographic signals in humans," *iopscience.iop.org*, 2004, doi: 10.1088/1741-2560/1/2/001.
[75] D. Khodagholy, T. Doublet, … P. Q.-N., and undefined 2013, "In vivo recordings of brain activity using organic transistors," *nature.com*, Accessed: Jun. 27, 2023. [Online]. Available: https://www.nature.com/articles/ncomms2573
[76] M. A. Escabí et al., "A high-density, high-channel count, multiplexed μECoG array for auditory-cortex recordings," *journals.physiology.org*, vol. 112, no. 6, pp. 1566–1583, Sep. 2014, doi: 10.1152/jn.00179.2013.
[77] C.-H. Chiang et al., "A low-cost, multiplexed μECoG system for high-density recordings in freely moving rodents," *iopscience.iop.org*, vol. 13, pp. 26030–26043, 2016, doi: 10.1088/1741-2560/13/2/026030.
[78] M. P. Branco, Z. V. Freudenburg, E. J. Aarnoutse, M. G. Bleichner, M. J. Vansteensel, and N. F. Ramsey, "Decoding hand gestures from primary somatosensory cortex using high-density ECoG," *Neuroimage*, vol. 147, pp. 130–142, Feb. 2017, doi: 10.1016/J.NEUROIMAGE.2016.12.004.
[79] A. Obaid et al., "Massively parallel microwire arrays integrated with CMOS chips for neural recording," *Sci Adv*, vol. 6, no. 12, 2020, doi: 10.1126/SCIADV.AAY2789.
[80] K. D. Wise, D. J. Anderson, J. F. Hetke, D. R. Kipke, and K. Najafi, "Wireless implantable microsystems: High-density electronic interfaces to the nervous system," *Proceedings of the IEEE*, vol. 92, no. 1, pp. 76–97, 2004, doi: 10.1109/JPROC.2003.820544.
[81] P. J. Rousche and R. A. Normann, "Chronic recording capability of the Utah Intracortical Electrode Array in cat sensory cortex," *J Neurosci Methods*, vol. 82, no. 1, pp. 1–15, Jul. 1998, doi: 10.1016/S0165-0270(98)00031-4.
[82] M. Abidian, D. M.-A. functional materials, and undefined 2009, "Multifunctional nanobiomaterials for neural interfaces," *Wiley Online Library*, vol. 19, no. 4, pp. 573–585, Feb. 2009, doi: 10.1002/adfm.200801473.
[83] M. A. L Nicolelis et al., "Chronic, multisite, multielectrode recordings in macaque monkeys," 2003, Accessed: Jun. 27, 2023. [Online]. Available: www.pnas.org
[84] J. Chung et al., "High-density, long-lasting, and multi-region electrophysiological recordings using polymer electrode arrays," *Elsevier*, Accessed: Jun. 27, 2023. [Online]. Available: https://www.sciencedirect.com/science/article/pii/S0896627318309930
[85] T. L. Hanson, C. A. Diaz-Botia, V. Kharazia, M. M. Maharbiz, and P. N. Sabes, "The 'sewing machine' for minimally invasive neural recording," *biorxiv.org*, doi: 10.1101/578542.
[86] K. Tanaka et al., "A high-density carbon fiber neural recording array technology," *iopscience.iop.org*, 2019, doi: 10.1088/1741-2552/aae8d9.
[87] S. M. Lawrence, G. S. Dhillon, and K. W. Horch, "Fabrication and characteristics of an implantable, polymer-based, intrafascicular electrode.," *J Neurosci Methods*, vol. 131, no. 1–2, pp. 9–26, Dec. 2003, doi: 10.1016/S0165-0270(03)00231-0.
[88] T. Boretius et al., "A transverse intrafascicular multichannel electrode (TIME) to interface with the peripheral nerve," *Biosens Bioelectron*, vol. 26, no. 1, pp. 62–69, 2010, doi: 10.1016/J.BIOS.2010.05.010.

[89] S. Raspopovic, "Advancing limb neural prostheses," *Science (1979)*, vol. 370, no. 6514, pp. 290–291, Oct. 2020, doi: 10.1126/SCIENCE.ABB1073/ASSET/A59CC366-7A10-49C0-A13C-A7FB951DD316/ASSETS/GRAPHIC/370_290_F1.JPEG.

[90] N. A. Steinmetz et al., "Neuropixels 2.0: A miniaturized high-density probe for stable, long-term brain recordings," *Science (1979)*, vol. 372, no. 6539, Apr. 2021, doi: 10.1126/SCIENCE.ABF4588/SUPPL_FILE/ABF4888_MDAR_REPRODUCIBILITY_CHECKLIST.PDF.

[91] D. Khodagholy et al., "NeuroGrid: recording action potentials from the surface of the brain," 2014, doi: 10.1038/nn.3905.

[92] G. Hong and C. M. Lieber, "Novel electrode technologies for neural recordings," *Nat Rev Neurosci*, vol. 20, no. 6, pp. 330–345, Jun. 2019, doi: 10.1038/S41583-019-0140-6.

[93] T. H. Kim et al., "Long-Term Optical Access to an Estimated One Million Neurons in the Live Mouse Cortex," *Cell Rep*, vol. 17, no. 12, pp. 3385–3394, Dec. 2016, doi: 10.1016/J.CELREP.2016.12.004.

[94] N. T. Ersaro, C. Yalcin, and R. Muller, "The future of brain–machine interfaces is optical," *Nature Electronics 2023 6:2*, vol. 6, no. 2, pp. 96–98, Feb. 2023, doi: 10.1038/s41928-023-00926-y.

[95] S. Schaffelhofer, A. Agudelo-Toro, and H. Scherberger, "Decoding a wide range of hand configurations from macaque motor, premotor, and parietal cortices," *Journal of Neuroscience*, vol. 35, no. 3, pp. 1068–1081, Jan. 2015, doi: 10.1523/JNEUROSCI.3594-14.2015.

[96] C. E. Bouton et al., "Restoring cortical control of functional movement in a human with quadriplegia," *Nature 2016 533:7602*, vol. 533, no. 7602, pp. 247–250, Apr. 2016, doi: 10.1038/nature17435.

[97] G. Lajoie, N. I. Krouchev, J. F. Kalaska, A. L. Fairhall, and E. E. Fetz, "Correlation-based model of artificially induced plasticity in motor cortex by a bidirectional brain-computer interface," *PLoS Comput Biol*, vol. 13, no. 2, p. e1005343, Feb. 2017, doi: 10.1371/JOURNAL.PCBI.1005343.

[98] D. L. K. Yamins, H. Hong, C. F. Cadieu, E. A. Solomon, D. Seibert, and J. J. DiCarlo, "Performance-optimized hierarchical models predict neural responses in higher visual cortex," *Proc Natl Acad Sci U S A*, vol. 111, no. 23, pp. 8619–8624, Jun. 2014, doi: 10.1073/PNAS.1403112111/SUPPL_FILE/PNAS.201403112SI.PDF.

[99] M. R. Cohen and J. H. R. Maunsell, "A Neuronal Population Measure of Attention Predicts Behavioral Performance on Individual Trials," *Journal of Neuroscience*, vol. 30, no. 45, pp. 15241–15253, Nov. 2010, doi: 10.1523/JNEUROSCI.2171-10.2010.

[100] M. Siegel, T. J. Buschman, and E. K. Miller, "Cortical information flow during flexible sensorimotor decisions," *Science (1979)*, vol. 348, no. 6241, pp. 1352–1355, Jun. 2015, doi: 10.1126/SCIENCE.AAB0551/SUPPL_FILE/SIEGEL-SM.PDF.

[101] R. Q. Quiroga, "Concept cells: the building blocks of declarative memory functions," *Nature Reviews Neuroscience 2012 13:8*, vol. 13, no. 8, pp. 587–597, Jul. 2012, doi: 10.1038/nrn3251.

[102] H. Gelbard-Sagiv, R. Mukamel, M. Harel, R. Malach, and I. Fried, "Internally generated reactivation of single neurons in human hippocampus during free recall," *Science (1979)*, vol. 322, no. 5898, pp. 96–101, Oct. 2008, doi: 10.1126/SCIENCE.1164685/SUPPL_FILE/GELBARD-SAGIV1164685S2.REVISION.1.MOVIE.MOV.

[103] R. Q. Quiroga, L. Reddy, G. Kreiman, C. Koch, and I. Fried, "Invariant visual representation by single neurons in the human brain," *Nature 2005 435:7045*, vol. 435, no. 7045, pp. 1102–1107, Jun. 2005, doi: 10.1038/nature03687.

[104] M. J. Vansteensel et al., "Fully Implanted Brain–Computer Interface in a Locked-In Patient with ALS," *New England Journal of Medicine*, vol. 375, no. 21, pp. 2060–2066, Nov. 2016, doi: 10.1056/NEJMOA1608085/SUPPL_FILE/NEJMOA1608085_DISCLOSURES.PDF.

[105] M. Sharifshazileh, K. Burelo, J. Sarnthein, and G. Indiveri, "An electronic neuromorphic system for real-time detection of high frequency oscillations (HFO) in intracranial EEG," *Nature Communications 2021 12:1*, vol. 12, no. 1, pp. 1–14, May 2021, doi: 10.1038/s41467-021-23342-2.

[106] W. Park, G. H. Kwon, Y. H. Kim, J. H. Lee, and L. Kim, "EEG response varies with lesion location in patients with chronic stroke," *J Neuroeng Rehabil*, vol. 13, no. 1, pp. 1–10, Mar. 2016, doi: 10.1186/S12984-016-0120-2/FIGURES/4.

[107] S. I. Gonçalves et al., "Correlating the alpha rhythm to BOLD using simultaneous EEG/fMRI: Inter-subject variability," *Neuroimage*, vol. 30, no. 1, pp. 203–213, Mar. 2006, doi: 10.1016/J.NEUROIMAGE.2005.09.062.

[108] I. Käthner, S. C. Wriessnegger, G. R. Müller-Putz, A. Kübler, and S. Halder, "Effects of mental workload and fatigue on the P300, alpha and theta band power during operation of an ERP (P300) brain–computer interface," *Biol Psychol*, vol. 102, no. 1, pp. 118–129, Oct. 2014, doi: 10.1016/J.BIOPSYCHO.2014.07.014.

[109] N. N. Johnson et al., "Combined rTMS and virtual reality brain–computer interface training for motor recovery after stroke," *J Neural Eng*, vol. 15, no. 1, p. 016009, Jan. 2018, doi: 10.1088/1741-2552/AA8CE3.

[110] S. Saha et al., "Wavelet Entropy-Based Inter-subject Associative Cortical Source Localization for Sensorimotor BCI," *Front Neuroinform*, vol. 13, Jul. 2019, doi: 10.3389/FNINF.2019.00047/FULL.

[111] D. Sand, Z. Peremen, D. Haor, D. Arkadir, H. Bergman, and A. Geva, "Optimization of deep brain stimulation in STN among patients with Parkinson's disease using a novel EEG-based tool," *Brain Stimul*, vol. 10, no. 2, p. 510, Mar. 2017, doi: 10.1016/J.BRS.2017.01.490.

[112] U. Chaudhary, N. Birbaumer, and M. R. Curado, "Brain-Machine Interface (BMI) in paralysis," *Ann Phys Rehabil Med*, vol. 58, no. 1, pp. 9–13, Feb. 2015, doi: 10.1016/J.REHAB.2014.11.002.

[113] R. Fukuma et al., "Real-time control of a neuroprosthetic hand by magnetoencephalographic signals from paralysed patients," *Sci Rep*, vol. 6, Feb. 2016, doi: 10.1038/SREP21781.

[114] V. D. Calhoun and T. Adali, "Time-Varying Brain Connectivity in fMRI Data: Whole-brain data-driven approaches for capturing and characterizing dynamic states," *IEEE Signal Process Mag*, vol. 33, no. 3, pp. 52–66, May 2016, doi: 10.1109/MSP.2015.2478915.

[115] M. T. DeBettencourt, J. D. Cohen, R. F. Lee, K. A. Norman, and N. B. Turk-Browne, "Closed-loop training of attention with real-time brain imaging," *Nat Neurosci*, vol. 18, no. 3, pp. 470–478, Mar. 2015, doi: 10.1038/NN.3940.

[116] A. Rezazadeh Sereshkeh, R. Yousefi, A. T. Wong, F. Rudzicz, and T. Chau, "Development of a ternary hybrid fNIRS-EEG brain–computer interface based on imagined speech," *Brain-Computer Interfaces*, vol. 6, no. 4, pp. 128–140, Oct. 2019, doi: 10.1080/2326263X.2019.1698928.

[117] A. M. Chiarelli, P. Croce, A. Merla, and F. Zappasodi, "Deep learning for hybrid EEG-fNIRS brain-computer interface: application to motor imagery classification," *J Neural Eng*, vol. 15, no. 3, Apr. 2018, doi: 10.1088/1741-2552/AAAF82.

[118] C. Li, M. Su, J. Xu, H. Jin, and L. Sun, "A Between-Subject fNIRS-BCI Study on Detecting Self-Regulated Intention during Walking," *IEEE Transactions on Neural Systems and Rehabilitation Engineering*, vol. 28, no. 2, pp. 531–540, Feb. 2020, doi: 10.1109/TNSRE.2020.2965628.

[119] L. Rejc et al., "Longitudinal evaluation of neuroinflammation and oxidative stress in a mouse model of Alzheimer disease using positron emission tomography," *Alzheimers Res Ther*, vol. 14, no. 1, pp. 1–14, Dec. 2022, doi: 10.1186/S13195-022-01016-5/FIGURES/8.

[120] A. Kaas, R. Goebel, G. Valente, and B. Sorger, "Topographic Somatosensory Imagery for Real-Time fMRI Brain-Computer Interfacing," *Front Hum Neurosci*, vol. 13, p. 427, Dec. 2019, doi: 10.3389/FNHUM.2019.00427/BIBTEX.

[121] M. Pahwa, M. Kusner, C. D. Hacker, D. T. Bundy, K. Q. Weinberger, and E. C. Leuthardt, "Optimizing the Detection of Wakeful and Sleep-Like States for Future Electrocorticographic Brain Computer Interface Applications.," *PLoS One*, vol. 10, no. 11, p. e0142947, Jan. 2015, doi: 10.1371/JOURNAL.PONE.0142947.

[122] F. Sauter-Starace et al., "Long-Term Sheep Implantation of WIMAGINE®, a Wireless 64-Channel Electrocorticogram Recorder," *Front Neurosci*, vol. 13, Aug. 2019, doi: 10.3389/FNINS.2019.00847/PDF.

[123] G. K. Anumanchipalli, J. Chartier, and E. F. Chang, "Speech synthesis from neural decoding of spoken sentences," *Nature 2019 568:7753*, vol. 568, no. 7753, pp. 493–498, Apr. 2019, doi: 10.1038/s41586-019-1119-1.

[124] J. G. Makin, D. A. Moses, and E. F. Chang, "Machine translation of cortical activity to text with an encoder–decoder framework," *Nature Neuroscience 2020 23:4*, vol. 23, no. 4, pp. 575–582, Mar. 2020, doi: 10.1038/s41593-020-0608-8.

[125] M. Yin et al., "Wireless Neurosensor for Full-Spectrum Electrophysiology Recordings during Free Behavior," *Neuron*, vol. 84, no. 6, pp. 1170–1182, Dec. 2014, doi: 10.1016/J.NEURON.2014.11.010.

[126] E. W. Keefer, B. R. Botterman, M. I. Romero, A. F. Rossi, and G. W. Gross, "Carbon nanotube coating improves neuronal recordings," *Nature Nanotechnology 2008 3:7*, vol. 3, no. 7, pp. 434–439, Jun. 2008, doi: 10.1038/nnano.2008.174.

[127] J. E. Downey et al., "Blending of brain-machine interface and vision-guided autonomous robotics improves neuroprosthetic arm performance during grasping," *J Neuroeng Rehabil*, vol. 13, no. 1, pp. 1–12, Mar. 2016, doi: 10.1186/S12984-016-0134-9/TABLES/2.

[128] T. J. Oxley et al., "Motor neuroprosthesis implanted with neurointerventional surgery improves capacity for activities of daily living tasks in severe paralysis: first in-human experience," *J Neurointerv Surg*, vol. 13, no. 2, pp. 102–108, Feb. 2021, doi: 10.1136/NEURINTSURG-2020-016862.

[129] T. J. Oxley et al., "Minimally invasive endovascular stent-electrode array for high-fidelity, chronic recordings of cortical neural activity," *Nature Biotechnology 2016 34:3*, vol. 34, no. 3, pp. 320–327, Feb. 2016, doi: 10.1038/nbt.3428.

[130] R. M. Neely, D. K. Piech, S. R. Santacruz, M. M. Maharbiz, and J. M. Carmena, "Recent advances in neural dust: towards a neural interface platform," *Curr Opin Neurobiol*, vol. 50, pp. 64–71, Jun. 2018, doi: 10.1016/J.CONB.2017.12.010.

[131] T. Mondal, K. Laursen, S. Hosseini, A. Rashidi, F. Moradi, and B. Corbett, "Realization of high efficiency ultrasound-powered micro-LEDs for optogenetics," *https://doi.org/10.1117/12.2555762*, vol. 11364, pp. 126–133, Apr. 2020, doi: 10.1117/12.2555762.

[132] K. Laursen, A. Rashidi, S. Hosseini, T. Mondal, B. Corbett, and F. Moradi, "Ultrasonically Powered Compact Implantable Dust for Optogenetics," *IEEE Trans Biomed Circuits Syst*, vol. 14, no. 3, pp. 583–594, Jun. 2020, doi: 10.1109/TBCAS.2020.2984921.

[133] S. Hosseini, K. Laursen, A. Rashidi, T. Mondal, B. Corbett, and F. Moradi, "S-MRUT: Sectored-Multiring Ultrasonic Transducer for Selective Powering of Brain Implants," *IEEE Trans Ultrason Ferroelectr Freq Control*, vol. 68, no. 1, pp. 191–200, Jan. 2021, doi: 10.1109/TUFFC.2020.3001084.

[134] A. Rashidi, S. Hosseini, K. Laursen, and F. Moradi, "STARDUST: Optogenetics, electrophysiology and pharmacology with an ultrasonically powered DUST for parkinson's disease," *2019 26th IEEE International Conference on Electronics, Circuits and Systems, ICECS 2019*, pp. 109–110, Nov. 2019, doi: 10.1109/ICECS46596.2019.8965042.

[135] M. N. Christensen, M. Zamani, A. Rashidi, and F. Moradi, "Ultrasonic Backscatter Communication for Brain Implants: Mathematical Model, Simulation, and Measurement," *BioCAS 2021 - IEEE Biomedical Circuits and Systems Conference, Proceedings*, 2021, doi: 10.1109/BIOCAS49922.2021.9644968.

[136] A. Rashidi, K. Laursen, S. Hosseini, H. A. Huynh, and F. Moradi, "An Implantable Ultrasonically Powered System for Optogenetic Stimulation with Power-Efficient Active Rectifier and Charge-Reuse Capability," *IEEE Trans Biomed Circuits Syst*, vol. 13, no. 6, pp. 1362–1371, Dec. 2019, doi: 10.1109/TBCAS.2019.2949154.

[137] A. Paulk, Y. Kfir, A. Khanna, … M. M.-N., and undefined 2022, "Large-scale neural recordings with single neuron resolution using Neuropixels probes in human cortex," *nature.com*, Accessed: Jun. 30, 2023. [Online]. Available: https://www.nature.com/articles/s41593-021-00997-0

[138] A. Paolo Buccino *et al.*, "3D printed guide tube system for acute Neuropixels probe recordings in non-human primates," *iopscience.iop.org*, 2023, doi: 10.1088/1741-2552/acd0d7.

[139] N. A. Steinmetz *et al.*, "Neuropixels 2.0: A miniaturized high-density probe for stable, long-term brain recordings," *Science (1979)*, vol. 372, no. 6539, Apr. 2021, doi: 10.1126/SCIENCE.ABF4588.

[140] R. R.-C. opinion in neurobiology and undefined 2019, "Towards neural co-processors for the brain: combining decoding and encoding in brain–computer interfaces," *Elsevier*, Accessed: Sep. 05, 2023. [Online]. Available: https://www.sciencedirect.com/science/article/pii/S0959438818301843

[141] M. Asgharpour, R. Foodeh, M. D.-J. of N. Methods, and undefined 2021, "Regularized Kalman filter for brain-computer interfaces using local field potential signals," *Elsevier*, Accessed: Sep. 06, 2023. [Online]. Available: https://www.sciencedirect.com/science/article/pii/S0165027020304453

[142] R. Q. Quiroga, S. P.-N. R. Neuroscience, and undefined 2009, "Extracting information from neuronal populations: information theory and decoding approaches," *nature.comR Quian Quiroga, S PanzeriNature Reviews Neuroscience, 2009•nature.com*, Accessed: Sep. 06, 2023. [Online]. Available: https://www.nature.com/articles/nrn2578

[143] J. I. Glaser, A. S. Benjamin, R. H. Chowdhury, M. G. Perich, L. E. Miller, and K. P. Kording, "Machine Learning for Neural Decoding," *eNeuro*, vol. 7, no. 4, pp. 1–16, Jul. 2020, doi: 10.1523/ENEURO.0506-19.2020.

[144] A. Ghanbari, C. Lee, … H. R.-J. of N., and undefined 2019, "Modeling stimulus-dependent variability improves decoding of population neural responses," *iopscience.iop.orgA Ghanbari, CM Lee, HL Read, IH StevensonJournal of Neural Engineering, 2019•iopscience.iop.org*, 2019, doi: 10.1088/1741-2552/ab3a68.

[145] M. Kashefi and M. R. Daliri, "A stack LSTM structure for decoding continuous force from local field potential signal of primary motor cortex (M1)," *BMC Bioinformatics*, vol. 22, no. 1, Dec. 2021, doi: 10.1186/S12859-020-03953-0.

[146] D. Liu *et al.*, "Intracranial brain-computer interface spelling using localized visual motion response," *Neuroimage*, vol. 258, p. 119363, Sep. 2022, doi: 10.1016/J.NEUROIMAGE.2022.119363.

[147] F. Lotte *et al.*, "A review of classification algorithms for EEG-based brain–computer interfaces: a 10 year update," *J Neural Eng*, vol. 15, no. 3, p. 031005, Apr. 2018, doi: 10.1088/1741-2552/AAB2F2.

[148] H. Wen, J. Shi, Y. Zhang, K.-H. Lu, J. Cao, and Z. Liu, "Neural Encoding and Decoding with Deep Learning for Dynamic Natural Vision," *Cerebral Cortex*, vol. 28, pp. 4136–4160, 2018, doi: 10.1093/cercor/bhx268.

[149] A. Valero-Cabré, J. L. Amengual, C. Stengel, A. Pascual-Leone, and O. A. Coubard, "Transcranial magnetic stimulation in basic and clinical neuroscience: A comprehensive review of fundamental principles and novel insights," *Neurosci Biobehav Rev*, vol. 83, pp. 381–404, Dec. 2017, doi: 10.1016/J.NEUBIOREV.2017.10.006.

[150] W. Paulus, M. A. Nitsche, and A. Antal, "Application of Transcranial Electric Stimulation (tDCS, tACS, tRNS)," *http://dx.doi.org/10.1027/1016-9040/a000242*, vol. 21, no. 1, pp. 4–14, Mar. 2016, doi: 10.1027/1016-9040/A000242.

[151] W. Legon *et al.*, "Transcranial focused ultrasound modulates the activity of primary somatosensory cortex in humans," *Nature Neuroscience 2013 17:2*, vol. 17, no. 2, pp. 322–329, Jan. 2014, doi: 10.1038/nn.3620.

[152] L. Osborn, B. Christie, … D. M.-2021 43rd annual, and undefined 2021, "Intracortical microstimulation of somatosensory cortex enables object identification through perceived sensations," *ieeexplore.ieee.org*, Accessed: Jun. 27, 2023. [Online]. Available: https://ieeexplore.ieee.org/abstract/document/9630450/

[153] C. Hughes, S. Flesher, … J. W.-J. of N., and undefined 2021, "Neural stimulation and recording performance in human sensorimotor cortex over 1500 days," *iopscience.iop.org*, 2021, doi: 10.1088/1741-2552/ac18ad.

[154] G. W. V. Vidal, M. L. Rynes, Z. Kelliher, and S. J. Goodwin, "Review of Brain-Machine Interfaces Used in Neural Prosthetics with New Perspective on Somatosensory Feedback through Method of Signal Breakdown," *Scientifica (Cairo)*, vol. 2016, 2016, doi: 10.1155/2016/8956432.

[155] J. A. Camacho-Conde, M. del R. Gonzalez-Bermudez, M. Carretero-Rey, and Z. U. Khan, "Brain stimulation: a therapeutic approach for the treatment of neurological disorders.," *CNS Neurosci Ther*, vol. 28, no. 1, pp. 5–18, Dec. 2021, doi: 10.1111/CNS.13769.

[156] M. Bucur and C. Papagno, "Deep Brain Stimulation in Parkinson Disease: A Meta-analysis of the Long-term Neuropsychological Outcomes," *Neuropsychol Rev*, Jun. 2022, doi: 10.1007/S11065-022-09540-9.

[157] E. Zrenner, "Will retinal implants restore vision?," *Science (1979)*, vol. 295, no. 5557, pp. 1022–1025, Feb. 2002, doi: 10.1126/SCIENCE.1067996/ASSET/536722EB-3EAD-4368-8969-A46F209B4E48/ASSETS/GRAPHIC/SE0520182003.JPEG.

[158] V. Busskamp *et al.*, "Genetic reactivation of cone photoreceptors restores visual responses in retinitis pigmentosa," *Science (1979)*, vol. 329, no. 5990, pp. 413–417, Jul. 2010, doi: 10.1126/SCIENCE.1190897/SUPPL_FILE/BUSSKAMP.SOM.PDF.

[159] I. Tochitsky *et al.*, "Restoring visual function to blind mice with photoswitches that selectively target the degenerated retina.," *Invest Ophthalmol Vis Sci*, vol. 55, no. 13, pp. 4616–4616, Apr. 2014.

[160] S. Moeller, T. Crapse, L. Chang, and D. Y. Tsao, "The effect of face patch microstimulation on perception of faces and objects," *Nat Neurosci*, vol. 20, no. 5, pp. 743–752, May 2017, doi: 10.1038/NN.4527.

[161] P. Mégevand et al., "Seeing scenes: topographic visual hallucinations evoked by direct electrical stimulation of the parahippocampal place area," *J Neurosci*, vol. 34, no. 16, pp. 5399–5405, 2014, doi: 10.1523/JNEUROSCI.5202-13.2014.

[162] V. Mendez, F. Iberite, S. Shokur, and S. Micera, "Annual Review of Control, Robotics, and Autonomous Systems Current Solutions and Future Trends for Robotic Prosthetic Hands Neuroprosthesis: a device that connects to the nervous system and either replaces missing parts of it or improves it," *Annu. Rev. Control Robot. Auton. Syst*, vol. 4, pp. 595–627, 2021, doi: 10.1146/annurev-control-071020.

[163] B. T. Nghiem et al., "Providing a Sense of Touch to Prosthetic Hands," *Plast Reconstr Surg*, vol. 135, no. 6, pp. 1652–1663, Jun. 2015, doi: 10.1097/PRS.0000000000001289.

[164] G. A. Tabot et al., "Restoring the sense of touch with a prosthetic hand through a brain interface," *Proc Natl Acad Sci U S A*, vol. 110, no. 45, pp. 18279–18284, Nov. 2013, doi: 10.1073/PNAS.1221113110/-/DCSUPPLEMENTAL/PNAS.201221113SI.PDF.

[165] R. Vallejo, K. Bradley, and L. Kapural, "Spinal Cord Stimulation in Chronic Pain," *Spine (Phila Pa 1976)*, vol. 42, pp. S53–S60, Jul. 2017, doi: 10.1097/BRS.0000000000002179.

[166] B. Lundeland, M. Toennis, M. Züchner, L. Janerås, A. Stubhaug, and P. Hansson, "Spinal cord stimulation for the treatment of peripheral neuropathic pain," *Tidsskrift for Den norske legeforening*, vol. 141, no. 9, Jun. 2021, doi: 10.4045/TIDSSKR.20.1010.

[167] M. Figee et al., "Deep brain stimulation restores frontostriatal network activity in obsessive-compulsive disorder," *Nature Neuroscience 2013 16:4*, vol. 16, no. 4, pp. 386–387, Feb. 2013, doi: 10.1038/nn.3344.

[168] I. O. Bergfeld et al., "Deep Brain Stimulation of the Ventral Anterior Limb of the Internal Capsule for Treatment-Resistant Depression: A Randomized Clinical Trial," *JAMA Psychiatry*, vol. 73, no. 5, pp. 456–464, May 2016, doi: 10.1001/JAMAPSYCHIATRY.2016.0152.

[169] M. Figee and H. Mayberg, "The future of personalized brain stimulation," *Nature Medicine 2021 27:2*, vol. 27, no. 2, pp. 196–197, Feb. 2021, doi: 10.1038/s41591-021-01243-7.

[170] S. Kumar and M. Khammash, "Platforms for Optogenetic Stimulation and Feedback Control," *Front Bioeng Biotechnol*, vol. 10, Jun. 2022, doi: 10.3389/FBIOE.2022.918917.

[171] T. Lan, L. He, Y. Huang, Y. Z.-T. in Genetics, and undefined 2022, "Optogenetics for transcriptional programming and genetic engineering," *Elsevier*, Accessed: Jun. 27, 2023. [Online]. Available: https://www.sciencedirect.com/science/article/pii/S0168952522001408

[172] K. Deisseroth, "Optogenetics," *Nature Methods 2011 8:1*, vol. 8, no. 1, pp. 26–29, Dec. 2010, doi: 10.1038/nmeth.f.324.

[173] V. Gradinaru, M. Mogri, K. R. Thompson, J. M. Henderson, and K. Deisseroth, "Optical deconstruction of parkinsonian neural circuitry," *Science (1979)*, vol. 324, no. 5925, pp. 354–359, Apr. 2009, doi: 10.1126/SCIENCE.1167093/SUPPL_FILE/GRADINARU.SOM.PDF.

[174] S. A. Guillory and K. A. Bujarski, "Exploring emotions using invasive methods: review of 60 years of human intracranial electrophysiology," *Soc Cogn Affect Neurosci*, vol. 9, no. 12, pp. 1880–1889, Dec. 2014, doi: 10.1093/SCAN/NSU002.

[175] W. Han et al., "Integrated Control of Predatory Hunting by the Central Nucleus of the Amygdala," *Cell*, vol. 168, no. 1–2, pp. 311-324.e18, Jan. 2017, doi: 10.1016/J.CELL.2016.12.027.

[176] C. A. Zimmerman et al., "Thirst neurons anticipate the homeostatic consequences of eating and drinking," *Nature 2016 537:7622*, vol. 537, no. 7622, pp. 680–684, Aug. 2016, doi: 10.1038/nature18950.

[177] Y. Ezzyat et al., "Closed-loop stimulation of temporal cortex rescues functional networks and improves memory," *Nature Communications 2018 9:1*, vol. 9, no. 1, pp. 1–8, Feb. 2018, doi: 10.1038/s41467-017-02753-0.

[178] J. Dai, D. I. Brooks, and D. L. Sheinberg, "Optogenetic and Electrical Microstimulation Systematically Bias Visuospatial Choice in Primates," *Current Biology*, vol. 24, no. 1, pp. 63–69, Jan. 2014, doi: 10.1016/J.CUB.2013.11.011.

[179] R. J. Krauzlis, L. P. Lovejoy, and A. Zénon, "Superior Colliculus and Visual Spatial Attention," *Annu Rev Neurosci*, vol. 36, pp. 165–182, Jul. 2013, doi: 10.1146/ANNUREV-NEURO-062012-170249.

[180] J. Jacobs et al., "Direct Electrical Stimulation of the Human Entorhinal Region and Hippocampus Impairs Memory," *Neuron*, vol. 92, no. 5, pp. 983–990, Dec. 2016, doi: 10.1016/J.NEURON.2016.10.062.

[181] N. Lipsman et al., "Deep brain stimulation of the subcallosal cingulate for treatment-refractory anorexia nervosa: 1 year follow-up of an open-label trial," *Lancet Psychiatry*, vol. 4, no. 4, pp. 285–294, Apr. 2017, doi: 10.1016/S2215-0366(17)30076-7.

[182] H. C. Tsai et al., "Phasic firing in dopaminergic neurons is sufficient for behavioral conditioning," *Science (1979)*, vol. 324, no. 5930, pp. 1080–1084, May 2009, doi: 10.1126/SCIENCE.1168878/SUPPL_FILE/TSAI.SOM.REV1.PDF.

[183] C. D. Proulx, O. Hikosaka, and R. Malinow, "Reward processing by the lateral habenula in normal and depressive behaviors," *Nature Neuroscience 2014 17:9*, vol. 17, no. 9, pp. 1146–1152, Aug. 2014, doi: 10.1038/nn.3779.

[184] S. Cadoni, "Sonogenetic stimulation for vision restoration.," 2020, Accessed: Jul. 06, 2023. [Online]. Available: https://cnrs.hal.science/tel-03862202/

[185] S. Cadoni et al., "Ectopic expression of a mechanosensitive channel confers spatiotemporal resolution to ultrasound stimulations of neurons for visual restoration," *Nature Nanotechnology 2023 18:6*, vol. 18, no. 6, pp. 667–676, Apr. 2023, doi: 10.1038/s41565-023-01359-6.

[186] B. Kotchoubey, S. Busch, U. Strehl, and N. Birbaumer, "Changes in EEG Power Spectra During Biofeedback of Slow Cortical Potentials in Epilepsy," *Applied Psychophysiology and Biofeedback 1999 24:4*, vol. 24, no. 4, pp. 213–233, 1999, doi: 10.1023/A:1022226412991.

[187] Y. Yu, Y. Liu, J. Jiang, E. Yin, Z. Zhou, and D. Hu, "An Asynchronous Control Paradigm Based on Sequential Motor Imagery and Its Application in Wheelchair Navigation," *IEEE Transactions on Neural Systems and Rehabilitation Engineering*, vol. 26, no. 12, pp. 2367–2375, Dec. 2018, doi: 10.1109/TNSRE.2018.2881215.

[188] H. Cecotti, "Adaptive Time Segment Analysis for Steady-State Visual Evoked Potential Based Brain-Computer Interfaces," *IEEE Transactions on Neural Systems and Rehabilitation Engineering*, vol. 28, no. 3, pp. 552–560, Mar. 2020, doi: 10.1109/TNSRE.2020.2968307.

[189] C. Hughes, A. Herrera, R. Gaunt, and J. Collinger, "Bidirectional brain-computer interfaces," *Handb Clin Neurol*, vol. 168, pp. 163–181, Jan. 2020, doi: 10.1016/B978-0-444-63934-9.00013-5.

[190] L. Kros *et al.*, "Cerebellar output controls generalized spike-and-wave discharge occurrence," *Ann Neurol*, vol. 77, no. 6, pp. 1027–1049, Jun. 2015, doi: 10.1002/ANA.24399.

[191] G. Panuccio, M. Semprini, … L. N.-B. and, and undefined 2018, "Progress in Neuroengineering for brain repair: New challenges and open issues," *journals.sagepub.com*, vol. 2, p. 239821281877647, Jan. 2018, doi: 10.1177/2398212818776475.

[192] X. Zhang *et al.*, "The combination of brain-computer interfaces and artificial intelligence: applications and challenges," *Ann Transl Med*, vol. 8, no. 11, pp. 712–712, Jun. 2020, doi: 10.21037/ATM.2019.11.109.

[193] J. P. Nguyen, J. Nizard, Y. Keravel, and J. P. Lefaucheur, "Invasive brain stimulation for the treatment of neuropathic pain," *Nature Reviews Neurology 2011 7:12*, vol. 7, no. 12, pp. 699–709, Sep. 2011, doi: 10.1038/nrneurol.2011.138.

[194] C. D. Salzman, C. M. Murasugi, K. H. Britten, and W. T. Newsome, "Microstimulation in visual area MT: Effects on direction discrimination performance," *Journal of Neuroscience*, vol. 12, no. 6, pp. 2331–2355, 1992, doi: 10.1523/JNEUROSCI.12-06-02331.1992.

[195] E. Fernández *et al.*, "Visual percepts evoked with an intracortical 96-channel microelectrode array inserted in human occipital cortex," *Journal of Clinical Investigation*, vol. 131, no. 23, Dec. 2021, doi: 10.1172/JCI151331.

[196] X. Chen, F. Wang, E. Fernandez, and P. R. Roelfsema, "Shape perception via a high-channel-count neuroprosthesis in monkey visual cortex," *Science (1979)*, vol. 370, no. 6521, Dec. 2020, doi: 10.1126/SCIENCE.ABD7435.

[197] V. T. Lehman *et al.*, "MRI and tractography techniques to localize the ventral intermediate nucleus and dentatorubrothalamic tract for deep brain stimulation and MR-guided focused ultrasound: a narrative review and update," *Neurosurg Focus*, vol. 49, no. 1, p. E8, Jul. 2020, doi: 10.3171/2020.4.FOCUS20170.

[198] Q. Xian *et al.*, "Modulation of deep neural circuits with sonogenetics," *Proc Natl Acad Sci U S A*, vol. 120, no. 22, May 2023, doi: 10.1073/PNAS.2220575120.

[199] C.-H. Fan *et al.*, "Sonogenetic-Based Neuromodulation for the Amelioration of Parkinson's Disease," *Cite This: Nano Lett*, vol. 21, pp. 5967–5976, 2021, doi: 10.1021/acs.nanolett.1c00886.

[200] T. Liu *et al.*, "Sonogenetics: Recent advances and future directions," *Elsevier*, Accessed: Jul. 03, 2023. [Online]. Available: https://www.sciencedirect.com/science/article/pii/S1935861X22002066

[201] X. Hong *et al.*, "Brain plasticity following MI-BCI training combined with tDCS in a randomized trial in chronic subcortical stroke subjects: a preliminary study," *Scientific Reports 2017 7:1*, vol. 7, no. 1, pp. 1–12, Aug. 2017, doi: 10.1038/s41598-017-08928-5.

[202] B. S. Baxter, B. J. Edelman, A. Sohrabpour, and B. He, "Anodal transcranial direct current stimulation increases bilateral directed brain connectivity during motor-imagery based brain-computer interface control," *Front Neurosci*, vol. 11, no. DEC, p. 691, Dec. 2017, doi: 10.3389/FNINS.2017.00691/BIBTEX.

[203] R. P. N. Rao *et al.*, "A Direct Brain-to-Brain Interface in Humans," *PLoS One*, vol. 9, no. 11, p. e111332, Nov. 2014, doi: 10.1371/JOURNAL.PONE.0111332.

[204] C. Grau *et al.*, "Conscious Brain-to-Brain Communication in Humans Using Non-Invasive Technologies," *Brain Stimulation: Basic, Translational, and Clinical Research in Neuromodulation*, vol. 8, no. 2, p. 323, Mar. 2015, doi: 10.1016/J.BRS.2015.01.047.

[205] V. Walsh and A. Cowey, "Transcranial magnetic stimulation and cognitive neuroscience," *Nature Reviews Neuroscience 2000 1:1*, vol. 1, no. 1, pp. 73–80, 2000, doi: 10.1038/35036239.

[206] N. Schaworonkow, J. Triesch, U. Ziemann, and C. Zrenner, "EEG-triggered TMS reveals stronger brain state-dependent modulation of motor evoked potentials at weaker stimulation intensities," *Brain Stimul*, vol. 12, no. 1, pp. 110–118, Jan. 2019, doi: 10.1016/J.BRS.2018.09.009.

[207] J. Jin, Z. Chen, R. Xu, Y. Miao, X. Wang, and T. P. Jung, "Developing a Novel Tactile P300 Brain-Computer Interface with a Cheeks-Stim Paradigm," *IEEE Trans Biomed Eng*, vol. 67, no. 9, pp. 2585–2593, Sep. 2020, doi: 10.1109/TBME.2020.2965178.

[208] V. Guy, M. H. Soriani, M. Bruno, T. Papadopoulo, C. Desnuelle, and M. Clerc, "Brain computer interface with the P300 speller: Usability for disabled people with amyotrophic lateral sclerosis," *Ann Phys Rehabil Med*, vol. 61, no. 1, pp. 5–11, Jan. 2018, doi: 10.1016/J.REHAB.2017.09.004.

[209] L. Botrel, E. M. Holz, and A. Kübler, "Brain Painting V2: evaluation of P300-based brain-computer interface for creative expression by an end-user following the user-centered design," *https://doi.org/10.1080/2326263X.2015.1100038*, vol. 2, no. 2–3, pp. 135–149, Apr. 2015, doi: 10.1080/2326263X.2015.1100038.

[210] R. E. Alcaide-Aguirre, S. A. Warschausky, D. Brown, A. Aref, and J. E. Huggins, "Asynchronous brain–computer interface for cognitive assessment in people with cerebral palsy," *J Neural Eng*, vol. 14, no. 6, p. 066001, Oct. 2017, doi: 10.1088/1741-2552/AA7FC4.

[211] L. A. Farwell, D. C. Richardson, G. M. Richardson, and J. J. Furedy, "Brain fingerprinting classification concealed information test detects US Navy military medical information with P300," *Front Neurosci*, vol. 8, no. DEC, p. 410, 2014, doi: 10.3389/FNINS.2014.00410/BIBTEX.

[212] I. Iturrate, J. M. Antelis, A. Kübler, and J. Minguez, "A noninvasive brain-actuated wheelchair based on a P300 neurophysiological protocol and automated navigation," *IEEE Transactions on Robotics*, vol. 25, no. 3, pp. 614–627, 2009, doi: 10.1109/TRO.2009.2020347.

[213] K. Kasahara, C. S. DaSalla, M. Honda, and T. Hanakawa, "Neuroanatomical correlates of brain–computer interface performance," *Neuroimage*, vol. 110, pp. 95–100, Apr. 2015, doi: 10.1016/J.NEUROIMAGE.2015.01.055.

[214] S. Darvishi, A. Gharabaghi, C. B. Boulay, M. C. Ridding, D. Abbott, and M. Baumert, "Proprioceptive feedback facilitates motor imagery-related operant learning of sensorimotor ß-band modulation," *Front Neurosci*, vol. 11, no. FEB, p. 60, Feb. 2017, doi: 10.3389/FNINS.2017.00060/BIBTEX.

[215] T. Elbert, B. Rockstroh, W. Lutzenberger, and N. Birbaumer, "Biofeedback of slow cortical potentials. I," *Electroencephalogr Clin Neurophysiol*, vol. 48, no. 3, pp. 293–301, Mar. 1980, doi: 10.1016/0013-4694(80)90265-5.

[216] Y. M. Chi, Y. Te Wang, Y. Wang, C. Maier, T. P. Jung, and G. Cauwenberghs, "Dry and noncontact EEG sensors for mobile brain-computer interfaces," *IEEE Transactions on Neural Systems and Rehabilitation Engineering*, vol. 20, no. 2, pp. 228–235, Mar. 2012, doi: 10.1109/TNSRE.2011.2174652.

[217] X. Zhao, Y. Chu, J. Han, and Z. Zhang, "SSVEP-Based Brain-Computer Interface Controlled Functional Electrical Stimulation System for Upper Extremity Rehabilitation," *IEEE Trans Syst Man Cybern Syst*, vol. 46, no. 7, pp. 947–956, Jul. 2016, doi: 10.1109/TSMC.2016.2523762.

[218] M. Van Vliet, A. Robben, N. Chumerin, N. V. Manyakov, A. Combaz, and M. M. Van Hulle, "Designing a brain-computer interface controlled video-game using consumer grade EEG hardware," *2012 ISSNIP Biosignals and Biorobotics Conference: Biosignals and Robotics for Better and Safer Living, BRC 2012*, 2012, doi: 10.1109/BRC.2012.6222186.

[219] J. J. S. Norton *et al.*, "Soft, curved electrode systems capable of integration on the auricle as a persistent brain-computer interface," *Proc Natl Acad Sci U S A*, vol. 112, no. 13, pp. 3920–3925, Mar. 2015, doi: 10.1073/PNAS.1424875112.

[220] Y. Petrov, J. Nador, C. Hughes, S. Tran, O. Yavuzcetin, and S. Sridhar, "Ultra-dense EEG sampling results in two-fold increase of functional brain information," *Neuroimage*, vol. 90, pp. 140–145, Apr. 2014, doi: 10.1016/J.NEUROIMAGE.2013.12.041.

[221] D. Silver *et al.*, "Mastering the game of Go without human knowledge," *Nature 2017 550:7676*, vol. 550, no. 7676, pp. 354–359, Oct. 2017, doi: 10.1038/nature24270.

[222] N. R. B. Martins, W. Erlhagen, and R. A. Freitas, Jr., "Human Connectome Mapping and Monitoring Using Neuronanorobots," *Journal of Ethics and Emerging Technologies*, vol. 26, no. 1, pp. 1–25, Jan. 2016, doi: 10.55613/jeet.v26i1.49.

[223] N. R. B. Martins *et al.*, "Human brain/cloud interface," *Front Neurosci*, vol. 13, no. March, 2019, doi: 10.3389/fnins.2019.00112.

[224] D. D. Cox and T. Dean, "Neural Networks and Neuroscience-Inspired Computer Vision," *Current Biology*, vol. 24, no. 18, pp. R921–R929, Sep. 2014, doi: 10.1016/J.CUB.2014.08.026.

[225] S. L. Jackman and W. G. Regehr, "The Mechanisms and Functions of Synaptic Facilitation," *Neuron*, vol. 94, no. 3, pp. 447–464, May 2017, doi: 10.1016/J.NEURON.2017.02.047.

[226] J. J. Moore *et al.*, "Dynamics of cortical dendritic membrane potential and spikes in freely behaving rats," *Science (1979)*, vol. 355, no. 6331, Mar. 2017, doi: 10.1126/SCIENCE.AAJ1497.

[227] K. Fukushima, S. Miyake, T. I.-I. transactions on systems, and undefined 1983, "Neocognitron: A neural network model for a mechanism of visual pattern recognition," *ieeexplore.ieee.org*, Accessed: Jun. 28, 2023. [Online]. Available: https://ieeexplore.ieee.org/abstract/document/6313076/

[228] K. Fukushima, "Neocognitron: A self-organizing neural network model for a mechanism of pattern recognition unaffected by shift in position," *Biol Cybern*, vol. 36, no. 4, pp. 193–202, Apr. 1980, doi: 10.1007/BF00344251.

[229] S. Venkataramani, K. Roy, and A. Raghunathan, "Efficient embedded learning for IoT devices," *Proceedings of the Asia and South Pacific Design Automation Conference, ASP-DAC*, vol. 25-28-January-2016, pp. 308–311, Mar. 2016, doi: 10.1109/ASPDAC.2016.7428029.

[230] C. Mead, "How we created neuromorphic engineering," *Nature Electronics 2020 3:7*, vol. 3, no. 7, pp. 434–435, Jul. 2020, doi: 10.1038/s41928-020-0448-2.

[231] T. Potok, C. Schuman, R. Patton, T. Hylton, H. Li, and R. Pino, "Neuromorphic Computing, Architectures, Models, and Applications. A Beyond-CMOS Approach to Future Computing, June 29-July 1, 2016, Oak Ridge, TN," Dec. 2016, doi: 10.2172/1341738.

[232] C. D. Schuman *et al.*, "A Survey of Neuromorphic Computing and Neural Networks in Hardware," May 2017, doi: 10.48550/arxiv.1705.06963.

[233] K. Berggren *et al.*, "Neuromorphic neural interfaces: from neurophysiological inspiration to biohybrid coupling with nervous systems," *J Neural Eng*, vol. 14, no. 4, p. 041002, Jun. 2017, doi: 10.1088/1741-2552/AA67A9.

[234] V. Sze, Y. H. Chen, J. Einer, A. Suleiman, and Z. Zhang, "Hardware for machine learning: Challenges and opportunities," *Proceedings of the Custom Integrated Circuits Conference*, vol. 2017-April, Jul. 2017, doi: 10.1109/CICC.2017.7993626.

[235] C. Mayr, S. Höppner, and S. Furber, "SpiNNaker 2: A 10 Million Core Processor System for Brain Simulation and Machine Learning," *Concurrent Systems Engineering Series*, vol. 70, pp. 277–280, Nov. 2019, doi: 10.48550/arxiv.1911.02385.

[236] S. B. Furber, F. Galluppi, S. Temple, and L. A. Plana, "The SpiNNaker project," *Proceedings of the IEEE*, vol. 102, no. 5, pp. 652–665, 2014, doi: 10.1109/JPROC.2014.2304638.

[237] M. Davies et al., "Loihi: A Neuromorphic Manycore Processor with On-Chip Learning," *IEEE Micro*, vol. 38, no. 1, pp. 82–99, Jan. 2018, doi: 10.1109/MM.2018.112130359.

[238] S. Moradi, N. Qiao, F. Stefanini, and G. Indiveri, "A Scalable Multicore Architecture with Heterogeneous Memory Structures for Dynamic Neuromorphic Asynchronous Processors (DYNAPs)," *IEEE Trans Biomed Circuits Syst*, vol. 12, no. 1, pp. 106–122, Feb. 2018, doi: 10.1109/TBCAS.2017.2759700.

[239] N. Qiao et al., "A reconfigurable on-line learning spiking neuromorphic processor comprising 256 neurons and 128K synapses," *Front Neurosci*, vol. 9, no. APR, 2015, doi: 10.3389/FNINS.2015.00141.

[240] H. Mostafa, L. K. Müller, and G. Indiveri, "An event-based architecture for solving constraint satisfaction problems," *Nature Communications 2015 6:1*, vol. 6, no. 1, pp. 1–10, Dec. 2015, doi: 10.1038/ncomms9941.

[241] P. A. Merolla et al., "A million spiking-neuron integrated circuit with a scalable communication network and interface," *Science (1979)*, vol. 345, no. 6197, pp. 668–673, Aug. 2014, doi: 10.1126/SCIENCE.1254642/SUPPL_FILE/MEROLLA.SM.REV1.PDF.

[242] J. Schemmel, S. Billaudelle, P. Dauer, and J. Weis, "Accelerated Analog Neuromorphic Computing," *Analog Circuits for Machine Learning, Current/Voltage/Temperature Sensors, and High-speed Communication*, pp. 83–102, 2022, doi: 10.1007/978-3-030-91741-8_6/FIGURES/12.

[243] J. Schemmel, D. Brüderle, A. Grübl, M. Hock, K. Meier, and S. Millner, "A wafer-scale neuromorphic hardware system for large-scale neural modeling," *ISCAS 2010 - 2010 IEEE International Symposium on Circuits and Systems: Nano-Bio Circuit Fabrics and Systems*, pp. 1947–1950, 2010, doi: 10.1109/ISCAS.2010.5536970.

[244] B. V. Benjamin et al., "Neurogrid: A mixed-analog-digital multichip system for large-scale neural simulations," *Proceedings of the IEEE*, vol. 102, no. 5, pp. 699–716, 2014, doi: 10.1109/JPROC.2014.2313565.

[245] C. S. Thakur et al., "Large-Scale Neuromorphic Spiking Array Processors: A Quest to Mimic the Brain," *Frontiers in Neuroscience*, vol. 12. Frontiers Media S.A., Dec. 03, 2018. doi: 10.3389/fnins.2018.00891.

[246] P. Blouw, X. Choo, E. Hunsberger, and C. Eliasmith, "Benchmarking keyword spotting efficiency on neuromorphic hardware," *ACM International Conference Proceeding Series*, vol. 19, Mar. 2019, doi: 10.1145/3320288.3320304.

[247] N. Getty, T. Brettin, D. Jin, R. Stevens, and F. Xia, "Deep medical image analysis with representation learning and neuromorphic computing," *Interface Focus*, vol. 11, no. 1, Feb. 2021, doi: 10.1098/RSFS.2019.0122.

[248] J. Pei et al., "Towards artificial general intelligence with hybrid Tianjic chip architecture," *Nature 2019 572:7767*, vol. 572, no. 7767, pp. 106–111, Jul. 2019, doi: 10.1038/s41586-019-1424-8.

[249] "Training a single AI model can emit as much carbon as five cars in their lifetimes | MIT Technology Review." https://www.technologyreview.com/2019/06/06/239031/training-a-single-ai-model-can-emit-as-much-carbon-as-five-cars-in-their-lifetimes/ (accessed Sep. 23, 2022).

[250] J. Kim et al., "Spin-based computing: Device concepts, current status, and a case study on a high-performance microprocessor," *Proceedings of the IEEE*, vol. 103, no. 1, pp. 106–130, Jan. 2015, doi: 10.1109/JPROC.2014.2361767.

[251] Y. Li, Z. Wang, R. Midya, Q. Xia, and J. Joshua Yang, "Review of memristor devices in neuromorphic computing: materials sciences and device challenges," *J Phys D Appl Phys*, vol. 51, no. 50, p. 503002, Sep. 2018, doi: 10.1088/1361-6463/AADE3F.

[252] H. Ghanatian, M. Ronchini, H. Farkhani, and F. Moradi, "STDP implementation using multi-state spin−orbit torque synapse," *Semicond Sci Technol*, vol. 37, no. 2, p. 024004, Dec. 2021, doi: 10.1088/1361-6641/AC419C.

[253] H. Farkhani, M. Tohidi, S. Farkhani, J. K. Madsen, and F. Moradi, "A Low-Power High-Speed Spintronics-Based Neuromorphic Computing System Using Real-Time Tracking Method," *IEEE J Emerg Sel Top Circuits Syst*, vol. 8, no. 3, pp. 627–638, Sep. 2018, doi: 10.1109/JETCAS.2018.2813389.

[254] H. Farkhani, I. L. Prejbeanu, and F. Moradi, "LAS-NCS: A laser-assisted spintronic neuromorphic computing system," *IEEE Transactions on Circuits and Systems II: Express Briefs*, vol. 66, no. 5, pp. 838–842, May 2019, doi: 10.1109/TCSII.2019.2908077.

[255] J. Torrejon et al., "Neuromorphic computing with nanoscale spintronic oscillators," *Nature 2017 547:7664*, vol. 547, no. 7664, pp. 428–431, Jul. 2017, doi: 10.1038/nature23011.

[256] H. Farkhani et al., "LAO-NCS: Laser Assisted Spin Torque Nano Oscillator-Based Neuromorphic Computing System," *Front Neurosci*, vol. 13, p. 1429, Jan. 2020, doi: 10.3389/FNINS.2019.01429/BIBTEX.

[257] F. Moradi et al., "Spin-Orbit-Torque-based Devices, Circuits and Architectures," Dec. 2019, doi: 10.48550/arxiv.1912.01347.

[258] M. Zahedinejad et al., "Two-dimensional mutually synchronized spin Hall nano-oscillator arrays for neuromorphic computing," *Nature Nanotechnology 2019 15:1*, vol. 15, no. 1, pp. 47–52, Dec. 2019, doi: 10.1038/s41565-019-0593-9.

[259] J. S. Najem et al., "Memristive Ion Channel-Doped Biomembranes as Synaptic Mimics," *ACS Nano*, vol. 12, no. 5, pp. 4702–4711, May 2018, doi: 10.1021/ACSNANO.8B01282/ASSET/IMAGES/LARGE/NN-2018-01282F_0005.JPEG.

[260] R. Islam et al., "Device and materials requirements for neuromorphic computing," *J Phys D Appl Phys*, vol. 52, no. 11, p. 113001, Jan. 2019, doi: 10.1088/1361-6463/AAF784.

[261] S. H. Jo, T. Chang, I. Ebong, B. B. Bhadviya, P. Mazumder, and W. Lu, "Nanoscale memristor device as synapse in neuromorphic systems," *Nano Lett*, vol. 10, no. 4, pp. 1297–1301, Apr. 2010, doi: 10.1021/NL904092H/SUPPL_FILE/NL904092H_SI_001.PDF.

[262] W. Maass, "Networks of Spiking Neurons: The Third Generation of Neural Network Models," *Neural Networks*, vol. 10, no. 9, pp. 1659–1671, Dec. 1997, doi: 10.1016/S0893-6080(97)00011-7.

[263] M. Abeles, H. Bergman, E. Margalit, and E. Vaadia, "Spatiotemporal firing patterns in the frontal cortex of behaving monkeys," *https://doi.org/10.1152/jn.1993.70.4.1629*, vol. 70, no. 4, pp. 1629–1638, 1993, doi: 10.1152/JN.1993.70.4.1629.

[264] N. K. Kasabov, "Time-Space, Spiking Neural Networks and Brain-Inspired Artificial Intelligence," vol. 7, 2019, doi: 10.1007/978-3-662-57715-8.

[265] L. Deng et al., "Rethinking the performance comparison between SNNS and ANNS," *Neural Networks*, vol. 121, pp. 294–307, Jan. 2020, doi: 10.1016/J.NEUNET.2019.09.005.

[266] W. Gerstner and W. M. Kistler, "Spiking Neuron Models," *Spiking Neuron Models*, Aug. 2002, doi: 10.1017/CBO9780511815706.

[267] A. L. Hodgkin and A. F. Huxley, "A quantitative description of membrane current and its application to conduction and excitation in nerve," *J Physiol*, vol. 117, no. 4, p. 500, Aug. 1952, doi: 10.1113/JPHYSIOL.1952.SP004764.

[268] E. M. Izhikevich, "Simple model of spiking neurons," *IEEE Trans Neural Netw*, vol. 14, no. 6, pp. 1569–1572, Nov. 2003, doi: 10.1109/TNN.2003.820440.

[269] W. E. Sherwood, "FitzHugh-Nagumo Model", doi: 10.1007/978-1-4614-7320-6_147-1.

[270] F. Walter, F. Röhrbein, A. Knoll, and B. Florian Walter, "Computation by Time," *Neural Process Lett*, vol. 44, pp. 103–124, 2016, doi: 10.1007/s11063-015-9478-6.

[271] B. Petro, N. Kasabov, and R. M. Kiss, "Selection and Optimization of Temporal Spike Encoding Methods for Spiking Neural Networks," *IEEE Trans Neural Netw Learn Syst*, vol. 31, no. 2, pp. 358–370, Feb. 2020, doi: 10.1109/TNNLS.2019.2906158.

[272] D. O. Hebb, "The Organization of Behavior : A Neuropsychological Theory," *The Organization of Behavior*, Apr. 2005, doi: 10.4324/9781410612403.

[273] W. S. McCulloch and W. Pitts, "A logical calculus of the ideas immanent in nervous activity," *The bulletin of mathematical biophysics 1943 5:4*, vol. 5, no. 4, pp. 115–133, Dec. 1943, doi: 10.1007/BF02478259.

[274] L. F. Abbott and S. B. Nelson, "Synaptic plasticity: taming the beast," *Nature Neuroscience 2000 3:11*, vol. 3, no. 11, pp. 1178–1183, 2000, doi: 10.1038/81453.

[275] W. Gerstner, M. Lehmann, V. Liakoni, D. Corneil, and J. Brea, "Eligibility Traces and Plasticity on Behavioral Time Scales: Experimental Support of NeoHebbian Three-Factor Learning Rules," *Front Neural Circuits*, vol. 12, Jul. 2018, doi: 10.3389/FNCIR.2018.00053/FULL.

[276] N. Frémaux and W. Gerstner, "Neuromodulated spike-timing-dependent plasticity, and theory of three-factor learning rules," *Front Neural Circuits*, vol. 9, no. JAN2016, Jan. 2015, doi: 10.3389/FNCIR.2015.00085/FULL.

[277] R. Legenstein, C. Naeger, and W. Maass, "Communicated by Wulfram Gerstner What Can a Neuron Learn with Spike-Timing-Dependent Plasticity?," 2005.

[278] Z. Yi et al., "Learning rules in spiking neural networks: A survey," *Elsevier*, Accessed: Jun. 30, 2023. [Online]. Available: https://www.sciencedirect.com/science/article/pii/S0925231223001662

[279] P. U. Diehl, G. Zarrella, A. Cassidy, B. U. Pedroni, and E. Neftci, "Conversion of artificial recurrent neural networks to spiking neural networks for low-power neuromorphic hardware," *2016 IEEE International Conference on Rebooting Computing, ICRC 2016 - Conference Proceedings*, Nov. 2016, doi: 10.1109/ICRC.2016.7738691.

[280] B. Rueckauer, I. A. Lungu, Y. Hu, M. Pfeiffer, and S. C. Liu, "Conversion of continuous-valued deep networks to efficient event-driven networks for image classification," *Front Neurosci*, vol. 11, no. DEC, p. 682, Dec. 2017, doi: 10.3389/FNINS.2017.00682/BIBTEX.

[281] M. Pfeiffer and T. Pfeil, "Deep Learning With Spiking Neurons: Opportunities and Challenges," *Front Neurosci*, vol. 12, p. 774, Oct. 2018, doi: 10.3389/FNINS.2018.00774/BIBTEX.

[282] A. Sengupta, Y. Ye, R. Wang, C. Liu, and K. Roy, "Going Deeper in Spiking Neural Networks: VGG and Residual Architectures," *Front Neurosci*, vol. 13, p. 95, Mar. 2019, doi: 10.3389/FNINS.2019.00095/BIBTEX.

[283] P. U. Diehl, D. Neil, J. Binas, M. Cook, S. C. Liu, and M. Pfeiffer, "Fast-classifying, high-accuracy spiking deep networks through weight and threshold balancing," *Proceedings of the International Joint Conference on Neural Networks*, vol. 2015-September, Sep. 2015, doi: 10.1109/IJCNN.2015.7280696.

[284] M. Abadi et al., "TensorFlow: A system for large-scale machine learning," 2016, Accessed: Jun. 13, 2022. [Online]. Available: https://tensorflow.org.

[285] Y. Cao, Y. Chen, and D. Khosla, "Spiking Deep Convolutional Neural Networks for Energy-Efficient Object Recognition," *Int J Comput Vis*, vol. 113, no. 1, pp. 54–66, May 2015, doi: 10.1007/S11263-014-0788-3/TABLES/9.

[286] E. Hunsberger and C. Eliasmith, "Spiking Deep Networks with LIF Neurons," Oct. 2015, Accessed: Jun. 30, 2023. [Online]. Available: http://arxiv.org/abs/1510.08829

[287] C. Stöckl and W. Maass, "Optimized spiking neurons can classify images with high accuracy through temporal coding with two spikes," *Nature Machine Intelligence 2021 3:3*, vol. 3, no. 3, pp. 230–238, Mar. 2021, doi: 10.1038/s42256-021-00311-4.
[288] F. Ponulak and A. Kasiński, "Introduction to spiking neural networks: Information processing, learning and applications," *Acta Neurobiol Exp (Wars)*, vol. 71, no. 4, pp. 409–433, 2011.
[289] X. Wang, X. Lin, and X. Dang, "Supervised learning in spiking neural networks: A review of algorithms and evaluations," *Neural Networks*, vol. 125, no. May, pp. 258–280, 2020, doi: 10.1016/j.neunet.2020.02.011.
[290] F. Ponulak and A. Kasiński, "Supervised Learning in Spiking Neural Networks with ReSuMe: Sequence Learning, Classification, and Spike Shifting," *Neural Comput*, vol. 22, no. 2, pp. 467–510, Feb. 2010, doi: 10.1162/NECO.2009.11-08-901.
[291] R. Gütig and H. Sompolinsky, "The tempotron: a neuron that learns spike timing–based decisions," *Nature Neuroscience 2006 9:3*, vol. 9, no. 3, pp. 420–428, Feb. 2006, doi: 10.1038/nn1643.
[292] S. Ghosh-Dastidar and H. Adeli, "A new supervised learning algorithm for multiple spiking neural networks with application in epilepsy and seizure detection," *Neural Networks*, vol. 22, no. 10, pp. 1419–1431, Dec. 2009, doi: 10.1016/J.NEUNET.2009.04.003.
[293] S. M. Bohte, J. N. Kok, and H. La Poutré, "Error-backpropagation in temporally encoded networks of spiking neurons," *Neurocomputing*, vol. 48, no. 1–4, pp. 17–37, Oct. 2002, doi: 10.1016/S0925-2312(01)00658-0.
[294] J. Van Campenhout and B. Schrauwen, "Improving spikeprop: Enhancements to an error-backpropagation rule for spiking neural networks," 2004, Accessed: Jun. 14, 2022. [Online]. Available: https://www.researchgate.net/publication/228571840
[295] E. O. Neftci, H. Mostafa, and F. Zenke, "Surrogate Gradient Learning in Spiking Neural Networks: Bringing the Power of Gradient-based optimization to spiking neural networks," *IEEE Signal Process Mag*, vol. 36, no. 6, pp. 51–63, Nov. 2019, doi: 10.1109/MSP.2019.2931595.
[296] S. Yin *et al.*, "Algorithm and hardware design of discrete-time spiking neural networks based on back propagation with binary activations," *2017 IEEE Biomedical Circuits and Systems Conference, BioCAS 2017 - Proceedings*, vol. 2018-January, pp. 1–4, Mar. 2018, doi: 10.1109/BIOCAS.2017.8325230.
[297] J. H. Lee, T. Delbruck, and M. Pfeiffer, "Training deep spiking neural networks using backpropagation," *Front Neurosci*, vol. 10, no. NOV, 2016, doi: 10.3389/fnins.2016.00508.
[298] H. Mostafa, "Supervised learning based on temporal coding in spiking neural networks," *IEEE Trans Neural Netw Learn Syst*, vol. 29, no. 7, pp. 3227–3235, Jul. 2018, doi: 10.1109/TNNLS.2017.2726060.
[299] Y. , C. C. & B. C. J. C. LeCun, "THE MNIST DATABASE of handwritten digits | CiNii Research," 1998. https://cir.nii.ac.jp/crid/1571417126193283840 (accessed Jun. 14, 2022).
[300] F. Zenke and E. O. Neftci, "Brain-Inspired Learning on Neuromorphic Substrates," *Proceedings of the IEEE*, vol. 109, no. 5, pp. 935–950, May 2021, doi: 10.1109/JPROC.2020.3045625.
[301] B. Cramer, Y. Stradmann, J. Schemmel, and F. Zenke, "The Heidelberg Spiking Data Sets for the Systematic Evaluation of Spiking Neural Networks," *IEEE Trans Neural Netw Learn Syst*, 2020, doi: 10.1109/TNNLS.2020.3044364.
[302] P. U. Diehl and M. Cook, "Unsupervised learning of digit recognition using spike-timing-dependent plasticity," *Front Comput Neurosci*, vol. 9, no. AUGUST, p. 99, Aug. 2015, doi: 10.3389/FNCOM.2015.00099/BIBTEX.
[303] A. Tavanaei, M. Ghodrati, S. R. Kheradpisheh, T. Masquelier, and A. Maida, "Deep learning in spiking neural networks," *Neural Networks*, vol. 111, pp. 47–63, Mar. 2019, doi: 10.1016/J.NEUNET.2018.12.002.
[304] J. Wu, Y. Chua, M. Zhang, Q. Yang, G. Li, and H. Li, "Deep Spiking Neural Network with Spike Count based Learning Rule," *Proceedings of the International Joint Conference on Neural Networks*, vol. 2019-July, Jul. 2019, doi: 10.1109/IJCNN.2019.8852380.
[305] A. Alemi, C. K. Machens, S. Denève, and J.-J. Slotine, "Learning Nonlinear Dynamics in Efficient, Balanced Spiking Networks Using Local Plasticity Rules," *Proceedings of the AAAI Conference on Artificial Intelligence*, vol. 32, no. 1, Apr. 2018, doi: 10.1609/AAAI.V32I1.11320.
[306] D. Kappel, B. Nessler, and W. Maass, "STDP Installs in Winner-Take-All Circuits an Online Approximation to Hidden Markov Model Learning," *PLoS Comput Biol*, vol. 10, no. 3, p. e1003511, 2014, doi: 10.1371/JOURNAL.PCBI.1003511.
[307] M. Oster, R. Douglas, and S. C. Liu, "Computation with Spikes in a Winner-Take-All Network," *Neural Comput*, vol. 21, no. 9, pp. 2437–2465, Sep. 2009, doi: 10.1162/NECO.2009.07-08-829.
[308] P. Wijesinghe, G. Srinivasan, P. Panda, and K. Roy, "Analysis of liquid ensembles for enhancing the performance and accuracy of liquid state machines," *Front Neurosci*, vol. 13, no. MAY, p. 504, 2019, doi: 10.3389/FNINS.2019.00504/BIBTEX.
[309] C. Du, F. Cai, M. A. Zidan, W. Ma, S. H. Lee, and W. D. Lu, "Reservoir computing using dynamic memristors for temporal information processing," *Nature Communications 2017 8:1*, vol. 8, no. 1, pp. 1–10, Dec. 2017, doi: 10.1038/s41467-017-02337-y.
[310] M. Pfeiffer *et al.*, "Design and Analysis of a Neuromemristive Reservoir Computing Architecture for Biosignal Processing.," *Front Neurosci*, vol. 9, pp. 502–502, Jan. 2015, doi: 10.3389/FNINS.2015.00502.
[311] K. E. Hamilton, T. M. Mintz, and C. D. Schuman, "Spike-based primitives for graph algorithms," Mar. 2019, doi: 10.48550/arxiv.1903.10574.

[312] W. Severa, R. Lehoucq, O. Parekh, and J. B. Aimone, "Spiking Neural Algorithms for Markov Process Random Walk," *Proceedings of the International Joint Conference on Neural Networks*, vol. 2018-July, Oct. 2018, doi: 10.1109/IJCNN.2018.8489628.

[313] K. Hamilton, P. Date, B. Kay, and C. Schuman D., "Modeling epidemic spread with spike-based models," *ACM International Conference Proceeding Series*, Jul. 2020, doi: 10.1145/3407197.3407219.

[314] J. D. Smith et al., "Neuromorphic scaling advantages for energy-efficient random walk computations," *Nature Electronics 2022 5:2*, vol. 5, no. 2, pp. 102–112, Feb. 2022, doi: 10.1038/s41928-021-00705-7.

[315] N. et al et al., "Wireless closed-loop smart bandage for chronic wound management and accelerated tissue regeneration," *bioRxiv*, p. 2022.01.16.476432, Jan. 2022, doi: 10.1101/2022.01.16.476432.

[316] S. He et al., "Closed-Loop Deep Brain Stimulation for Essential Tremor Based on Thalamic Local Field Potentials," *Movement Disorders*, vol. 36, no. 4, pp. 863–873, Apr. 2021, doi: 10.1002/MDS.28513.

[317] A. Segato, A. Marzullo, F. Calimeri, and E. De Momi, "Artificial intelligence for brain diseases: A systematic review," *APL Bioeng*, vol. 4, no. 4, 2020, doi: 10.1063/5.0011697.

[318] M. Zhang et al., "A One-Shot Learning, Online-Tuning, Closed-Loop Epilepsy Management SoC with 0.97μJ/Classification and 97.8% Vector-Based Sensitivity," *IEEE Symposium on VLSI Circuits, Digest of Technical Papers*, vol. 2021-June, Jun. 2021, doi: 10.23919/VLSICIRCUITS52068.2021.9492429.

[319] G. O'Leary, D. M. Groppe, T. A. Valiante, N. Verma, and R. Genov, "NuriP: Neural interface processor for brain-state classification and programmable-waveform neurostimulation," *IEEE J Solid-State Circuits*, vol. 53, no. 11, pp. 3150–3162, 2018, doi: 10.1109/JSSC.2018.2869579.

[320] Y. Wang et al., "A Closed-Loop Neuromodulation Chipset with 2-Level Classification Achieving 1.5-Vpp CM Interference Tolerance, 35-dB Stimulation Artifact Rejection in 0.5ms and 97.8%-Sensitivity Seizure Detection," *IEEE Trans Biomed Circuits Syst*, vol. 15, no. 4, pp. 802–819, Aug. 2021, doi: 10.1109/TBCAS.2021.3102261.

[321] A. R. Aslam, T. Iqbal, M. Aftab, W. Saadeh, and M. A. Bin Altaf, "A10.13uJ/classification 2-channel Deep Neural Network-based SoC for Emotion Detection of Autistic Children," *Proceedings of the Custom Integrated Circuits Conference*, vol. 2020-March, Mar. 2020, doi: 10.1109/CICC48029.2020.9075952.

[322] S. Y. Chang et al., "An ultra-low-power dual-mode automatic sleep staging processor using neural-network-based decision tree," *IEEE Transactions on Circuits and Systems I: Regular Papers*, vol. 66, no. 9, pp. 3504–3516, Sep. 2019, doi: 10.1109/TCSI.2019.2927839.

[323] T. Dewolf, T. C. Stewart, J.-J. Slotine, and C. Eliasmith, "A spiking neural model of adaptive arm control", doi: 10.1098/rspb.2016.2134.

[324] Z. Bing, C. Meschede, F. Röhrbein, K. Huang, and A. C. Knoll, "A survey of robotics control based on learning-inspired spiking neural networks," *Front Neurorobot*, vol. 12, p. 35, 2019, doi: 10.3389/FNBOT.2018.00035/BIBTEX.

[325] D. Gamez, A. K. Fidjeland, and E. Lazdins, "iSpike: a spiking neural interface for the iCub robot," *Bioinspir Biomim*, vol. 7, no. 2, p. 025008, May 2012, doi: 10.1088/1748-3182/7/2/025008.

[326] E. Capecci, Z. G. Doborjeh, N. Mammone, F. La Foresta, F. C. Morabito, and N. Kasabov, "Longitudinal study of Alzheimer's disease degeneration through EEG data analysis with a NeuCube spiking neural network model," *Proceedings of the International Joint Conference on Neural Networks*, vol. 2016-October, pp. 1360–1366, Oct. 2016, doi: 10.1109/IJCNN.2016.7727356.

[327] S. Ghosh-Dastidar and H. Adeli, "Improved spiking neural networks for EEG classification and epilepsy and seizure detection," *Integr Comput Aided Eng*, vol. 14, no. 3, pp. 187–212, Jan. 2007, doi: 10.3233/ICA-2007-14301.

[328] H. Wang et al., "An Approach of One-vs-Rest Filter Bank Common Spatial Pattern and Spiking Neural Networks for Multiple Motor Imagery Decoding," *IEEE Access*, vol. 8, pp. 86850–86861, 2020, doi: 10.1109/ACCESS.2020.2992631.

[329] P. Knag, J. K. Kim, T. Chen, and Z. Zhang, "A Sparse Coding Neural Network ASIC With On-Chip Learning for Feature Extraction and Encoding," *IEEE J Solid-State Circuits*, vol. 50, no. 4, pp. 1070–1079, Apr. 2015, doi: 10.1109/JSSC.2014.2386892.

[330] J. Seo, B. Brezzo, Y. Liu, … B. P.-2011 I. C., and U. 2011, "A 45nm CMOS neuromorphic chip with a scalable architecture for learning in networks of spiking neurons," *ieeexplore.ieee.org*, 2011, Accessed: Jun. 12, 2022. [Online]. Available: https://ieeexplore.ieee.org/abstract/document/6055293/

[331] Y. Wu, X. Wang, W. D. Lu, J. M. Cruz-Albrecht, T. Derosier, and N. Srinivasa, "A scalable neural chip with synaptic electronics using CMOS integrated memristors," *Nanotechnology*, vol. 24, no. 38, p. 384011, Sep. 2013, doi: 10.1088/0957-4484/24/38/384011.

[332] N. K. Kasabov, "NeuCube: A spiking neural network architecture for mapping, learning and understanding of spatio-temporal brain data," *Neural Networks*, vol. 52, pp. 62–76, Apr. 2014, doi: 10.1016/J.NEUNET.2014.01.006.

[333] S. Budhraja, B. Bhattacharya, … S. D.-… J. C. on, and undefined 2020, "Sleep Stage Classification using NeuCube on SpiNNaker: a Preliminary Study," *ieeexplore.ieee.org*, 2020, Accessed: Jun. 12, 2022. [Online]. Available: https://ieeexplore.ieee.org/abstract/document/9207369/

[334] K. Kumarasinghe, M. Owen, D. Taylor, N. Kasabov, and C. Kit, "FaNeuRobot: A Framework for Robot and Prosthetics Control Using the NeuCube Spiking Neural Network Architecture and Finite Automata Theory," *Proc IEEE Int Conf Robot Autom*, pp. 4465–4472, 2018, doi: 10.1109/ICRA.2018.8460197.

[335] F. Corradi and G. Indiveri, "A Neuromorphic Event-Based Neural Recording System for Smart Brain-Machine-Interfaces," *IEEE Trans Biomed Circuits Syst*, vol. 9, no. 5, pp. 699–709, Oct. 2015, doi: 10.1109/TBCAS.2015.2479256.

[336] F. Boi *et al.*, "A bidirectional brain-machine interface featuring a neuromorphic hardware decoder," *Front Neurosci*, vol. 10, Dec. 2016, doi: 10.3389/FNINS.2016.00563.
[337] D. J. Guggenmos *et al.*, "Restoration of function after brain damage using a neural prosthesis," *Proc Natl Acad Sci U S A*, vol. 110, no. 52, pp. 21177–21182, Dec. 2013, doi: 10.1073/PNAS.1316885110.
[338] M. Ronchini *et al.*, "A CMOS-based neuromorphic device for seizure detection from LFP signals," *J Phys D Appl Phys*, vol. 55, no. 1, p. 014001, Oct. 2021, doi: 10.1088/1361-6463/AC28BB.
[339] J. H. Park, J. S. Y. Tan, H. Wu, Y. Dong, and J. Yoo, "1225-Channel Neuromorphic Retinal-Prosthesis SoC with Localized Temperature-Regulation," *IEEE Trans Biomed Circuits Syst*, vol. 14, no. 6, pp. 1230–1240, Dec. 2020, doi: 10.1109/TBCAS.2020.3036091.
[340] B. S. Mashford, A. J. Yepes, I. Kiral-Kornek, J. Tang, and S. Harrer, "Neural-network-based analysis of EEG data using the neuromorphic TrueNorth chip for brain-machine interfaces," *IBM J Res Dev*, vol. 61, no. 2–3, 2017, doi: 10.1147/JRD.2017.2663978.
[341] E. Nurse, B. S. Mashford, A. J. Yepes, I. Kiral-Kornek, S. Harrer, and D. R. Freestone, "Decoding EEG and LFP signals using deep learning: Heading truenorth," *2016 ACM International Conference on Computing Frontiers - Proceedings*, pp. 259–266, May 2016, doi: 10.1145/2903150.2903159.
[342] M. D. Murphy, D. J. Guggenmos, D. T. Bundy, and R. J. Nudo, "Current challenges facing the translation of brain computer interfaces from preclinical trials to use in human patients," *Front Cell Neurosci*, vol. 9, no. JAN2016, Jan. 2016, doi: 10.3389/FNCEL.2015.00497/FULL.
[343] L. F. Nicolas-Alonso and J. Gomez-Gil, "Brain computer interfaces, a review," *Sensors*, vol. 12, no. 2, pp. 1211–1279, 2012, doi: 10.3390/s120201211.
[344] M. J. Vansteensel and B. Jarosiewicz, "Brain-computer interfaces for communication," *Handb Clin Neurol*, vol. 168, pp. 67–85, Jan. 2020, doi: 10.1016/B978-0-444-63934-9.00007-X.
[345] S. Nagel and M. Spüler, "World's fastest brain-computer interface: Combining EEG2Code with deep learning," *PLoS One*, vol. 14, no. 9, Sep. 2019, doi: 10.1371/JOURNAL.PONE.0221909.
[346] F. D. Broccard, S. Joshi, J. Wang, and G. Cauwenberghs, "Neuromorphic Neural Interfaces", doi: 10.1007/978-981-15-2848-4_41-1.
[347] S. Soman, jayadeva, and M. Suri, "Recent trends in neuromorphic engineering," *Big Data Analytics 2016 1:1*, vol. 1, no. 1, pp. 1–19, Dec. 2016, doi: 10.1186/S41044-016-0013-1.
[348] J. Hasler and B. Marr, "Finding a roadmap to achieve large neuromorphic hardware systems," *Front Neurosci*, vol. 7, no. 7 SEP, pp. 1–29, 2013, doi: 10.3389/fnins.2013.00118.
[349] J.-Q. Yang *et al.*, "Neuromorphic Engineering: From Biological to Spike-Based Hardware Nervous Systems," *Advanced Materials*, vol. 32, no. 52, p. 2003610, Dec. 2020, doi: 10.1002/ADMA.202003610.
[350] J. Yoo and M. Shoaran, "Neural interface systems with on-device computing: machine learning and neuromorphic architectures," *Curr Opin Biotechnol*, vol. 72, pp. 95–101, Dec. 2021, doi: 10.1016/J.COPBIO.2021.10.012.
[351] M. Ronchini, Y. Rezaeiyan, M. Zamani, G. Panuccio, and F. Moradi, "NET-TEN: a silicon neuromorphic network for low-latency detection of seizures in local field potentials," Oct. 2022, doi: 10.48550/arxiv.2210.10565.
[352] D. S. Bassett and O. Sporns, "Network neuroscience," *Nature Neuroscience 2017 20:3*, vol. 20, no. 3, pp. 353–364, Feb. 2017, doi: 10.1038/nn.4502.
[353] Y. R. Tabar *et al.*, "At-home sleep monitoring using generic ear-EEG," *Front Neurosci*, vol. 17, Feb. 2023, doi: 10.3389/FNINS.2023.987578.
[354] S. W. Gangstad, "Ultra long-term subcutaneous EEG monitoring of brain functioning and disease," *Downloaded from orbit.dtu.dk on*. Technical University of Denmark, 2020. Accessed: Sep. 05, 2023. [Online]. Available: https://orbit.dtu.dk/en/publications/ultra-long-term-subcutaneous-eeg-monitoring-of-brain-functioning-
[355] E. Leuthardt, G. Schalk, … J. W.-J. of neural, and undefined 2004, "A brain–computer interface using electrocorticographic signals in humans," *iopscience.iop.orgEC Leuthardt, G Schalk, JR Wolpaw, JG Ojemann, DW MoranJournal of neural engineering, 2004•iopscience.iop.org*, 2004, doi: 10.1088/1741-2560/1/2/001.
[356] F. Kohler *et al.*, "Closed-loop interaction with the cerebral cortex: a review of wireless implant technology§," *http://dx.doi.org/10.1080/2326263X.2017.1338011*, vol. 4, no. 3, pp. 146–154, Jul. 2017, doi: 10.1080/2326263X.2017.1338011.
[357] J. Viventi, D. Kim, L. Vigeland, … E. F.-N., and undefined 2011, "Flexible, foldable, actively multiplexed, high-density electrode array for mapping brain activity in vivo," *nature.comJ Viventi, DH Kim, L Vigeland, ES Frechette, JA Blanco, YS Kim, AE Avrin, VR TiruvadiNature neuroscience, 2011•nature.com*, Accessed: Sep. 05, 2023. [Online]. Available: https://www.nature.com/articles/nn.2973
[358] E. H. Rijnbeek, N. Eleveld, and W. Olthuis, "Update on peripheral nerve electrodes for closed-loop neuroprosthetics," *Front Neurosci*, vol. 12, no. MAY, May 2018, doi: 10.3389/FNINS.2018.00350/FULL.